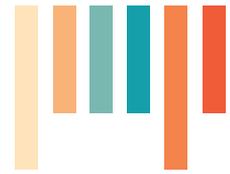

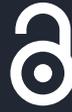

# Einstein Against Singularities: Analysis versus Geometry

**JOHN D. NORTON**

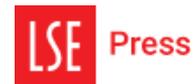

## ABSTRACT

Einstein identified singularities in spacetimes, such as at the Schwarzschild radius, where later relativists only find a coordinate system assigning multiple values to a single spacetime event. These differing judgments derive from differences in mathematical methods. Later relativists employ geometrical structures to correct anomalies in the coordinate systems used in analytic expressions. Einstein took the analytic expressions to be primary and the geometrical structures as mere heuristics that could be overruled if physical assumptions required it. Einstein's non-geometric methods had a firm base in the history of mathematical methods. They continued the non-geometric orientation of Christoffel, Ricci and Levi-Civita. Einstein's insistence that singularities must be eliminated marked a departure from earlier tolerance of singularities. It was founded upon his long-term project of eliminating arbitrariness from fundamental physical theories. However, Einstein was willing to theorize with singularities only temporarily if they were the least arbitrary approach then available.

**CORRESPONDING AUTHOR:**

**John D. Norton**

Department of History and Philosophy of Science, University of Pittsburgh, US

jdnorton@pitt.edu

# 1. INTRODUCTION

Einstein's treatment of spacetime singularities in his general theory of relativity has presented significant challenges to later commentators. His work elicits both admiration and exasperation. The work is admired for creating a new realm of physical theory within which a century's work on gravity and cosmology would unfold. It is exasperating for its treating of singularities in ways that later relativists judge to be novice errors. It is all the more exasperating since this treatment persisted even after the problems were pointed out to Einstein. John Earman and Jean Eisenstaedt's (1999) survey give the most thorough recounting and appraisal of Einstein's work on singularities. They summarize the tensions as:

> … Einstein was forced to fight a number of skirmishes with singularities, and more often than not he lost. The losses exhibit a strange asymmetry: he was unreasonably disturbed by (what today we would classify as) apparent singularities to the extent that he was prevented from embracing some of the most revolutionary consequences of his theory; but on the other hand he was not at all concerned about (what today we would call) real singularities, and as a result he pursued for over twenty years an ill-founded research programme on the problem of motion (186).

Their puzzlement persists throughout their narrative:

> …his 1935 paper with Rosen may strike the modern reader as bizarre and nearly incomprehensible (230).

It was not just Einstein. Even David Hilbert, under Einstein's sway, was apparently unable to discern when singularities are merely artefacts of a singular coordinate transformation:

> How Hilbert, one of the great mathematical minds of the century, could have failed to appreciate this elementary point defies rational explanation (193).

Klaas Landsman, in his survey of the foundations of general relativity, echoes their concern:

> Even Hilbert and Einstein were initially confused about the meaning of these apparent or real singularities, but today it is clear that $r = 2m$ [Schwarzschild radius] and $r = \rho$ [mass horizon] are just singularities of the coordinate systems in which the Schwarzschild and de Sitter solution are expressed (2021, 125).

This paper will seek to show that Einstein's treatment of spacetime singularities conformed with a consistent program of research. It may now appear willfully capricious, but only because the presumptions of Einstein's program are no longer ours. His analysis was not an instance of baffling mathematical incompetence. Rather, it employed mathematical methods different from those we now use and that now appear opaque to us. They were, in turn, guided by physical assumptions that were, in their time, cogent but are now discarded, often for solid empirical reasons. The ultimate failure was not a consequence of arbitrary opportunism, but the reverse: a dogged adherence to methods and assumptions that failed.

For concreteness, this paper will focus on two apparent anomalies in Einstein's treatments of spacetime singularities:

*1. Einstein's Apparent Misidentification of Mere Coordinate Singularities. The Wedge.*

If a coordinate system is poorly adapted to the spacetime geometry, quantities expressed as functions of the coordinates may become singular. They are not true pathologies of the spacetime geometry but merely artefacts of a coordinate system that has "gone bad" and can be remedied by adopting a new coordinate system. Three prominent cases in Einstein's work are his identification of singularities in a uniformly accelerated coordinate system in a Minkowski spacetime; at the "$r = 2m$" Schwarzschild radius of a Schwarzschild spacetime; and what was then called the "mass horizon" in a de Sitter spacetime. They all have the same form: Coordinate systems fail since they assign multiple coordinate values to the same event in a construction that appears as a wedge-like formation in a spacetime diagram.

*2. Einstein's Vacillations Over the Formal Admissibility of Spacetime Singularities.*

At times, Einstein portrayed spacetime singularities in calamitous terms. He wrote (Einstein and Rosen, 1935, 73) "For a singularity brings so much arbitrariness into the theory that it actually nullifies its laws." He was even willing to modify his celebrated gravitational field equations to eliminate them or to argue that physical processes prohibit their formation. Yet, on other occasions he entertained field singularities as models for light quanta; as essential to understanding de Sitter's solution to his $\lambda$-augmented gravitational field equations; and as a means to derive the equations of motion of a free particle from his gravitational field equations.

Einstein (1949, 684) famously described himself as a "type of unscrupulous opportunist" when compared to the systematic epistemologist. The temptation is to extend the description to his program in physics. The goal of this paper is to argue otherwise. There is, it will be argued, a deeper coherence underlying these apparent anomalies. It derives from two elements in Einstein's thinking. One is methodological. The other is an enduring and controlling heuristic within his worldview.

*1. Einstein privileged analytic expressions over synthetic geometry.*

It has long been recognized that Einstein was not a geometrical thinker, after the modern manner.[1] His privileging of analytic expressions was not a personal aberration. There were, in Einstein's time, two distinct mathematical traditions, which will be described in Section 6 below. One was a synthetic, geometrical tradition emerging from Gauss and Riemann's theory of curved surfaces. The other was analytic and derived from Christoffel and Riemann's work on the invariants of quadratic differential forms. In formulating his general theory of relativity Einstein aligned with the second, analytic tradition.

While both these traditions contributed to the mathematics of Einstein's general theory of relativity, they differ in which are the primary objects of study.

- For synthetic geometry in the tradition of Gauss, the primary object of study is the geometric surface or space. Quadratic differential forms are a means of describing their properties.
- For the methods in the analytic tradition of Christoffel, the primary objects of study are mathematical expressions or formulae, that is, quadratic differential forms defined over variables and their transforms.

That quadratic differential forms could be used in geometry was an incidental application of the analytic methods. It was a fact to be acknowledged but not allowed to control the method.

---

1 See, for example, Norton (1993, §2), Reich (1994, 210).

These differences were generally unimportant in Einstein's work on general relativity. The pertinent exception came when spacetime singularities arose from a misbehaving coordinate system. The synthetic, geometry tradition can escape them. The analytic tradition struggles with them. For each "wedge" example above, the pathology in the coordinate system is that, geometrically understood, it assigns *multiple* values to a single event. The new, regular coordinate system must assign a single value to each *single* event. That means that the new coordinate system cannot be generated from the pathological coordinate system by a one-to-one mapping. This presents no difficulty of principle for the synthetic, geometric tradition. The intrinsic geometry of the space guides construction of the new coordinate system. It does not need and cannot use a one-to-one mapping from the old coordinate systems.

That a one-to-one mapping cannot be used is, however, troublesome for the analytic approach within which Einstein worked. Its basic objects of study are analytic expressions and their one-to-one, differentiable transformations. Once an expression has a "wedge" type singularity in one set of variables, none of the admissible transformations can eradicate it. The analytic expressions are irreparably singular. The difficulty can be escaped in this analytic tradition only by weakening this basic conception of the primary objects of study. Einstein was, on occasion, willing to make the compromise. We shall see that, in a natural concession to the geometry, he allowed that the irregularities at the origin of polar coordinate systems are not pathological. However, absent the motivation of an underlying intrinsic geometry, he clearly felt no compulsion to make this same concession in other cases. Of course, the geometric picture has heuristic value. What would decide whether it was to be used was Einstein's non-geometric, physical interpretation of his equations. Prominent amongst these was an expectation that the analytic expressions are static in their time coordinate. These physical interpretations fitted better, we shall see, with the singularities.

It is surely surprising to modern relativists that Einstein would forgo the geometric picture that now proves so fertile. Here Dennis Lehmkuhl's (2014) study provides an essential recalibration. It documents how Einstein was dismissive and even scornful of the now familiar idea that general relativity had "geometrized" gravity. In correspondence in 1926, he even disparaged geometrical conceptions as an *Eselsbrücke* ["donkey bridge"], that is, roughly speaking an artificial crutch, a convenience for novices.

We now turn to the second element in Einstein's work.

### 2. Einstein used the elimination of arbitrariness as a guide in theory construction.

That we recover better theories by eliminating arbitrariness was an enduring heuristic that figured prominently in Einstein's opposition to spacetime singularities. While the heuristic had played a part in much of Einstein's earlier physical theorizing, its presence was easy to overlook, perhaps even by Einstein himself, since it found expression in many different forms, according to the physical problem addressed. It drove his search for the most general relativity of motion. Anything less required us to designate arbitrarily which is the ether state of rest or which are the inertial frames of reference. Indeterministic quantum mechanics introduced arbitrariness not present in deterministic theories. The ultimate physics, Einstein foretold, is one without any arbitrary constants.

We are now apprehensive of spacetime singularities, since they arise in extreme regimes where we expect our present theories to fail. Prior to Einstein's deprecation of them, their presence in physical theories was generally untroubling and required only a brief mention in passing. Einstein, however, made their elimination a core requirement of his long-sought unified field theory. His aversion to singularities was unlike our modern apprehensions. Rather he likened singularities to boundary conditions in field theories. They are elements that can be chosen freely. It is a harmful freedom since it expands the set of solutions of a theory's fundamental equations in a way that merely reflects our arbitrary choices. For this reason, in 1917, he rejected a cosmology that stipulates that special relativity obtains at spatial infinity. Admitting singularities, Einstein feared, allowed similar, unjustified freedoms in theorizing.

While finding a final theory without arbitrariness was Einstein's ultimate goal, his day-to-day work sought to move towards this goal by reducing arbitrariness wherever feasible, even if it could not be eliminated completely. General relativity, as originally formulated, had two independent laws: the gravitational field equation for the configuration of the metric field; and the geodesic law for the motion of free bodies or, more generally, an independent matter theory. To derive the geodesic law from the source-free gravitational field equations would eliminate the arbitrariness of two independent laws in favor of the lesser arbitrariness of one law. Such a derivation seemed within Einstein's reach if he represented particles as singularities in a source free field. It was a temporary expedient that replaced a greater by a lesser arbitrariness.

There are many more facets to Einstein's engagement with spacetime singularities than can be covered here. This paper is limited to reporting just those fragments of the history that bear directly on the two elements described above. A fuller accounting of the history may be found in the existing literature, such as the synoptic account of Earman and Eisenstaedt (1999) and in the editorial notes to Einstein's correspondence of 1917 and 1918 in Schulmann et al. (1998). Lehmkuhl (2017; 2019) provides an illuminating account of Einstein's earlier treatments of the problem of motion.

The three sections of Part I below deal directly with the three cases of singularities identified by Einstein, the modern reaction to them, and an account of how Einstein's views did not cohere with it. The related Appendix B defends with a simple example the viability of Einstein's prioritizing of analytic expressions. Part II provides an historical perspective on Einstein's mathematical methods. Sections 5 and 6 recount the mathematical tradition upon which Einstein drew. Section 7 recounts how this tradition was incorporated by Einstein and Grossmann into their early sketch of the general theory of relativity. Details of their novel use of the term "tensor" are in Appendix A. Section 8 reports how Einstein's suppression of geometrical notions appeared in his theorizing in general relativity. Section 9 of Part III recounts the presence of singularities in the literature prior to Einstein. Section 10 then reviews how Einstein's goal of eliminating arbitrariness from fundamental theories called for an elimination of singularities; and how his temporary use of singularities was a choice of the least arbitrary of the avenues available.

A final, concluding Section 11 summarizes the coherence of Einstein's program, but notes that it has not proven productive. The alternative of giving priority to the geometry is the basis of modern advances in both black hole physics and cosmology. These advances show that Einstein's program of research failed. It did not fail, however, because of elementary errors, but because Einstein, like the rest of us, cannot foresee what future research may bring.

# PART I: THREE SINGULARITIES

# 2. EINSTEIN'S SIMPLEST SPACETIME SINGULARITY

## 2.1 THE SINGULARITY

To grasp Einstein's understanding of spacetime singularities, we can do no better than to recount the elementary example that Einstein himself used. It was developed in his collaborative work with Rosen (Einstein and Rosen 1935) specifically for the purpose of explaining his view of the role of singularities in his physical theories. The example starts with the familiar line element of special relativity:

$$ds^2 = -d\xi_1^2 - d\xi_2^2 - d\xi_3^2 + d\xi_4^2, \tag{1}$$

where $\xi_1$, $\xi_2$, $\xi_3$ are the Cartesian coordinates for space and $\xi_4$ is the time coordinate, with $c = 1$. They introduce a uniformly accelerated coordinate system, adapted to the hyperbolic motion of uniform acceleration in the $+\xi_1$ direction, by the transformation:

$$\xi_1 = x_1 \cosh(\alpha x_4) \quad \xi_2 = x_2 \quad \xi_3 = x_3 \quad \xi_4 = x_1 \sinh(\alpha x_4). \tag{2}$$

Under this transformation, the line element (1) becomes:

$$ds^2 = -dx_1^2 - dx_2^2 - dx_3^2 + \alpha^2 x_1^2 dx_4^2. \tag{3}$$

Free bodies moving on geodesics have a non-zero coordinate acceleration in the -x1 direction. Recalling Einstein's principle of equivalence, this acceleration is to be interpreted as due to a homogeneous gravitational field acting in the $-x_1$ direction. Here is Einstein and Rosen's formulation of this interpretive principle:

> ... "Principle of Equivalence": If in a space free from gravitation a reference system is uniformly accelerated, the reference system can be treated as being "at rest," provided one interprets the condition of the space with respect to it as a homogeneous gravitational field (1935, 74).

Einstein and Rosen identify the hyperplane $x_1 = 0$ as "a singularity of the field" so that Einstein's source free gravitational field equations, $R_{kl} = 0$, for $R_{kl}$ the Ricci tensor, fail to be satisfied. This conclusion is justified by:

> ... the fact that the determinant of $g$ of the $g_{\mu\nu}$ vanishes for $x_1 = 0$. The contravariant $g^{\mu\nu}$ therefore become infinite and the tensors $R^i_{klm}$ and $R_{kl}$ take on the form 0/0. From the standpoint of Eqs.[(3)] the hyperplane $x_1 = 0$ then represents a singularity of the field (1935, 74).

They then offer an interpretation of the significance of the singularity. Einstein's original gravitational field equations, with stress-energy sources represented by the tensor $T_{ik}$, is

$$R_{ik} - (1/2)g_{ik}R = -T_{ik}. \tag{4}$$

To apply this equation to the line element (3), they consider a modification of the line element:

$$ds^2 = -dx_1^2 - dx_2^2 - dx_3^2 + (\alpha^2 x_1^2 + \sigma)dx_4^2.$$

Applying the field equations (4), the stress-energy tensor[2] can be computed and has non-zero terms they write as:

$$T_{22} = T_{23} = \alpha^2/\sigma/(1 + \alpha^2 x_1^2/\sigma)^2.$$

As $\sigma$ approaches zero, the original line element (3) is approached in the limit. In this limit, the non-zero terms of the stress-energy tensor exhibit a singularity at $x_1 = 0$. These terms diverge at $x_1 = 0$ but converge to zero for non-zero $x_1$.[3] This divergence is interpreted as:

> ... the solution [(3)] contains a singularity which corresponds to an energy or mass concentrated in the surface $x_1 = 0$ (1935).

## 2.2 THE MODERN REACTION

It surely understates matters to say that modern readers find this analysis "astonishing," to use Earman and Eisenstaedt's (1999, 215) word. The entirety of the results Einstein and Rosen report is due to artifacts of the coordinate system introduced by the transformation (2). A mere change of coordinates introduces no singularities in the spacetime geometry. It remains everywhere flat with vanishing curvature and Ricci tensor. All this is made transparent by plotting the new coordinate system in a spacetime diagram, shown in Figure 1, where coordinates $\xi_2$ and $\xi_3$ have been suppressed:

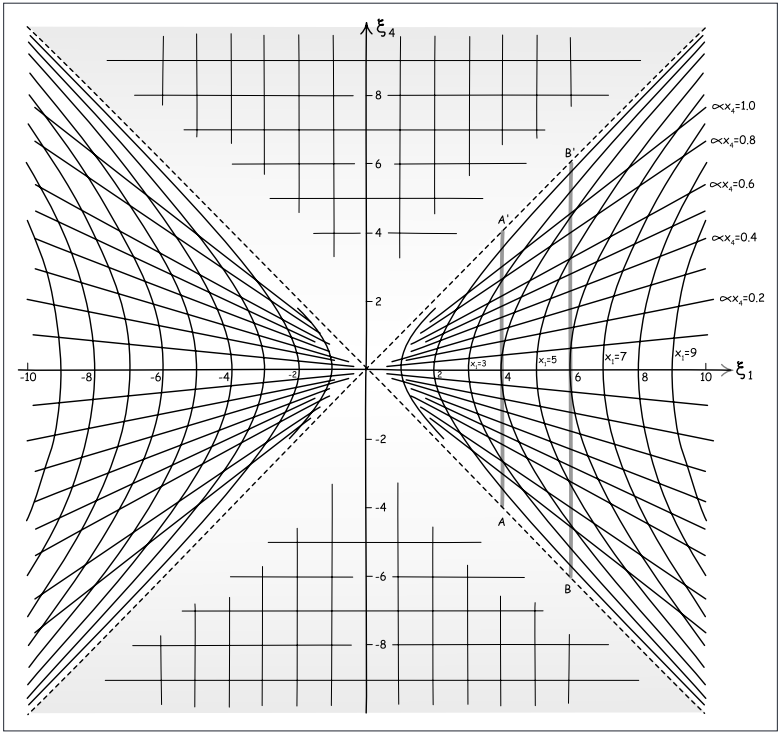

**Figure 1** Einstein and Rosen's accelerated coordinate system.

The diagram shows that the accelerated coordinate system "goes bad" at $x_1 = 0$. That is, for fixed $x_2 = \xi_2$ and $x_3 = \xi_3$, the infinitely many events $x_1 = 0$, $-\infty < x_4 < \infty$ are mapped, many-to-one to the single event at $\xi_1 = \xi_4 = 0$. The singularity Einstein and Rosen report is due entirely to this pathology of the coordinate system.

The diagram also shows the worldlines of two geodesics, $AA'$ and $BB'$. Since their worldlines have vanishing four-acceleration, that invariant fact remains so even after the

---

2    It is designated parenthetically as "fictitious."

3    The order of operations in the expression for $T_{22}$ as given by Einstein and Rosen is ambiguous. To recover the singularity claimed, the expression should be read as $\alpha^2/[\sigma(1 + \alpha^2 x_1^2/\sigma)^2] = \alpha^2/[(\sqrt{\sigma} + \alpha^2 x_1^2/\sqrt{\sigma})^2]$ we adopt a new coordinate system. The "falling" is an illusion arising as a coordinate effect. Another coordinate effect is that the falling point would require infinite $x_4$ coordinate time to complete its fall through all values of position $x_1 > 0$; and infinite $x_4$ coordinate time to rise through all values of position $x_1 > 0$. The coordinate surface $x_1 = 0$ seems to represent an unachievable horizon. It is a coordinate effect only, since both motions are completed in finite proper time.

## 2.3 WHY WE MUST TAKE EINSTEIN'S EXAMPLE SERIOUSLY

This modern appraisal now seems so natural and so at variance with Einstein and Rosen's claims that is easy to suppose that Einstein and Rosen were poorly informed or simply hasty or both. Surely even a brief acquaintance with modern views would have elicited a retraction from them.[4] This supposition must be resisted. There is ample evidence that the view presented was Einstein's well-considered and enduring view.

To begin, the example was not hastily chosen. It is the basis of his formulation of the principle of equivalence, which was, for Einstein, one of the most enduring foundational elements of his general theory of relativity. It initiated his first steps towards the theory in 1907 and remained unaltered throughout all his writings.[5] It was and remained at the core of Einstein's understanding of his theory.

Next, Einstein and Rosen's paper came two decades after the completion of Einstein's theory and after the nature of singularities in his theory had been closely examined by the physics community. The sorts of considerations we now consider as modern had been developed and understood. They were communicated to Einstein by correspondents. We shall see below that no lesser figure than Felix Klein had pointed out to Einstein the pathology of his coordinate system in the case of the mass horizon of the de Sitter solution. The considerations had also appeared in the physics literature from the pen of a figure known personally to Einstein. We shall see again that Lemaitre had shown that the singularity Einstein and Rosen presumed in a Schwarzschild solution was, likewise, an artefact of a pathology in the coordinate system. We may infer that Einstein chose not to adopt this understanding, but the idea that he did not know of it in 1935 is unsustainable.

Further, the notion that a singularity is a surrogate for matter had already been explored seriously in the same way in the context of the de Sitter solution, to be discussed below. With Einstein's assent and approval, Hermann Weyl had there sought to justify this interpretation of a singularity by the same artifice used by Einstein and Rosen in 1935. That is, he had sought to understand a singularity as the singular limit of a regular matter distribution. Einstein and Rosen's artifice was not a hasty, ad hoc expedient, but an application of approach already present in the literature. Finally, the physics and geometry of the example is transparently simple. Einstein and Rosen acknowledged explicitly that their new coordinate system covers only the two wedges shown in Figure 1. They wrote

> It is worth pointing out that this metric field [(1)] does not represent the whole of the Minkowski space but only a part of it ... only those points for which $\xi_1^2 \geq \xi_4^2$ correspond to points for which [(1)] is the metric (1935, 74).

---

4    Even if one does not know that the line element (3) was produced by a coordinate transformation from a familiar Minkowskian line element, Wald (1984, 149–153) shows how efforts to geodesically extend the line element (3) would lead back to the full Minkowski line element (1).

5    For a survey, see Norton (1985).

They clearly knew how their accelerated coordinate system is distributed over the Minkowski spacetime. They knew that the surfaces of constant $x_4$ intersect at $\xi_1 = \xi_4 = 0$. There, in those coordinates, the Riemann curvature tensor of the Minkowski spacetime vanishes. Einstein had emphasized that, if a tensor is a zero tensor in one coordinate system, it remains so for all. In his review article of 1916, written upon completion of the general theory, his synopsis of the "Mathematical Aids to the Formation of Generally Covariant Equations" included this fact as fundamental for the development of the laws of his theory:

> The things hereafter called tensors are further characterized by the fact that the equations of the transformation for their components are linear and homogeneous. Accordingly, all the components in the new system vanish, if they all vanish in the original system. If, therefore, a law of nature is expressed by equating all the components of a tensor to zero, it is generally covariant (1916a, 121).

In short, there was no simple error in the illustrative example given by Einstein and Rosen. It was transparent, well-informed, and quite purposefully chosen. They meant what they said.

## 2.4 WHAT EINSTEIN MEANT: ANALYTIC EXPRESSIONS OVER GEOMETRY

The key to discerning what Einstein meant is that his methods were not those of modern treatments. There are two elements in the treatments: the geometry intrinsic to a spacetime, and the analytic expressions that can be used to describe that geometry.

The modern approach privileges the geometry over the analytic expressions. When there is a mismatch, the analytic expressions are to be corrected against the geometry. Einstein and Rosen's accelerated coordinate system produces the line element (3) that fails adequately to represent the intrinsic geometry of the Minkowski spacetime at $x_1 = 0$. The geometry is upheld and the line element is regarded as defective at $x_1 = 0$. Its coordinates have "gone bad" there.

In Einstein's approach, the analytic expression (3) itself is primary. The Minkowski spacetime played a heuristic role in its construction. Einstein can draw on its geometry to discern the properties of the analytic expression. The geometry is pertinent only in so far as it serves that purpose. When there is a mismatch, such as at $x_1 = 0$, there is no automatic presumption that the analytic expression is at fault. Which of the geometry or analytic expressions requires a correction is a matter to be decided by the intended physical interpretation.

The principle of equivalence provided that interpretation: The line element (3) is to be associated with a homogeneous gravitational field in which free bodies undergo accelerated fall. From this perspective, the singularity identified at $x_1 = 0$ proves to be congenial. Einstein and Rosen's analysis interprets it as an energy or mass concentration. It would, presumably, serve as a natural source of the homogeneous field. That the singularity should be interpreted in this way fits with Einstein's larger research agenda. As we shall see below, these sorts of singularities were routinely interpreted as proxies for the matter distributions that a successful unified field theory would put in their place.

This alternative is only sustainable as long as Einstein and Rosen do not treat their accelerated coordinates as pathological at $x_1 = 0$. Rather, for fixed coordinate values $x_2$ and $x_3$, they must treat as distinct events all those with $x_1 = 0$, but differing $x_4$. To collapse all these events to a single event so that the original geometry of the Minkowski spacetime is respected would render untenable the identification of energy or mass at $x_1 = 0$.

It is quite incongruous with Einstein's mode of presentation to draw a spacetime diagram for this alternative interpretation. However, we can better picture Einstein and Rosen's interpretation if we see how such a diagram, given in Figure 2, differs from the familiar one of Figure 1.

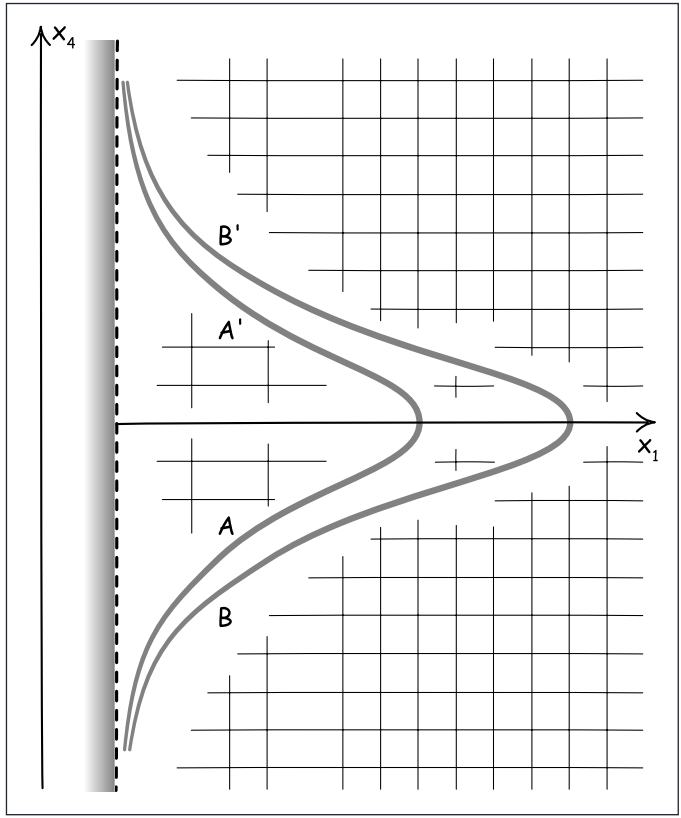

The figure shows the spacetime for $x_1 \geq 0$, with the singularity at $x_1 = 0$ represented not by a point but a line with all possible $x_4$ coordinates. The geodesics $\xi_1 = k$, for various constants $k$, of Figure 1 are now represented by bell-like curves. It follows from the transformation equations (2) that their trajectories in $x_1$, $x_4$ coordinates are

$$x_1 = k/\cosh(\alpha x_4).$$

A small amount of algebra shows that for small $x_4$ times, each trajectory exhibits, in close approximation, the expected parabola of free fall:

$$x_1 = k(1 - (1/2)(\alpha x_4)^2).$$

For large $x_4$ times, positive and negative, the trajectory approaches asymptotically the horizon at $x_1 = 0$ that is unattainable in $x_4$ times. This effect is the origin of the gravitational red shift Einstein first inferred from the principle of equivalence in 1907.

Einstein and Rosen do not mention this horizon. However, since it will return explicitly in later examples, we should note that there is no elementary confusion here. It is transparent from the construction of the example that a freely falling body will traverse the wedge $\xi_1^2 \geq \xi_4^2$ and leave it in finite proper time. It would be a complete novice folly to think otherwise. The horizon exists in the different sense that it is the limit of events covered by the accelerated coordinate system $x_i$. Motions described by the line element (3) using these coordinates will never reach it.

Once the independence of the line element (3) from the Minkowski spacetime geometry is recognized, it is no longer puzzling that Einstein and Rosen report that tensors $R^i_{klm}$ and $R_{kl}$ take on the singular form 0/0. For, they are not the tensors of the original line element (1), but of the new line element (3) that differs from (1) at $x_1 = 0$. That there is no one-to-one mapping between the original $x_i$ coordinates and the new $x_i$ coordinates is essential for the difference. For otherwise, their vanishing in one coordinate system would necessitate their vanishing in the other.

How can Einstein interpret the line element (3) as associated with a homogeneous gravitational field when its spacetime curvature for all $x_1 > 0$ remains flat, just like the Minkowski spacetime? The question raises issues that cannot be fully addressed here. Briefly, once again it is a matter of how the equations are to be interpreted.

When this objection was put to him by Laue in correspondence of 1950, Einstein replied that

> … what characterizes the existence of a gravitational field from the empirical standpoint is the non-vanishing of the [non-generally covariant coefficients of the connection] $\Gamma^l_{ik}$, not the non-vanishing of the [Riemann curvature tensor] $R_{iklm}$.[6]

Thus, the line element (1) with vanishing $\Gamma^l_{ik}$, is interpreted as gravitation-free, whereas the line element (3) with non-vanishing $\Gamma^l_{ik}$, harbors a gravitational field.

What is noticeable here and in all of Einstein's writings on general relativity is that it is not geometrical in its expressions. Where the geometrical approach talks of a "Minkowski spacetime" and a "Schwarzschild spacetime," Einstein's analytic approach talks of "solutions" to equations, where the term designates analytic expressions such as (1) and (3). The term "solution" appears 40 times in the five pages of Einstein and Rosen's paper. The term "space-time" and "space" each appear only once.

## 3. SINGULARITY AT THE SCHWARZSCHILD RADIUS

This last example serves Einstein's and our purpose well. For, Einstein's understanding of the more complicated examples is the same, but just harder to see because the physics is less transparent. We shall see this in two cases: the singularity in the Schwarzschild solution in this section; and the "mass singularity" in the de Sitter solution in the next section.

### 3.1 THE RADIUS AS A SINGULARITY

Einstein and Rosen (1935, 75) wrote the Schwarzschild solution in a form that Eisenstaedt (1989, 216) identified as using "Droste coordinates" after the version of the solution given in Droste (1916):

> As is well known, Schwarzschild found the spherically symmetric static solution of the gravitational equations

$$ds^2 = -\frac{1}{(1 - \frac{2m}{r})}dr^2 - r^2(d\theta^2 + \sin^2\theta d\phi^2) + (1 - \frac{2m}{r})dt^2 \qquad (5)$$

---

6    See Norton (1985, §11) and Lehmkuhl (2022, §9.3.6) for details; and for further discussion, Section 8.2 below.

($r > 2m$, $\theta$ from 0 to $\pi$, $\phi$ from 0 to $2\pi$); ... The vanishing of the determinant of the $g_{\mu\nu}$ for $\theta = 0$ is unimportant, since the corresponding (spatial) direction is not preferred. On the other hand, $g_{11}$ for $r = 2m$ becomes infinite and hence we have there a singularity.

The character of the singularity is elaborated in Einstein's later paper that argues against the physical possibility of gravitational collapse beyond the Schwarzschild radius. He employed an isotropic coordinate system to present the Schwarzschild solution:

If one considers Schwarzschild's solution of the static gravitational field equations of spherical symmetry

(1)
$$ds^2 = -(1 + \frac{\mu}{2r})^4(dx_1^2 + dx_2^2 + dx_3^2) + \left(\frac{1 - \frac{\mu}{2r}}{1 + \frac{\mu}{2r}}\right)^2 dt^2$$

it is noted that

$$g_{44} = \left(\frac{1 - \frac{\mu}{2r}}{1 + \frac{\mu}{2r}}\right)^2$$

vanishes for $r = \mu/2$. This means that a clock kept at this place would go at the rate zero. Further it is easy to show that both light rays and material particles take an infinitely long time (measured in "coördinate time") in order to reach the point $r = \mu/2$ when originating from a point $r > \mu/2$. In this sense the sphere $r = \mu/2$ constitutes a place where the field is singular. ($\mu$ represents the gravitating mass.) (1939, 922).

These remarks from the 1930s reflect what had already become a standard view amongst physicists after the completion of Einstein's theory in 1915 and Schwarzschild's presentation of his solution. There is a singularity at the Schwarzschild radius "$r = 2m$." In his influential, two-part memo, "Foundations of Physics," Hilbert (1917, 70) noted the irregularity in the solution. In his *Teubner Encyklopädie* article, Pauli (1921, 728) also mentioned it in passing; as did Laue (1921, 215) in his relativity textbook.

## 3.2 THE MODERN REACTION

This standard view did not persist. Lemaître (1933, 52, 82) showed how the singularity could be eradicated by a change of the coordinate system and was thus "fictive."[7] Almost three decades passed before the idea of eliminating the singularity by a mere coordinate transformation had attracted enough of a following to enter the mainstream. Finkelstein (1958), Szekeres (1959), and Kruskal (1960) showed how a single coordinate system could be found that covered the entire Schwarzschild solution such that there is no singularity at the Schwarzschild radius.

We can plot the Droste coordinates of Einstein's Schwarzschild's solution (5) in a diagram of the fully extended Schwarzschild spacetime based on Kruskal-Szekeres coordinates, as shown in Figure 3 below, with two angle coordinates suppressed. Kruskal (1960) replaced the $t$ and $r$ coordinates of (5) with coordinates $v$ and $u$ such that light propagates along the diagonals $dv/du = \pm 1$. (The trajectories of light are represented by the cross-hatching with diagonal lines in Figure 3.)

---

7    For a recounting of Lemaître's analysis, see Eisenstaedt (1993).

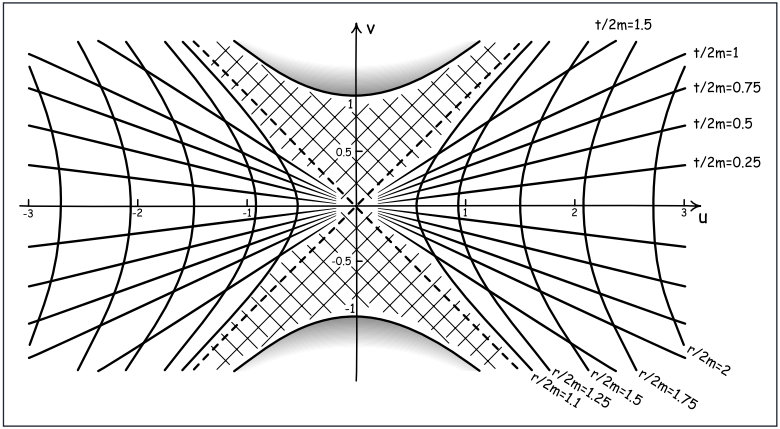

**Figure 3** Droste coordinate curves in the fully extended Schwarzschild spacetime in Kruskal coordinates.

The Droste coordinates cover a wedge in the figure and, as Einstein and Rosen recognized, a duplicate set of Droste coordinates cover a second wedge on the other side of the figure. There are further parts of the fully extended Schwarzschild solution not covered by these two Droste coordinate systems. They are the regions inside the Schwarzschild radius. They are bounded above and below by a future and a past curvature singularities at $v^2 - u^2 = 1$.

The modern reaction to Einstein and Rosen's identification of a singularity at the Schwarzschild radius can be read directly from the figure. The Droste coordinates map many values of the coordinate $t$ to a single event at the apex of the wedges. There is no singularity at that event in the spacetime; there is only a pathology of the coordinate system. That the spacetime is regular at this radius is commonly affirmed by noting, as does Kruskal (1960, 1743), that "the curvature invariants of the Schwarzschild metric are perfectly finite and well behaved at [$r = 2m$]."

There is a remarkable similarity to the accelerated coordinate system of Einstein and Rosen (1935) in a Minkowski spacetime. Both cover only a wedge of the full space; both have a special place for hyperbolas, which are invariant figures for hyperbolic rotations; and both have a coordinate pathology at the apex of the wedge. There, both coordinate systems map many values of the "$t$" coordinate to a single spacetime event. The analogy between the two cases was explored by Rindler (1977, §8.6) and became an important stimulus to further work on black holes and their horizons. For this reason, the wedge is often called a "Rindler wedge" and Einstein and Rosen's accelerated coordinates "Rindler coordinates."

### 3.3 WHAT EINSTEIN MEANT

The modern view privileges the spacetime geometry over the analytic expressions. From this perspective, Einstein's assertions about the Schwarzschild solution are incomprehensible. He has mistaken a pathology of his coordinate system for a true singularity of the spacetime geometry. This was not Einstein's view. He privileged the analytic expression over the geometry. There is no one-to-one mapping that takes the quadratic differential form (5) into the corresponding quadratic differential form for the Schwarzschild spacetime in Kruskal coordinates. One has to abandon such mappings and use the geometry of the fully extended Schwarzschild spacetime to correct the Droste coordinates. While that is the modern procedure, it was not Einstein's. He was not willing to compromise the quadratic differential form (5) by privileging the geometry and using it to eliminate the singularity.

Once we recognize that Einstein's remarks about the Schwarzschild solution refer specifically to this quadratic different form (5), they are not confusions, but quite consistent. First, we saw above that Einstein (1939, 922) remarked that a clock kept at the Schwarzschild radius "would go at a rate zero." That is read trivially from the vanishing of the "$g_{44}$" component the metric of (5). It also conforms with the spacetime geometry. For a clock, kept at the Schwarzschild radius in Droste coordinates, would be kept at a single event in the spacetime, where no proper time elapses, even though many different values of the time coordinate $t$ are assigned to the event.

Second, we have Einstein's remark (1939, 922) "that both light rays and material particles take an infinitely long time (measured in 'coördinate time') in order to reach [Schwarzschild radius]." This is not a confused assertion that these systems can never reach the Schwarzschild radius. Einstein is not confusing an infinite coordinate time with an infinity of proper time. There can be no doubt of that since Einstein explicitly specifies "coordinate time." The distinction is foundational in the theory he devised and he is, among all physicists, the least likely to confuse the two. Rather, Einstein is pointing out a limitation in the reach of the quadratic differential form (5). Its time coordinate $t$ can only cover motions outside the Schwarzschild radius. While it is a natural project to ask how the motion might continue past the reach of the coordinate system, it is one that likely did not interest Einstein. His 1939 paper argued that no spacetime could form with a bare Schwarzschild radius. Bodies falling towards it would first strike source matter.

This expectation about the limit of motions physically realizable by the Schwarzschild solution was not a novelty of 1939. Einstein had long presumed it. We learn from reporting by Charles Nordmann (1922, 154–156) that the presumption had figured prominently in discussions about relativity theory when Einstein had made a visit to the Collège de France in Paris.[8] In discussion on April 5, 1922, Hadamard has asked Einstein what would happen if a mass were condensed enough so that its exterior field included the singular Schwarzschild radius. Einstein seemed to be taken aback by the suggestion. "It would be," Nordmann quoted Einstein as saying, "an unimaginable misfortune [*malheur*] for theory…" The quote continued with Einstein seeming to suggest that even in this case the Schwarzschild solution would break down: "…it is very difficult to say what would happen physically, for then the formula ceases to be applicable." Nordmann continued his report that Einstein light-heartedly called the problem the "Hadamard catastrophe." It seems that Einstein had then simply assumed that a bare Schwarzschild radius was impossible, but he had not really worked through the physics needed to sustain its impossibility. For, he was otherwise unable to give a more cogent response. He needed time. That response came in the next meeting of April 7. Einstein argued that a mass, sufficiently condensed to bring about the Hadamard catastrophe, was physically impossible. He recalled Schwarzschild's earlier calculation that the pressure inside the mass would become infinite at the center, before the Hadamard catastrophe was realized, and also that clocks there would halt.

Once again, contrary to Einstein's methods, we can draw a spacetime diagram that represents how Einstein conceived the quadratic differential form (5). On the right, Figure 4 shows the many, distinct events it attributes at the Schwarzschild radius for the differing values of the time coordinate $t$. The left of the figure shows corresponding part of the Schwarzschild spacetime in Kruskal coordinates.

Both show the worldline of a body that maintains a constant $r/2m = 1.75$ coordinate position in the Droste coordinates of (5). The worldline is vertical in the Droste coordinates

---

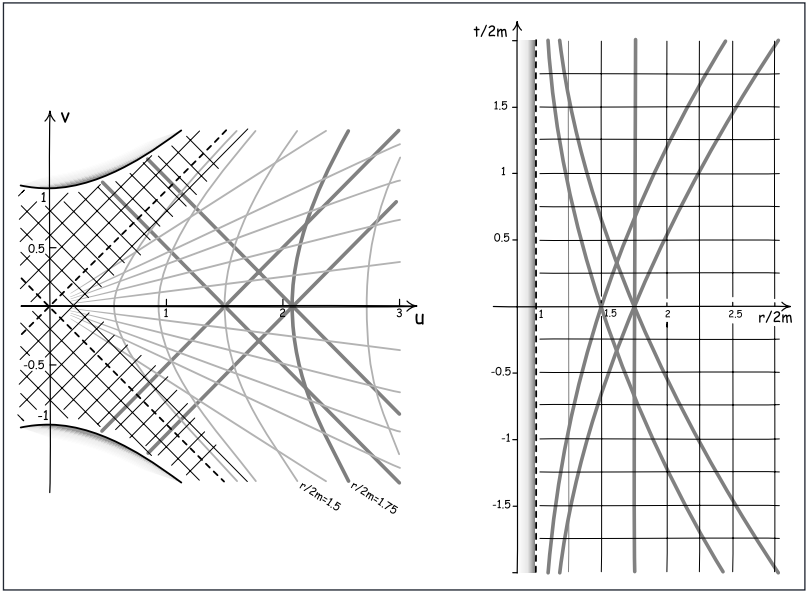

shown on the right and a hyperbola in Kruskal coordinates shown on the left. The worldlines of light propagating towards and away from the $r = 2m$ radius are shown on the right in Droste coordinates. They pass through events $t = 0$, $r/2m = 1.5, 1.75$. Both of the worldlines are asymptotic to the Schwarzschild radius $r = 2m$. The same trajectories are shown on the left in Kruskal coordinates as straight, diagonal lines. These fully extended trajectories pass through the Schwarzschild radius, shown as dashed diagonal lines, and continue to contact either the past or future singularity.

## 3.4 EINSTEIN'S DEFINITION OF A SINGULARITY

Einstein's remarks in these two papers (1935; 1939) enable us to identify more precisely what he meant by a singularity in these contexts.

First, we can say what his conception is not. It is not the modern conception of a failure of the intrinsic geometry, such as the divergence of curvature invariants. Is it a zero-valued component of the metric tensor? In (5) we have a zero-valued metrical component, $g_{44} = (1 - 2m/r) = 0$ at $r = 2m$, the events that Einstein and Rosen identify as singular. However, we shall see in Einstein and Rosen's treatment of their "bridge" that the mere fact of a zero-valued component of the metric tensor is insufficient for a singularity. This is not unexpected. A vanishing component of the metric tensor may be completely benign, as is commonly the case with off-diagonal components. Vanishing diagonal components, however, can be associated with a degeneracy of the metrical structure. A degeneracy is not a singular breakdown. The familiar example is the spatial metric of a Newtonian spacetime. Using ordinary Cartesian coordinates for space and absolute time as the time coordinate, the metric has the form

$$h_{ik} = \begin{bmatrix} 1 & 0 & 0 & 0 \\ 0 & 1 & 0 & 0 \\ 0 & 0 & 1 & 0 \\ 0 & 0 & 0 & 0 \end{bmatrix}.$$

It assigns ordinary Euclidean lengths to separations in space, but zero length to separations between events at one point in space but with different absolute time coordinates.

Correspondingly, we shall see that a diverging metrical component is also insufficient for there to be a singularity.[9] In Einstein and Rosen's analysis of their "bridge" (below), they show that the diverging $g_{11} = -1/(1 - 2m/r) = \infty$ at $r = 2m$ can be regularized by a one-to-one coordinate transformation.

What further conditions are needed for Einstein to identify a singularity? Einstein and Rosen specified them above in their discussion of the line element (3) in accelerated coordinate systems:

> ... the fact that the determinant of $g$ of the $g_{\mu\nu}$ vanishes for $x_1 = 0$. The contravariant $g^{\mu\nu}$ therefore become infinite and the tensors $R^i_{klm}$ and $R_{kl}$ take on the form 0/0.

It is not the vanishing or the divergence of components of the metric tensor, but the fact that their extreme values propagate through the theory, generating more infinite quantities or indeterminate quantities of the form 0/0 elsewhere. In particular, the quantities used to define the source-free field equations, the Riemann tensor $R^i_{klm}$ and its contraction $R_{kl}$ are compromised.

## 3.5 HILBERT'S DEFINITION OF A SINGULARITY

Einstein and Rosen's formulation appeared roughly two decades after Schwarzschild's solution was found. A similar but not identical characterization was published earlier by Hilbert in his 1916 "Foundations of Physics" and an equivalent version was given in his lectures from that summer. In the second part of his highly influential December 1916 communication on general relativity, Hilbert (1916) affirmed that the Schwarzschild line element (5) is singular at both radial coordinate positions, $r = 0$ and $r = \alpha$, where $r = \alpha$, corresponds to the Schwarzschild radius $r = 2m$ of (5). He then gave the applicable definition of regularity, that is, of non-singularity:

> That is, I call a metric or a gravitational field $g_{\mu\nu}$, at a position, *regular*, if it is possible, through reversible, one-to-one [*eindeutig*] transformation, to introduce such a coordinate system, that, for it, the corresponding functions $g'_{\mu\nu}$, are regular at that position, i.e. in it and in its neighborhood [they] are continuous and arbitrarily often differentiable and have a determinant $g'$ different from zero (1916, 70, his emphasis).

Earlier that year, in lectures given over Göttingen's summer semester, Hilbert gave another version of his characterization:

> We call a gravitational field or a metric "regular"—this definition was still to be added—if it is possible to introduce such a coordinate system that the functions $g_{\mu\nu}$ are regular at every position in the world and have a determinant different from zero. We further designate an individual function as regular if it is finite and continuous along with all its derivatives. This is, incidentally, always the definition of regularity in physics, while in mathematics it is required that a regular function be analytic. The metric [(5)] has singularities for $\alpha > 0$ and, correspondingly, $\alpha \leq 0$, at the position $r = 0$ and $r = \alpha$, and correspondingly at r = 0. If we consider that these singularities arise in the absence of masses, then it seems also plausible that they cannot be eliminated by coordinate transformation (1916, 252).

---

9     We saw above that Einstein and Rosen (1935, 75) found the Schwarzschild solution (5) in Droste coordinates to be singular since "$g_{11}$ for $r = 2m$ becomes infinite and hence we have there a singularity." I read this formulation as tacitly presuming that the infinity produces problems elsewhere when it propagates through to other quantities.

The singularity identified by Hilbert in these characterizations is specifically one arising in analytic expressions, such as the Schwarzschild line element (5), when they are understood to belong to a set of expressions closed under one-to-one coordinate transformations. The definition was not a means of impugning the Droste coordinate system as a preliminary for further analysis of the geometry of the spacetime at the Schwarzschild radius. The singular character of the line element at the Schwarzschild radius was, for Hilbert, a failure of the physics directly. Immediately after presenting the Schwarzschild solution in Droste coordinates and prior to giving the above definition of regularity in his 1916 lecture notes, Hilbert (1916, 251) had declared:

> According to our understanding of the nature of matter, we can consider as physically realizable solutions $g_{\mu\nu}$ of the [source-free gravitational] differential equations $K_{\mu\nu} = 0$ only those which are regular and free from singularities.

The same sentiment appears after the corresponding definition in his published communication "…in my opinion, only regular solutions of the physical fundamental equations represent reality immediately…" (1916a, 70).

These formulations are important. First, in the 1910s, when Einstein was completing his discovery of general relativity, there was no higher standard of mathematical excellence than David Hilbert in Göttingen. We might not now favor the characterization of singularities that he advanced. We cannot now dismiss it as a confusion of an Einstein, who, we would suggest, had brilliant physical intuitions but was prone to novice errors of mathematics.

Second, precisely because of his leading position, Hilbert's formulation was authoritative. When Laue identified the Schwarzschild radius as singular, it was in conformity with Hilbert's definition. He wrote,

> … the singularity of the metric ([5]) at [$r = 2m$] cannot be eliminated by the introduction of other coordinates, but lies entirely in the nature of things (1921, 215).

## 3.6 THEIR DIFFERENCES

Einstein and Hilbert's definitions of a singularity are similar but not identical. The main difference is that Hilbert adds an explicit consideration: "… it is possible, through reversible, one-to-one [*eindeutig*] transformation, to introduce such a coordinate system, that …". This condition fails at $r = 0$ for an ordinary Euclidean space with radial coordinates $r$ and $\theta$ and Euclidean line element $ds^2 = dr^2 + r^2 d\theta^2$. It certainly seems perverse now to offer a definition that would render the simplest of geometries singular. However, we should not presume that Hilbert was guilty of so trivial a mathematical oversight. As long as his criterion is applied to mathematical expressions, it may be uncongenial now, but it commits no mathematical blunder. It simply asserts the singularity of the expression $ds^2 = dr^2 + r^2 d\theta^2$ at $r = 0$, but allows the regularity everywhere of the expression $ds^2 = dx^2 + dy^2$, where $x$ and $y$ are the familiar Cartesian coordinates.

Einstein, however, was not committed to this level of mathematical purity. Presumably, Einstein would have agreed with Hilbert's definition in generic circumstances. However, he was willing to make exceptions, uncongenial to a rigorous mathematician, when his physical intuition called for it. Hilbert's definition of regularity fails for the Schwarzschild solution (5) at $\theta = 0$.[10] We saw, above, that Einstein and Rosen (1935, 75) dismissed this failure with:

> The vanishing of the determinant of the $g_{\mu\nu}$ for $\theta = 0$ is unimportant, since the corresponding (spatial) direction is not preferred.

---

10    The coefficient of $d\phi^2$, $r^2\sin^2\theta$, vanishes at $\theta = 0$ and leads to $g = 0$.

Here they are deferring to the geometry, but without giving a more precise mathematical foundation for their reasons. This is an interesting contrast. Hilbert's mathematical precision gives us a result we do not now like. Einstein's mathematical imprecision gives us the result we do now like.

We now almost automatically would regularize a line element like $ds^2 = dr^2 + r^2 d\theta^2$ by introducing a new coordinate system, such as $(x, y)$, that is not related by a one-to-one mapping with the original coordinate system $(r, \theta)$. Einstein supposes that this sort of regularization is not compelled mathematically, but must be supported by physical considerations. This supposition is central to Einstein's rejection of the geometric regularizations that we might now apply to all three examples discussed here. Appendix B offers a simple example to show that the regularization of the line element $ds^2 = dr^2 + r^2 d\theta^2$ is not a mathematical necessity, but requires a physical basis.

I set aside a complication with Hilbert's definition for regularity. He does not specify whether it is merely a sufficient condition or both necessary and sufficient. If it is merely sufficient, then it does not preclude regularity at the origin of radial coordinates in a Euclidean space. The natural reading is to include necessity: that systems failing his definition of regularity are singular. I will read it that way.

## 3.7 EINSTEIN-ROSEN BRIDGES

The principal function of Einstein and Rosen's paper was to offer an account of particles that would foreshadow Einstein's long-standing goal: a theory with a single unified field that embraces all atomic forms of matter. The Einstein-Rosen proposal was, at best, an intermediate on the way to this goal. It did not posit a single unified field, but employed two fields: the metric field $g_{ik}$ of general relativity and the Maxwell four-vector potential $\varphi_i$ of electrodynamics. The Maxwell field vector was needed to endow the particles with charge. However neutral particles could be represented with the metric tensor alone. The proposal was a development of the idea that the Schwarzschild radius in the Schwarzschild solution was an unrealizable surrogate for matter that would have a regular representation in a successor theory.

The core novelty of the Einstein-Rosen paper was a means of regularizing the singularity at the Schwarzschild radius so that the Schwarzschild solution could represent a particle. That singularity, it was noted above, did not arise from the vanishing or divergence of the coefficients of metric tensor at $r = 2m$. It arose from the vanishing of the determinant $g$. For division by its zero-value propagated through the formalism, yielding indeterminate quantities of the form 0/0. The contravariant metric $g^{ik}$ is equal to $[g_{ik}]/g$, where $[g_{ik}]$ is the co-factor of the metric tensor $g_{ik}$. This contravariant metric would then be used to contract the Riemann tensor to form the Ricci tensor, $R_{ik}$, for the gravitational field equations. Following a remark by W. Mayer, Einstein and Rosen realized that they could avoid division by the troublesome determinant $g$ if they replaced the source-free field equations, $R_{ik} = 0$, by $R_{ik}^* = g^2 R_{ik} = 0$.

An appeal of the new field equations was that the Schwarzschild line element (5) remained a solution and the line element ceased to be singular at $r = 2m$. To see its regularity, Einstein and Rosen rescaled the $r$ coordinate of (5) by introducing a new coordinate $u$, where $u^2 = r - 2m$. In the new coordinate system, the Schwarzschild solution became

$$ds^2 = -4(u^2 + 2m)du^2 - (u^2 + 2m)^2(d\theta^2 + \sin^2\theta d\phi^2) + \left(\frac{u^2}{u^2 + 2m}\right)dt^2. \qquad (6)$$

The troublesome $g_{11} = -1/(1 - 2m/r) = \infty$ at $r = 2m$ of (5) had been transformed to a benign $g_{11} = -4(u^2 + 2m) = -8m$ at the Schwarzschild radius $u = 0$. The corresponding zero-valued coefficient of (5), $g_{44} = (1 - 2m/r) = 0$ at $r = 2m$, remained zero-valued as $g_{44} = u^2/(u^2 + 2m) = 0$ at $u = 0$. While this zero value would lead to $g = 0$, its zero value was no longer troublesome for the modified field equations. Einstein and Rosen (1935, 75) could declare the solution "free from singularities."

In his more popular writing, "Physics and Reality," Einstein (1936) was more explicit about the elimination of the singularity. After reviewing his work with Rosen, he concluded of the Schwarzschild solution in the form (6):

> This solution behaves regularly for all values of [$u$]. The vanishing of the coefficient of $dt^2$ (i.e., $g_{44}$) for [$u$] = 0 results, it is true, in the consequence that the determinant $g$ vanishes for this value; but, with the methods of writing the field equations actually adopted, this does not constitute a singularity (321).

For our purposes, what matters is how the regularity of the Schwarzschild solution was established. The coordinate transformation $u^2 = r - 2m$ is a one-to-one mapping (assuming that the positive square root is taken for $u$). The requirement that coordinate transformations must be one-to-one had not been relaxed. Einstein and Rosen were still assigning all values of the $t$ coordinate to events with $u = 0$ (and each fixed $\theta$ and $\phi$). That is, they are still representing, as infinitely many events, what is treated as a single event in the fully extended Schwarzschild spacetime represented in Figure 3. This requirement of one-to-one mapping must be abandoned if the Kruskal coordinate system shown in Figure 3 is to be adopted.

With this in mind, we can review what Einstein and Rosen have established with their bridge. The structure they called a "bridge" arises when the Schwarzschild solution (6) is considered for all values of $-\infty < u < \infty$. It was their representation for a single, uncharged particle. Even though they allowed all values of their coordinates in (6), their bridge only covered the two wedges of the exterior Schwarzschild spacetime of Figure 3. The interior was untouched. While they did not draw any figures, it is now common for us to find, such as in Kruskal (1960), that their bridge represented by figures akin to Figure 5.

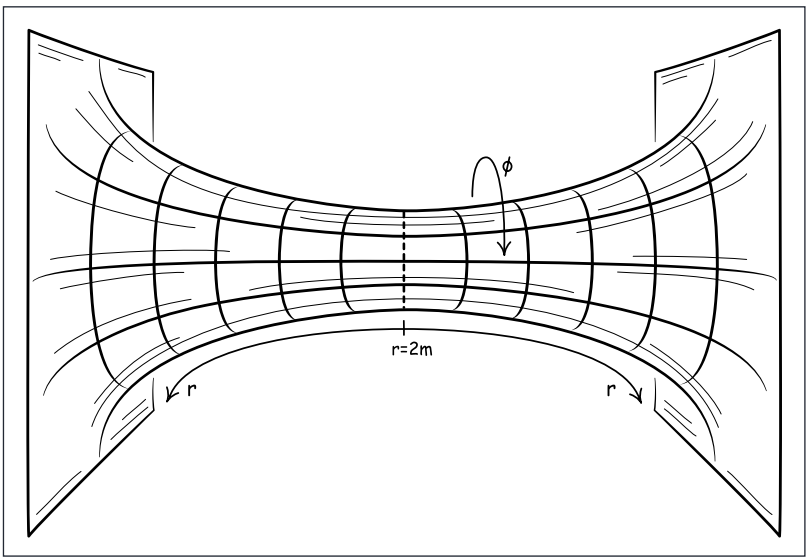

**Figure 5** Einstein-Rosen bridge.

This figure shows a spacelike surface within the solution, spanned by the coordinates $-\infty < u < \infty$ and $0 \le \phi < 2\pi$. In the fully extended Schwarzschild spacetime, an event with $u = 0$ (or equivalently, $r = 2m$) at the neck of the bridge corresponds to a single event at the apex of the two wedges shown in Figure 3.[11] For Einstein and Rosen, however, there were infinitely many such events, distinguished by different values of the $t$ coordinate. For them, a better picture would be something like Figure 6. It shows a single bridge, as Einstein and Rosen would have conceived it, extended according to the development of the $t$ time coordinate.

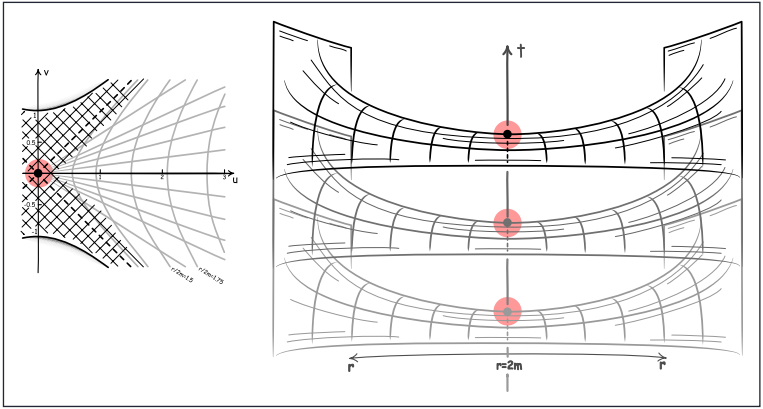

The key difference is how events at the Schwarzschild radius are treated. In the fully extended Schwarzschild spacetime, shown at the left, for a given set of angular coordinates, there is a single event picked out by the Droste coordinates at $r = 2m$, for all values of $t$. In Einstein and Rosen's conception, shown at the right, different values of $t$ pick out different events at $r = 2m$. That is, the set of events at $r = 2m$ for all $t$, forms a curve in the spacetime, for Einstein and Rosen, where it is just a single event in the fully extended Schwarzschild spacetime.

## 3.8 THE STATIC LINE ELEMENT

Finally, there is a single word throughout Einstein's discussions of the Schwarzschild solution that may appear innocent, but actually has a controlling influence on his analysis. All the forms of the solution Einstein discussed are identified as "spherically symmetric static solution of the gravitational equations." That they must be "static" means that Einstein sought a line element in which all the $g_{\mu\nu}$ are independent of the $x_4$ time coordinate; and $g_{14} = g_{24} = g_{34} = 0$.

The physical intuition behind the condition is plausible: the symmetric field surrounding an idealized mass like the sun should not change as time advances; and it should be possible to foliate the field into distinct moments of time (which would fail for constant time coordinate surfaces, if the $g_{14}$, $g_{24}$, $g_{34}$ were non-zero). Because of its naturalness, this condition was assumed automatically from the first moments. It was stated explicitly in Einstein's celebrated communication of November 19, 1915 (833). In it, he gave the first approximate solution for the spacetime of the sun in his near complete general relativity and used it to explain the anomalous 43 seconds of arc in Mercury's orbit. Schwarzschild (1916, 190), in his first derivation of his solution, repeats the conditions, citing Einstein's 1915 assertion of them. Droste (1916, 198) began with a stipulation of this static form for his line element. Both Weyl (1918, §30) and Laue (1921, §21) in their early texts on general relativity proceeded with the assumption of the same static form.

---

11   This familiar figure should be read with caution. It represents well the topology of the two-dimensional, spatial surface spanned by the coordinates $r$ and $\phi$. It is poorer at representing metrical relations, in so far as we expect them to be induced by embedding in a three-dimensional Euclidean space. For such an embedding, $r$ might be a Cartesian coordinate in the vertical direction and $x$ a Cartesian coordinate in the horizontal direction with an origin at the midpoint. A metrical distance $ds$ in the $r - x$ plane must satisfy $ds^2 = [1/(1 - 2m/r)]dr^2 = dr^2 + dx^2$. Solving the resulting differential equation for $x$, we find that the horizontal displacement of the surface from the midpoint is $x = 4m\sqrt{\frac{r}{2m} - 1}$. That is, the curve of bridge is a parabola. The asymptotically flat surfaces on either end are fictitious surfaces at infinite Euclidean distance from the midpoint.

The effect of requiring this static condition, however, is severe. It precludes Einstein and these later relativists from finding a non-singular representation of the fully-extended Schwarzschild spacetime. The coordinate systems that are capable of covering the fully-extended spacetime, such as Kruskal's, violate the condition.

### 3.9 IN RETROSPECT

What can we say in retrospect concerning Einstein's treatment of the Schwarzschild solution? In a limited sense, it was adequate for the practical issues of solar system astronomy that drew only on the Schwarzschild solution outside the $r = 2m$ event horizon. However, as research in gravitational physics progressed, Einstein's algebraic approach to the Schwarzschild solution and his insistence on a static line element proved infertile. The geometric approach proved productive. The full extension of the Schwarzschild spacetime, past the event horizon at $r = 2m$, has provided some of the most important developments. A black hole with its curvature singularity has become the favorite testing ground for theories of quantum gravity. Advances in observational astronomy include the detection through gravitational wave astronomy of the coalescence of two black holes and the visual imaging of the event horizon of a black hole. In this regard, Einstein's insistence on algebraic methods proved to be his own personal event horizon beyond which his gaze could not pass.

## 4. THE MASS HORIZON IN A DE SITTER SOLUTION

Einstein's work in cosmology of the late 1910s triggered an earlier debate over just what counts as a true singularity in general relativity. The debate is especially interesting for present concerns since, in its course, Einstein was forcefully presented with the geometric deflation of what he had identified as a singularity; we can be sure that Einstein knew of the extended geometry and we can follow how he argued against it.

The debate developed before and around Einstein's celebrated 1917 paper, "Cosmological Considerations on the General Theory of Relativity." Einstein's goal was to find a cosmological solution of his theory that conformed with what he came to call "Mach's principle." It is the requirement he later summarized, that "the *g*-field is determined by the masses of bodies *without residue*" (1918, 241, his emphasis). The principle has the consequence that "according to the gravitational field equations, no *G*-field is possible without matter" (243).

Einstein's gravitational field equations of 1915 admit the spacetime of special relativity as a matter-free solution, an impossibility according to Mach's principle.[12] Einstein's 1917 paper introduced a new cosmological solution that depicts a static, spherical space with a uniform matter distribution. Einstein found that he had to add a cosmological term $\lambda$ to his gravitational field equations so that they would admit the new solution. Mach's principle was upheld in the sense that the modified field equations no longer admitted the matter-free solution of special relativity.

---

12    This is given as a motivation for modifying his gravitational field equations in Einstein (1918, 243).

Einstein's proposal ran into grave difficulties. Almost immediately, De Sitter found a matter-free solution of the $\lambda$-augmented gravitational field equations. One of the great debates of modern physics erupted. At its core was the question of whether certain singular expressions identified by Einstein could be surrogates for source masses; and whether they were singularities at all.

The details of this episode and the resulting debate have been recounted elsewhere in greater detail and with greater insight than is possible here. A thorough recounting is found in editorial apparatus of the *Collected Papers*, Schulmann in the headnote "The Einstein-de Sitter-Weyl-Klein Debate" (1998, 351–357) and numerous footnotes elsewhere; in Earman and Eisenstaedt (1999, §3), in Janssen (2014, 198–208), and in Smeenk (2014). Below I will focus just on the specific aspects of the debate concerning the singularities pertinent to present concerns.

## 4.1 EINSTEIN ON THE SINGULARITY IN THE DE SITTER SOLUTION

Einstein (1918a) provided a synopsis of his then stable view of the de Sitter solution in a short note, "Critical Comment on a Solution of the Gravitational Field Equations Given by Hr. De Sitter," communicated to the Prussian Academy on March 7, 1918. He formulated the solution to his source free, $\lambda$-augmented gravitational field equations in a coordinate system $(r, \psi, \theta, t)$ as:

$$ds^2 = -dr^2 - R^2\sin^2(r/R)[d\psi^2 + \sin^2\psi d\theta^2] + \cos^2(r/R)c^2dt^2, \qquad (7)$$

where $R$ is the constant radius of curvature of a $t =$ constant space. De Sitter had presented this line element in exactly matching notation as one version of his solution in de Sitter (1917a, 7, eqn. 8B). Einstein had selected this version of de Sitter's solution because it has the key property of being static. That is, $g_{44} = \cos^2(r/R)$ is independent of $t$, which means that (in geometrical language)[13] the constant $t$ hypersurfaces all have the same spatial geometry, which Einstein presumed on astronomical evidence to be the case for our universe.

Einstein proceeded to specify with considerable care why he judged there to be a singularity at the value $r = \pi R/2$ of the $r$ coordinate for which $g_{44} = \cos^2(r/R)c^2$ vanishes. He first laid out the conditions required for non-singularity:[14]

> … it is a requirement of the theory that the [$\lambda$-augmented gravitation field] equations hold for all finite points. This can then only be the case if both the $g_{\mu\nu}$ as well as the associated contravariant $g^{\mu\nu}$ (together with their first derivatives) are continuous and differentiable; therefore in particular the determinant $g = |g_{\mu\nu}|$ must disappear nowhere in the finite (1918, 270).

This plausible condition, however, faced the same problem as the Schwarzschild solution in Droste coordinates: According to it, the de Sitter solution is singular at the origin $r = 0$ of the spherical coordinates. It is, presumably, with this difficulty in mind, that Einstein introduced a clarification:

---

13    And $g_{14} = g_{24} = g_{34} = 0$.

14    Einstein's concern with the finitude of the points derives from his disagreement with an assertion by de Sitter in a letter to Einstein of April 1, 2017 (Schulmann et al. 1998, Doc. 321) that the singular points are at infinity.

Further, the continuity condition for the $g_{\mu\nu}$ and $g^{\mu\nu}$ is not to be understood as requiring that there must be a choice of coordinates in which it [the continuity condition] is satisfied in the whole space. Obviously it must only be required that there is a choice of coordinates for the neighborhood of each point for which the condition of continuity is satisfied in this neighborhood; this restriction on the requirement of continuity arises naturally out of the general covariance of the [$\lambda$-augmented gravitation field equations] (1918, 271).

Einstein then used this weakened condition to eliminate the apparent singularity at the origin:

For the de Sitter solution now, according to [(7)]

$$g = -R^4 \sin^4(r/R)\sin^2\psi\cos^2(r/R)$$

Thus to begin with, $g$ vanishes for $r = 0$ and for $\psi = 0$. This relation, however, implies only an apparent violation of the condition of continuity, as can easily be shown through an appropriate change in the choice of coordinates (1918).

While Felix Klein would soon conclude that the singularity at $r = \pi R/2$ should be treated similarly, Einstein insisted otherwise. He continued:

However $g$ vanishes also for $r = \pi R/2$, and indeed here is a discontinuity that cannot be eliminated through any choice of coordinates. ... Until proven otherwise, it is therefore to be assumed that the de Sitter solution has a real singularity in the surface located at $r = \pi R/2$, that is does not conform with the [$\lambda$-augmented gravitational] field equations for any choice of coordinates (1918).

While Einstein took the analytic formula of line element (7) as the subject of analysis, for comparison purposes again, we can give a geometric picture of Einstein's understanding. For constant $t$, the line element (7) reduces to a three-dimensional spherically curved space with radius of curvature $R$, divided in half by a singular surface at $r = \pi R/2$. Suppressing the $\theta$ angle variable, one half of this space is shown in Figure 7 as a hemisphere with the singularity at its equator. The space is static, so that the hemisphere evolves forward, unchanged in time, as shown by the vertical $t$ axis, and carries the singularity with it through time.

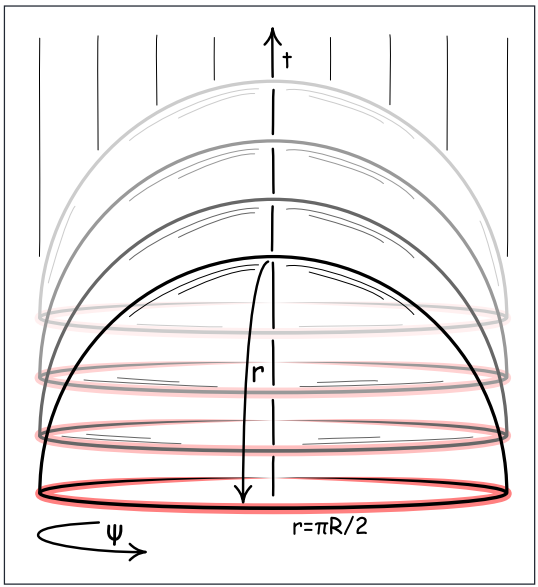

**Figure 7** Einstein's Conception of the de Sitter Solution.

De Sitter sent Einstein a summary of his new, proposed solution in a letter of March 20, 1917 (Schulmann et al. 1998, Doc. 313). Einstein responded four days later on March 24. De Sitter's new solution was inadmissible as a serious candidate for cosmology for physical reasons: the singularity itself was problematic, as well as the solution's matter-free character. Einstein summarized his concerns:

> Therefore it seems to me that your solution does not correspond to a physical possibility. The $g_{\mu\nu}$ and $g^{\mu\nu}$ (together with their first derivatives) must be continuous everywhere.

> In my opinion, it would be unsatisfactory if a world without matter were conceivable. Rather, the $g^{\mu\nu}$-field should be *fully determined by matter and not be able to exist without the matter*. This is the core of what I mean by the requirement of the relativity of inertia (Doc 317, Einstein's emphasis).

The "relativity of inertia" would shortly be given the more familiar title "Mach's principle."

Einstein's objections to de Sitter's solution continued in their correspondence. Einstein objected, for example, that at the singularity $r = \pi R/2$, since $g_{44} = 0$, the velocity of light drops to zero, the energy of a mass point would be zero and masses would have minimal gravitational potential, so they would accumulate there (Einstein to de Sitter July 22, 1917 Doc. 363; Schulmann et al. 1998). Such a singularity is "in my opinion to be excluded as physically not coming into consideration." For these reasons, Einstein remained recalcitrant. In a letter to de Sitter on July 31, 1917 (Schulmann et al. 1998, Doc. 366), he wrote that he could not "grant your solution any physical possibility" and:[15]

> For $r = \pi R/2$, the mass point has no energy; it does not exist there any more at all, but has eaten itself up completely on the way there. The admission of such cases appears to me to be absurd [*sinnwidrig*]. This will naturally always be true, regardless of how we may choose the variables.

## 4.2 THE MASS HORIZON

Einstein's opposition to de Sitter's solution soon softened, but did not disappear. He found a way to accommodate de Sitter's solution to Mach's principle. The idea was to identify the singularity as a concentration of matter at the singular surface. It was a "mass horizon," as it soon came to be called,[16] and responsible for the spatiotemporal properties of the world, in accordance with Mach's principle. The proposal concluded Einstein's March 7, 1918, with a critical comment on the de Sitter solution. He wrote:

> In fact, de Sitter's system [(71)] solves the [$\lambda$-augmented gravitational field equations] everywhere, only not in the surface $r = \pi R/2$. There—as in the immediate vicinity of a gravitating mass point—the component $g_{44}$ of the gravitational potential becomes zero. The de Sitter solution must in no way [be seen as] the case of a matter-free world, but much more as corresponding to the case of a world whose matter is all concentrated in the surface $r = \pi R/2$: this could indeed be demonstrated through a limiting process from a spatial to a surface-like distribution of matter (Einstein 1918a, 272).

---

15    Einstein use of the term "variables" and not "coordinates" conforms with his focus on analytic expressions formulated with variables as opposed to geometries with coordinates.

16    The term "mass horizon" appears, for example in Weyl (1918, 226; 1919, 34) as *Massenhorizont*.

In foreshadowing this proposal in a letter to de Sitter of August 8, 1917, (Schulmann et al. 1998, Doc. 370), Einstein explained more fully that "... the falling of the $g_{44}$ (to zero) in the approach to $r = \pi R/2$ is to be thought of as caused through matter, just like the falling of $g_{44}$ in the approach, f.[or] e.[xample] to the sun."

Einstein's proposal included a supposition that the mass horizon could be produced by a limiting process. Einstein reported with obvious satisfaction to de Sitter in a letter of April 15, 1918 (Schulmann et al. 1998, Doc. 506) that Hermann Weyl, in his new relativity text *Raum-Zeit-Materie*, had done just the requisite calculation.[17] Einstein, however, soon found he was not satisfied with the details of Weyl's calculation and the two exchanged letters and proposals in April to June 1918 until Weyl's corrections were acceptable to both.[18]

Hermann Weyl was a mathematician of the first rank. He was a student of Hilbert and became Hilbert's successor in Göttingen. His active engagement provided a mathematical ratification to Einstein's assessment of the singular character of de Sitter's solution.

## 4.3 THE GEOMETER'S REACTION

It took several decades before concerted opposition arose to Einstein and Hilbert's judgment of the singular nature of the Schwarzschild radius. In the case of the de Sitter solution, that repudiation arose more quickly. Einstein and de Sitter's proposals in cosmology attracted the attention of Felix Klein, a colleague of Hilbert's at Göttingen, and perhaps the foremost geometer of his era. His 1872 "Erlangen program" sought to unify many disparate geometries in one system. Its leading idea was that each geometry could be characterized by a group. His approach fared poorly with the Riemannian spacetimes of variable curvature of Einstein's general theory of relativity since the groups with which Klein worked there were, in general, trivial.[19] The exceptions were the geometrically more symmetric Minkowski spacetime of special relativity and the two cosmologies of Einstein and de Sitter. They were homogenous enough to admit a non-trivial group structure. Hence, it was not surprising that Klein took an interest in them. He repeatedly and clearly presented the geometer's objection that Einstein's singularity was a mere artefact of his choice of coordinate system.

Klein reported his geometric analysis of Einstein's and de Sitter's cosmological solutions in lectures in Göttingen of May 7 and June 11, 1918. Their content was summarized in Klein (1918) and he subsequently published more expansive versions with the same essential content. Klein (1919) reported a version communicated to the Amsterdam Academy of Science in a meeting of September 29, 1918. Klein (1918a) reported a still more elaborate version communicated in Göttingen on December 6, 1918.

Klein's analysis was set in a five-dimensional extension of the familiar four-dimensional Minkowski spacetime. Its line element was[20]

$$ds^2 = d\xi^2 + d\eta^2 + d\zeta^2 - d\nu^2 + d\omega^2. \tag{8}$$

The coordinates $\xi$, $\eta$, $\zeta$ and $\omega$ are spatial and the coordinate $\nu$ is temporal. Within this five-dimensional space, he introduced an hyperboloid with radius of curvature $R/c$ whose four-dimensional surface is the space of the de Sitter solution:

---

17    See Weyl (1918, §33) and a subsequent clarification in Weyl (1919).

18    See Docs. 511, 513, 525, 535, 544, 551 in Schulmann et al. 1998.

19    For discussion of the contrast, see Norton (1999).

20    Klein included considerably more geometry in his full discussion than is relevant to the treatment of the singularity. He made contact with his work in projective geometry and suggested that, spatially, the solution should not employ spherical geometry, but elliptical geometry in which antipodal points are identified.

$$\xi^2 + \eta^2 + \zeta^2 - \nu^2 + \omega^2 = R^2/c^2. \tag{9}$$

Klein then sought to recover the de Sitter solution in the form (7) employed at times by de Sitter and by Einstein in his communication of Einstein (1918a). To this end he introduced four spacetime coordinates, $r$, $\psi$, $\theta$ and $t$, through the relations[21]

$$\xi = R \sin(r/R)\cos\psi \tag{10}$$

$$\eta = R \sin(r/R)\sin\psi\cos\theta$$

$$\zeta = R \sin(r/R)\sin\psi\,sin\theta$$

$$\nu = R \cos(r/R)\sinh(ct/R)$$

$$\omega = R \cos(r/R)\cosh(ct/R)$$

Transformation (10) replaces Klein's original five coordinates by four. It follows that transformation (10) induces a relation among the new coordinates. Forming the sum of squares $\xi^2 + \eta^2 + \zeta^2 - \nu^2 + \omega^2$ using (10) affirms that the formula for the hyperboloid (9) holds as an identity among the four new coordinates. That is, these coordinates are restricted by their definitions to the surface of the hyperboloid.

This new coordinate system fails to cover the entire hyperboloid. To see this, form the ratio of the expressions for $\nu$ and $\omega$ in (10) and recover

$$\nu = \tanh(ct/R)\,\omega. \tag{11}$$

For a fixed value of the coordinate $t$, relation (11) identifies a plane in the five-dimensional space: $\nu = \text{constant}\omega$. Each of these planes intersects the hyperboloid in a three-dimensional geometrically spherical space of constant $t$. As $t$ varies from $-\infty$ to $+\infty$, the factor $\tanh(ct/R)$ varies from -1 to +1. It follows that these constant $t$ spaces all lie between the two planes $\nu = +\omega$ and $\nu = -\omega$, to which the hyperboloid is asymptotic. These planes are shown in Figure 8, in which two of the five dimensions have been suppressed. The new coordinates will cover the part of the hyperboloid lying within by the two-sided wedge formed by the planes $\nu = +\omega$ and $\nu = -\omega$.

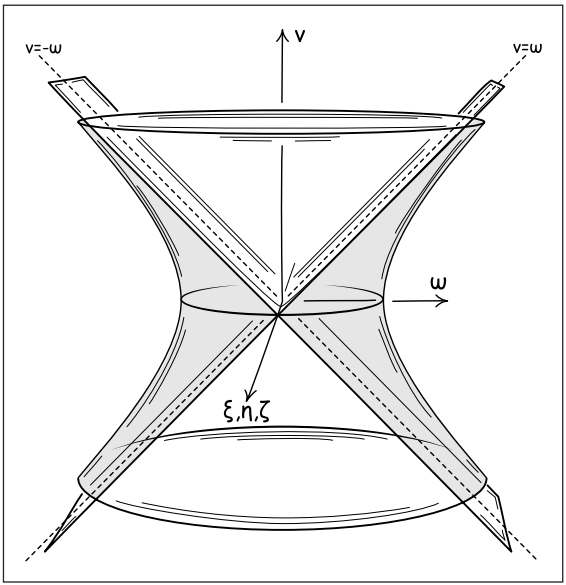

**Figure 8** Two-sided wedge in the de Sitter hyperboloid.

---

21    For continuity with Einstein's notation of (7), I have relabeled Klein's variables according to $\theta \to r/R$, $\phi \to \psi$ and $\psi \to \theta$.

It now follows that the new coordinates introduced in (10) become singular at antipodal points $\nu = \omega = 0$, where the two asymptotic planes intersect. The surfaces of constant $t$ and their intersection are shown in Figure 9. The circles of constant $t$ in the figure are, more fully, three-dimensional spherical spaces whose two angle coordinates, $\psi$ and $\theta$, are not shown. The intersection of all surfaces of constant $t$ appear as two antipodal points on each of these circles. These two points are the projection onto the circles of a two-dimensional spherical surface within the full three-dimensional spherical space spanned by the coordinates r, $\psi$, and $\theta$.

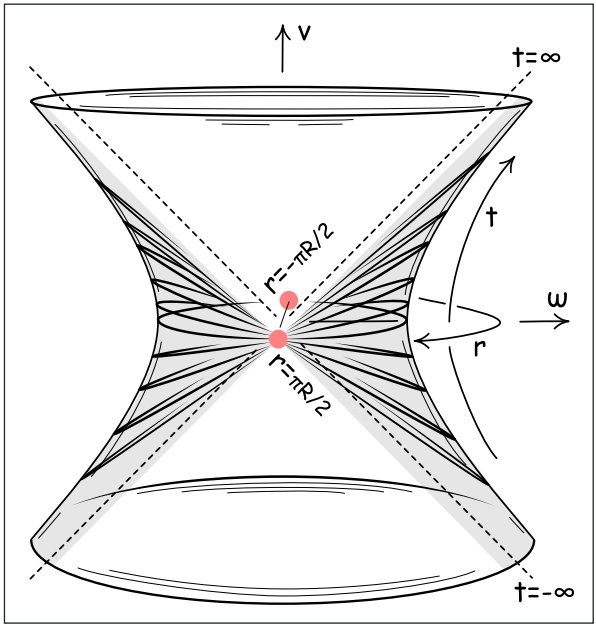

**Figure 9** Einstein's static coordinate system for the de Sitter solution.

This sphere of intersection of all spaces of constant $t$ is the singularity identified by Einstein that becomes the mass horizon. It is evident that, if the full spacetime is the geometrically homogeneous hyperboloid (9), then the singularity is simply an artifact of the new coordinate system introduced in (10). Indeed, if we take the hyperboloid to underlie de Sitter's solution, then any judgement of special, singular regions contradicts the homogeneity of the hyperboloid. From this perspective, Einstein's claim of a singularity must be mistaken.

In each of his presentations, Klein made this point of the spurious character of the singularity. He did not emphasize it, but was sure to make it. In his first reported publication, he concluded:

> Thereby, the criticism which Einstein recently leveled against the de Sitter $ds^2$ (Berlin Academy of March 7 1918) is reduced to its true value. The singularity, which the de Sitter $ds^2$ has at $\omega = 0$ and $\nu = 0$ depends only the arbitrariness with which the $t$ is introduced (Klein, 1918, 44).

In his 1919 paper, he remarked that, at $\omega = 0$ and $\nu = 0$, "$t$ becomes undetermined" (615). A footnote remarks: "However this is only a singularity of the coordinate $t$, not of the underlying manifold [(8), (9)]." Earlier, Klein had characterized the singularity at this point in the surface in a slightly different manner: "… it appears as something singular, that is, as the location of a world point for which $t$ adopts the value 0/0" (1918a, 421). Here, Klein presumably drew on the inverse of relation (11), which Klein had written as[22]

$$t = R/c \ \log[(\omega + \nu)/(\omega - \nu)] \quad (1918a, 420)$$

When $\omega = \nu = 0$, the argument of the log function takes the indeterminate form 0/0. Klein's 1918 writing was perhaps most dismissive:

> For the general analytic conception, one has thereby no singularity present other
> than that of the polar angle $\varphi$ at the origin of a common (polar-)coordinate system
> (1918a, 418).

Einstein's assertion of a singularity is here compared to a simple error in understanding how polar coordinates work at the origin of a polar coordinate system. The polar angle coordinate does take on all values or an indeterminate value at the origin without impugning the geometry of the surface to which it is applied. Only a geometrical novice, we might suppose, would confuse this with a singularity.

## 4.4 ASSESSING THE GEOMETER'S REACTION

Modern commentators have celebrated Klein's intervention as introducing the decisive insight into how singularities are to be treated in general relativity. Earman and Eisenstaedt assess it as:

> The importance of Klein's contribution can hardly be overemphasised: for the
> first time in the history of GTR, what had been taken to be a real singularity
> had been shown to be merely apparent; and as well Klein had provided a
> paradigm for demonstrating the fictitious character of a singularity by getting
> rid of it by extending the spacetime; and his paradigm contained the realisation
> that the extension may have to be done by means of a non-regular coordinate
> transformation to non-static coordinates (1999, 195).

That Klein was an accomplished geometer intervening on a novice is suggested by Klein's closing remarks in his letter of May 31, 1918. There, he emphasized the geometrical foundation of his understanding of the de Sitter solution:

> With this, I would like to close today. My entire letter is intended to be only
> a precisification of my earlier, related communication. If I now give definite
> formulae in the place of geometrical considerations, that is done since with their
> help clearer expression is possible. However, geometrical considerations remain
> still the source of the whole train of thought (Schulmann et al. 1998, Doc. 552).

While this may now be the modern assessment, it was not shared then widely by mathematicians of the first tier who worked on de Sitter's solution. They fully understood Klein's analysis and merely treated it as an alternative to Einstein's. If anything, they favored Einstein's view. For example, in his discussion of "de Sitter's world," Lanczos reported Klein's view:

> Now, since all points of this pseudosphere are equivalent and regular, any singularity
> of the line element can only derive from the coordinate system used (1921, 540).

However, he immediately retreated from Klein's position by recalling Weyl's supposition of a layer of matter around the singularity.

> It is a completely independent assertion and involves a singularity that cannot
> be removed by a transformation (1921).

---

22   Klein's formula omits a factor of 1/2 found in standard expressions for the inverse of the function $\tanh(x)$. The factor was added silently in the reprinting of the paper in Klein's collected papers (1921, 609).

The primary burden of Lanczos' note was a close analysis of Weyl's proposal.

Similarly, Weyl's (1921) treatment of cosmology in his 4th expanded edition of *Raum-Zeit-Materie* elaborated his own proposal for a mass horizon in the de Sitter solution. It then only briefly sketched Klein's alternative in the closing pages of the section (256), without naming Klein. The alternative was introduced with a neutral "One recovers a metrically homogeneous world if …" Weyl left open the decision between the two systems. It is to be decided by their differing physical properties.

> The question arises whether it is the first or the second co-ordinate system that serves to represent the whole world in a regular manner. In the first case, the world would not be static as a whole, and it is consistent with the laws of nature that it is empty of mass: de Sitter proceeds with this assumption. In the second case, we have a static world that is not possible without a mass-horizon; this assumption, which we have discussed more fully here, is preferred by Einstein (1921).

## 4.5 EINSTEIN'S TOPOLOGICAL RESPONSE

We can be sure that Einstein was well-informed of Klein's geometrical analysis from the outset. Klein informed him of his work in letters of May 31 and June 16, 1918, roughly coincident with Klein's 1918 presentations of May 7 and June 11 in Göttingen on de Sitter's solution.[23] (Schulmann 1998, Docs. 552, 566). In the second, Klein gave a compact summary of the main analysis, comparable to the content of his communication of September 26 to the Amsterdam Academy (Klein, 1919). There was enough of the formulae for a reader of the letter to reconstruct the full analysis. Reporting on his recent communication in Göttingen, Klein introduced the analysis with a simple summary,

> I arrived at the result that the singularity you have noted in fact can be simply transformed away (1919).

Einstein's reply came four days later in a letter to Klein of June 20, 1918 (Schulmann et al. 1998, Doc. 567). His tone was conciliatory and unruffled. His letter began:

> You are quite right. De Sitter's world is, in itself, free of singularities and its world points are all equivalent. A singularity only comes about through the substitution that affords the transition to the static form of the line element. This substitution alters the *analysis-situs* relations. That is, two hypersurfaces
>
> $$t = t_1$$
>
> and  $t = t_2$
>
> intersect in the original representation, while they do *not* intersect in the static [representation]. This is related to the fact that one must have masses for the static conception, but not in contrast in the first [one].

Einstein's direct reply to Klein's assertion is illustrated in the difference between Figures 7 and 9 above. Einstein made a point in topology, a field whose older name was "analysis situs." In the static de Sitter solution of (7) depicted in Figure 7, the singularity consists of many two-dimensional spheres, each distinguished by a different t coordinate. In the hyperboloid of Figure 9 recounted by Klein, the singularity consists of a single two-dimensional sphere. All the t coordinates of (7) are assigned to this one sphere, since the hypersurfaces of each value of t intersect in the one sphere. It follows that the two singularities differ in their topology: Einstein's is three-dimensional and Klein' is two-dimensional. They cannot be related by a one-one coordinate transformation. Klein's full hyperboloid, Einstein argued here, is distinct from the de Sitter solution of (7).

---

23    In the versions of (8) and (9) in the June 16 letter, in a minor aberration in his notation, Klein assigned the minus sign to $\omega$ and not $\nu$.

Einstein's response to Klein was brief given what later assessments judged to be devastating criticism. However, it was brief since Einstein needed only to add the topological consideration. In an earlier letter to Klein, sometime before June 3, he had already explained in greater detail why he dismissed de Sitter's solution. He summarized his dismissal in that earlier letter as:

> From a physical point of view, I believe that I can assert quite definitively that this mathematically more elegant, four-dimensionally unified conception of the world does not correspond to reality. That is, the world seems to be constituted such that its finely distributed matter can remain at rest in a suitably choice of the coordinate system (Schulmann et al. 1998, Doc. 556).

He then gave the details. De Sitter's line element (7) at least gave a static spacetime. However, its matter must be located in the mass horizon that Weyl had analyzed. In contrast, Einstein's own cosmology of 1917 had the requisite uniform matter distribution. Einstein was eager to be understood. He drew small pictures illustrating the matter distributions in space in his 1917 cosmology and in the mass horizon of de Sitter's solution.

To later commentators who interpret Klein's letter as the intervention of an expert in geometry pointing out the error of a novice in geometry, Einstein's response is fragile and unconvincing. Einstein's patient tone indicates that he did not see it that way. He had no need to see it that way. Einstein regarded the primary element of his theorizing to be analytic expressions, such as (7). That the expression can be generated as a solution to his field equations by means of the hyperboloid is merely an artefact of its construction. It does not establish that the full hyperboloid is the physically applicable solution of Einstein's gravitational field equations. Which that is, whether expression (7) or not, is a matter of physics. As we shall see below, the physics, in Einstein's view, directed the selection of (7).

Klein accepted Einstein's response. The full hyperboloid was unobjectionable, mathematically, as a solution to Einstein's gravitational field equations. However, its physical applicability to our universe was to be decided by physical considerations. Klein deferred to Einstein's assertion that these considerations spoke against the full hyperboloid. Klein reported this assessment at the end of his December 6 communication in Göttingen. He noted that his analysis contradicted Weyl's proposal of a mass horizon and continued:

> I have not checked the correctness of Weyl's calculations. However, I am happy to adopt the understanding that Einstein has expressed in correspondence that the difference in our mutual results must be based on the difference in the coordinates used. What I designate as a single point of intersection with the application of the [coordinates] $\xi, \eta, \zeta, \nu$ and $\omega$, is simply extended region with the use of the [coordinates][24] $\theta, \phi, \psi, t$ (because of the undetermined, remaining values of t). It should not be hard to arrive at a full elucidation here (Klein, 1918a, 423).

Having accepted Einstein's narrow topological argument, Klein then gave his final assessment:

> My concluding vote on de Sitter's claims, however, is that mathematically—up to that one still not fully clarified point—everything is in order. However, one is led to physical consequences that contradict our familiar way of thinking and, in any case, the agenda pursued by Einstein with the introduction of the spatially closed world (1918a).

---

24  [JDN] Corresponding to r/R, $\psi, \theta, t$ in (7)

If Einstein had committed some novice blunder, Klein did not see it. Indeed, Klein's criticism seems hasty. His barb was that Einstein's singularity is no different from the singularity at the origin of polar coordinates suggests a negligence on Einstein's part. The barb is incompatible with Einstein's earlier writing. For, as we saw above, Einstein (1918a) had addressed just this problem in his critical comment on de Sitter's solution. He had noted the presence of just this sort of apparent singularity at the origin of the coordinates of the line element (7) and had explained why he found it to be eliminable in a way that the singularity at $r = \pi R/2$ was not eliminable.

Einstein's topological argument was a minor consideration in his thinking. In the next two subsections, we shall see that he was well acquainted with the issues Klein had raised and had well developed means to address them.

## 4.6 EINSTEIN AND THE HYPERBOLOID

It is perhaps tempting to imagine that, with his fixation on analytic expressions like (7), Einstein was unaware that de Sitter's solution was represented most simply by the hyperboloid. The view is unsustainable. That de Sitter's solution derives from the hyperboloid had been present, prominently, from the start. The hyperboloid is the second most simple relativistic spacetime and the simplest solution of Einstein's augmented gravitational field equations.

To see its simplicity, it is helpful to consider the analogous case of an ordinary three-dimensional space. The simplest case is just a flat, Euclidean space of vanishing curvature. The next simplest case is of a homogeneous, spherical space with the same positive curvature elsewhere. This second case can be constructed by the artifice of considering a four-dimensional Euclidean space with Cartesian coordinates $\xi_1, \xi_2, \xi_3,$ and $\xi_4$ and the line element

$$d\sigma^2 = d\xi_1^2 + d\xi_2^2 + d\xi_3^2 + d\xi_4^2. \tag{12}$$

This line element induces a constant, positive metrical curvature on a spherical, three-dimensional hypersurface defined by

$$R^2 = \xi_1^2 + \xi_2^2 + \xi_3^2 + \xi_4^2. \tag{13}$$

An explicit expression for the metric on this hypersurface is recovered by using (13) to eliminate[25] $\xi_4$ from the line element (12) and to arrive at the line element and metric of the hypersurface[26]

$$d\sigma^2 = \gamma_{ik} d\xi_i d\xi_k \qquad \gamma_{ik} = \delta_{ik} + \xi_i \xi_k / (R^2 - \rho^2), \tag{14}$$

where $i, k = 1, 2, 3$ and $\rho^2 = \xi_1^2 + \xi_2^2 + \xi_3^2$.

This construction, if promoted to a five-dimensional Minkowski spacetime, is the construction Klein employed for the de Sitter hyperboloid. Line element (13) and hypersphere (14) are the direct analogs of Klein's line element (8) and hyperboloid (9). The main difference, aside from dimensionality, is that a hyperbola in a Minkowski spacetime is the curve of constant curvature analogous to the circle of constant curvature in a Euclidean space. Moving to higher dimensions, the hypersphere is a three-dimensional homogenous surface in a four-dimensional Euclidean space. Its analog is the de Sitter hyperboloid, which is a four-dimensional, homogenous hypersurface of constant curvature in a five-dimensional Minkowski spacetime.

---

25    By substituting $\xi_4 d\xi_4 = -\xi_1 d\xi_1 - \xi_2 d\xi_2 - \xi_3 d\xi_3$ and $\xi_4^2 = R^2 - \rho^2$ from (13).

26    No single coordinate system can be mapped one-one to the surface of a sphere, so there must be some anomalies. Writing (13) as $\xi_4^2 = R^2 - \xi_1^2 + \xi_2^2 + \xi_2$, we find that each set of values of $\xi_1, \xi_2$ and $\xi_3$ corresponds to two points on the sphere, according to whether positive or negative root of $\xi_4^2$ is taken. The metric expression (14) is singular at the equatorial sphere where $\xi_4 = 0$ and $R^2 = \xi_1^2 + \xi_2^2 + \xi_3^2$, so that $R^2 - \rho^2 = 0$.

We can have no doubt that Einstein was fully aware of the construction of the spherical space through (12), (13), and (14). These are the equations Einstein used in his 1917 cosmology paper (149–150) to introduce the spherical space of his 1917 cosmology. It is not possible that Einstein could have overlooked the analogy between his work of 1917 and the construction of the de Sitter hyperboloid, for de Sitter (1917) motivated his original derivation of his solution precisely by means of this analogy. Lest his readers could have any doubt of his use of the analogy, de Sitter employed the presentation (1917, 1219), typographically taxing in the era of hot metal typesetting, of displaying the two derivations side by side, as shown in Figure 10.

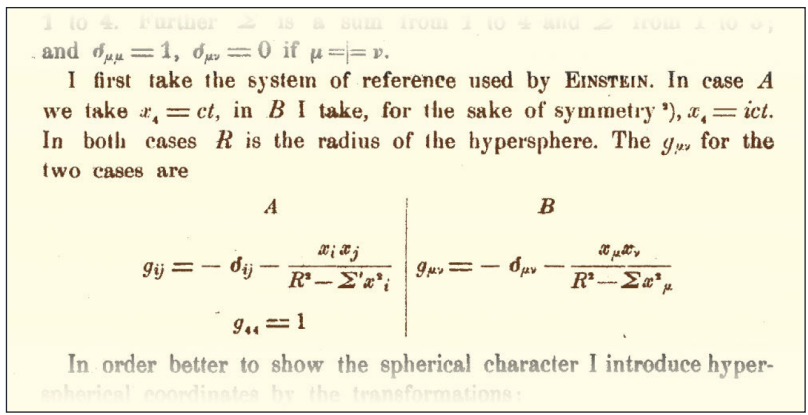

Case A is the three-dimensional space of Einstein's 1917 cosmology. The closely analogous case B is de Sitter's first presentation of his new solution to Einstein's augmented gravitational field equations. Klein's analysis employed a Minkowskian line element (8) with an indefinite signature. To bring the two cases even closer together, de Sitter employed the then familiar device of an imaginary time coordinate $x_4 = ict$ so that his line element and metric were positive definite and his hypersurface was a hypersphere. De Sitter recognized that the artifice of an imaginary time coordinate was dispensable and did not alter the results. His footnote at the word "symmetry" is "We can also take $x_4 = ct$. Then the four-dimensional world is hyperbolical—instead of spherical, but the results remain the same."

Finally, we can be sure that Einstein was fully informed of the analogy used by de Sitter. De Sitter's results were communicated to the Amsterdam Academy on March 31, 1917. Days earlier, on March 20, de Sitter had already communicated his results to Einstein in a letter (Schulmann et al. 1998, Doc. 313). The letter employed the same side-by-side presentation of Einstein's case A and de Sitter's case B and reported the presentation of the hyperboloid in two further coordinate systems. This letter and de Sitter's (1917) paper marked the start of an energetic correspondence between Einstein and de Sitter on the physical cogency of the new solution, and de Sitter followed up with several more publications elaborating on the solution.

For our purposes, what matters is that Klein, in 1918, had little new to tell Einstein. for Einstein was already well aware that the de Sitter solution was generated by means of a fully homogeneous hyperboloid whose geometry was everywhere regular. It is not credible that Einstein had somehow overlooked the fact that the singularity of the mass horizon arose as an artifact of the way the coordinates were applied to the hyperboloid. Einstein's (1918a) earlier published critique of de Sitter's solution has already compared the coordinate singularity at the origin $r = 0$ with that at $r = \pi R/2$ and decided that they had a different character. That decision suggests that Einstein already had in hand his topological rejoinder to Klein. In any case, the whole issue was incidental to Einstein for his interest was the line element (7), independently of how it related to the hyperboloid.

The modern diagnosis of a geometrically sophisticated Klein correcting a geometrically naïve Einstein would appear to have the history inverted. In its place, we have Klein the geometer, whose vision is restricted to purely mathematical matters of geometry, corrected by an Einstein, who was willing to see beyond pure matters of geometry to a larger class of solutions of his gravitational field equations in the service of his physics.

## 4.7 EINSTEIN'S PHYSICAL OBJECTIONS TO THE FULL HYPERBOLOID

In Einstein's view, at best, de Sitter's analysis was an exercise in mathematics that uncovered new solutions of Einstein's augmented gravitational field equations. What mattered to Einstein was not just the abstract mathematics, but which solutions satisfied physical conditions appropriate to our universe. To this end, Einstein provided two physical conditions that precluded de Sitter's solution in all their forms. These two conditions controlled his analysis and he insisted repeatedly on them.

The first was the Machian intuition that had underpinned Einstein's investigations into general relativity from the start and had figured prominently in his cosmological paper of 1917: The metric field is determined completely by the masses of bodies, without which no spacetime is possible. In 1918, he gave the condition the name Mach's principle, where he reaffirmed forcefully his commitment to it with

> ... the necessity of retaining it is in no way supported by all colleagues.
> However, I myself view its satisfaction as unconditionally necessary (1918, 242).

That commitment immediately ruled out de Sitter's (1917) proposal of the full hyperboloid since it is a matter-free spacetime.

Einstein recalled his commitment to Mach's principle in many places. We saw above that, only four days after de Sitter had informed Einstein of his new solution, on March 24 Einstein gave it as a reason for rejecting the physical admissibility of de Sitter's solution (Schulmann 1998, Doc 317, Einstein's emphasis):

> In my opinion, it would be unsatisfactory if a world without matter were conceivable. Rather, the $g^{\mu\nu}$-field should be *fully determined by matter and not able to exist without the matter*. This is the core of what I mean by the requirement of the relativity of inertia.

De Sitter found Einstein's remark so significant that he included it in a postscript to his 1917 presentation (1225). Only two days later, in a letter of March 26, Einstein repeated the point to Klein, this time most likely in the context of Einstein's 1917 cosmology (Schulmann et al. 1998, Doc. 319):

> Whoever is not disturbed if the existence of a $g_{\mu\nu}$ field without field producing matter is possible according to the theory, and if a single mass (conceived as alone in the world) can possess inertia—that is someone not to be convinced of the necessity of the new step.

Einstein's commitment to Mach's principle remained firm after all the deliberations concerning the de Sitter solution had been settled. He reaffirmed it in his 1921 Princeton lectures (Einstein 1923). Over several pages, he repeated his Machian-inspired cosmological analysis at length (110–119). It included his formulation of Mach's principle:

> If we think these ideas consistently through to the end we must expect the *whole* inertia, that is, the *whole* $g_{\mu\nu}$-field, to be determined by the matter of the universe, and not mainly by the boundary conditions at infinity (113, his emphasis).

The second physical condition was that an admissible cosmology must have a static line element in which, in particular, $g_{44}$ is constant. This condition had been carefully deduced in his 1917 cosmology paper from what he took to be the most basic of astronomical observations. He gave its starting point as

> The most important fact that we draw from experience as to the distribution of matter is that the relative velocities of the stars are very small as compared with the velocity of light (Einstein, 1917a, 184).

Proceeding step by step from this observational foundation, Einstein deduced that, to close approximation, the metric for spacetime on a cosmic scale must be static, that is, in some coordinate systems, satisfy

$$g_{44} = 1 \quad and \quad g_{14} = g_{24} = g_{34} = 0.$$

The added condition of the constancy of $g_{44}$ follows since, in the first approximation, it is the variability of $g_{44}$ alone that controls the acceleration of free bodies.[27] A constant $g_{44}$ in the first approximation precludes any significant accelerations for the stars. The vanishing of the remaining metrical coefficient is not explained beyond "as always with static problems." Presumably the intent is to preclude rotating coordinate systems in which Coriolis forces would appear.

This second condition completed the case for the rejection of the de Sitter hyperboloid as a viable cosmology. With the singularity of (7) reinterpreted as a mass concentration, this form of the de Sitter solution was compatible with Mach's principle. The line element (7) was static in the sense that the metrical coefficients were independent of the time coordinate $x_4$ and conformed with $g_{14} = g_{24} = g_{34} = 0$. However, its $g_{44} = \cos^2(r/R)c^2$ was not constant and dropped to zero as the mass horizon was approached. The outcome was that free masses, such as stars, would be accelerated towards the cosmic masses at the mass horizon, just as we would expect from Newtonian intuitions.

This second condition was repeatedly stressed by Einstein as precluding the de Sitter line element of (7) as a viable cosmology. He wrote to de Sitter on June 14, 1917:

> It seems to me that a reasonable interpretation of the world before us necessarily requires the approximate spatial constancy of the $g_{44}$, on account of the fact of the small relative motion of the stars (Schulmann et al. 1998, Doc. 351).

Einstein gave the same argument against the de Sitter line element (7) and in favor of his original 1917 cosmology in his letter of early June 1918 to Klein (Schulmann et al. 1998, Doc. 556):[28]

> If the world were really so [as (7) depicts], then the fixed stars must have tremendous speeds, so that their statistical distribution could be maintained [*sich erhalten*], on account of the colossal differences of gravitational potential which must be present between the different points of such a world. The non-

---

27   As shown for example in Einstein (1923, 89).

28   It is unclear to me how Einstein imagines that a rapid motion could *maintain* the observed uniform distribution.

existence of great star speeds compels us to believe that matter in the large is
not distributed unduly non-uniformly over the world.

Einstein did not waver from this second condition. A careful recapitulation of it forms the
second part of the case Einstein made for his cosmology in his 1921 Princeton lectures
(Einstein, 1923, 113–119).

The combination of these two conditions finally elucidates a puzzle in Einstein's (1918)
discussion of March 6 of Mach's principle and his $\lambda$-augmented gravitation field equations.
He there expressed his expectation that (243):

> A singularity-free space-time-continuum with everywhere vanishing energy
> tensor of matter appears not to exist according to [$\lambda$-augmented gravitational
> field equations].

This is a curious remark for Einstein to make a year after de Sitter had shown that just such
a solution exists in the full hyperboloid. We might charitably guess that Einstein was tacitly
including a further condition that the solution sought must be static, for then the full de
Sitter hyperboloid would be excluded.

That this was the case is indicated by the correction to this remark that Einstein made
to Klein in his letter of June 20, 1918 (Schulmann et al. 1998, Doc. 567). We saw above
his concession in that letter that "de Sitter's world is, in itself, free of singularities." That
concession was immediately followed by recalling the condition that imposes a singularity
onto the solution: "A singularity only comes about through the substitution that affords
the transition to the static form of the line element." A few lines later, Einstein elaborated:

> My critical remark about the de Sitter solution requires a correction; there is in fact
> a singularity-free solution of the gravitational equations without matter. However,
> this world may in no way come into consideration as a physical possibility. For
> there can be no time $t$ specified so that the three-dimensional slices $t =$ const. do
> not intersect, and so that these slices are equal (metrically) to one another.

## 4.8 IN RETROSPECT

For all his certainty, Einstein's position proved fragile. The astronomical affirmation of the
expansion of the nebulae came within a decade by the end of the 1920s. It is fitting that
Einstein and de Sitter (1932) co-authored a note in which the requirement of a static line
element and the added $\lambda$ term in the gravitational field equations were both retracted. They
posited a non-static, expanding Einstein-de Sitter spacetime, distinct from the 1917 de
Sitter solution. Einstein's defense of Mach's principle collapsed by the time of the writing
of his "Autobiographical Notes." He dismissed Mach's analysis (1949, 29) as inappropriate
for a consistent field theory. Einstein's $\lambda$ outlived him. More recent work in cosmology
has found that it is needed to accommodate our astronomical observations. Our current $\lambda$
CDM model of cosmology indicates that, in the distant future, a $\lambda$ driven expansion of the
universe will bring our cosmos ever closer to a rapidly expanding, de Sitter spacetime with
asymptotically depleted matter. Einstein would surely not be pleased.

# PART II: THE HISTORICAL CONTEXT

To examine Einstein's treatment of these three singularities in isolation gives an incomplete
picture of Einstein's understanding of them. His work, as we shall see in this Part II, cohered
with an analytic tradition in mathematics that has long fallen from favor in modern work
in relativity; we shall see in the next Part III that his treatment of singularities conformed
with one of the most enduring themes of Einstein's work in physics, his sustained efforts
to eliminate arbitrariness from physical theories.

# 5. ANALYTIC AND SYNTHETIC GEOMETRY

Einstein's privileging of the analytic approach was not a capricious partition of a homogeneous body of mathematics in distinct parts. For centuries, two traditions—the analytic and the synthetic—had cohabited in geometrical writings in a fertile synergy. The existence of the two traditions was a commonplace of mathematics when Einstein began his work on general relativity. The leading geometer of his time, Felix Klein, wrote a series of volumes intended for mathematics teachers in the high schools. The second volume on geometry derived from lectures Klein gave in the summer semester of 1908. In them, he wrote:[29]

> However, I should like to add to this account an explanation of the *difference between analytic and synthetic geometry*, which always plays a part in such discussions. According to their original meaning, synthesis and analysis are different methods of presentation. Synthesis begins with details, and builds up from them more general, and finally the most general, notions. Analysis, on the contrary, starts with the most general, and separates out more and more the details. It is precisely this difference in meaning which finds its expression in the designations synthetic and analytic chemistry. Likewise, in school geometry, we speak of the *analysis of geometric constructions*: we assume there that the desired triangle has been found, and we then dissect the given problem into separate partial problems.
>
> In higher mathematics, however, these words have, curiously, taken on an entirely different meaning. *Synthetic geometry is that which studies figures as such, without recourse to formulas, whereas analytic geometry consistently makes use of such formulas as can be written down after the adoption of an appropriate system of coordinates.* Rightly understood, there exists only a *difference of gradation* between these two kinds of geometry, according as one gives *more prominence to the figures or to the formulas*. Analytic geometry which dispenses entirely with geometric representation can hardly be called geometry; synthetic geometry does not get very far unless it makes use of a suitable language of formulas to give precise expression to its results. Our procedure, in this course, has been to recognize this, for we used formulas from the start and we then inquired into their geometric meaning (1909, 110–112, emphasis in original).

The difference sketched here by Klein is the age-old difference between Euclid's and Descartes' geometries. It is so important to the present paper that we give a banal example to illustrate it and to show how it can become problematic.

We recover the same result in Cartesian geometry by means of the two variables $x$ and $y$ and a linear relation over them:

$$Ax + By = 0 \tag{15}$$

for real constants $A$ and $B$. If the variables $x$ and $y$ are understood to be the coordinates of a Euclidean two-dimensional surface, this is the equation of a straight line through the origin (0,0). The isotropy of the possible dispositions of such lines is recovered from the leading idea of Klein's "Erlangen Program": each geometry is defined by the invariants of its characteristic group. The group of affine transformations:

$$x' = ax + by \quad y' = cx + dy \tag{16}$$

---

for real constants $a$, $b$, $c$ and $d$, is sufficient to characterize the isotropy of the disposition of these lines. Under the action of this group, the line represented by (15) is mapped to every other possible straight line passing through the origin. The totality of the lines is an invariant structure that is mapped to itself by the affine transformations (16). That fact is expressed most compactly as a formal property of a symbolic equation: the covariance of the equation (15) under the affine transformations of the variables (16).

We can see how easily the compatibility of the two approaches can be disturbed. Assume that, instead of representing straight lines by the linear relations (15), we had proposed a functional dependence of the variable $y$ on $x$:

$$y = Cx \tag{17}$$

for a real constant C. In this representation, the line along the $y$-axis corresponding to $x = 0$ becomes anomalous. It corresponds to something like the constant C taking an infinite value, or, perhaps, more carefully, all possible values of y must correspond to the single value of the variable $x = 0$. The condition that (17) is a function is violated. The anomaly persists, even allowing for the covariance of (17) under affine transformations, in so far as an affine transformation would need to contradict the functional property of (17) to eradicate the anomaly. There is, of course, no corresponding anomaly in the Euclidean representation.

For definiteness, we summarize the key ideas of the two approaches. They share the same elements: the notion of a geometric space or surface; and algebraic equations in variables adapted to the surfaces. The key difference is which takes priority:

> *Synthetic approach*: The geometric surfaces or spaces are the primary subject of investigation. The equations serve as an analytic means of describing them and anomalies in the equations may be neglected if they do not correspond to anomalies in the spaces.
>
> *Analytic approach*: The equations and their transformations are the primary subject of investigation. The geometric surfaces and spaces are supplementary and sometimes merely have an heuristic role in aiding in the construction of suitable equations.

The two approaches are complementary. It would be difficult to pursue modern work in spacetime theories without employing both. However different theorists will differ in the emphasis placed on each. In the most extreme example of the analytic approach, the geometric connections are all but completely suppressed. This happens so routinely that we now scarcely notice it. The volume $V$ and temperature $T$ of an ideal gas are related linearly, with the constant of proportionality fixed by the mass of the gas sample. Geometrically, each of these linear relations are represented by a straight line in $VT$ space. This geometric fact is relegated to elementary didactics, if it is noticed at all. We would dismiss tensions between the geometry and analysis as anomalies of the geometry. The straight line along the $T$-axis where $V = 0$ does not represent a sample of gas of zero mass that retains its zero volume at all temperatures. Rather it represents no gas at all.

Einstein, we have seen, favored the analytic approach heavily and gave its equations priority over the geometrical surfaces investigated in the synthetic approach.

# 6. TWO TRADITIONS IN EINSTEIN'S TIME

In the early twentieth century, when Einstein was developing his general theory of relativity, he drew on two manifestations of the analytic and the synthetic approaches.

Synthetic geometry contributed to his theory through Gauss' theory of surfaces, and the analytic approach contributed through Christoffel's theory of the invariants of quadratic differential forms. While the connections between them were obvious, the two had developed as independent mathematical traditions.[30]

## 6.1 GAUSS' THEORY OF SURFACES

Gauss' theory of surfaces provided the geometrical component of the foundations upon which general relativity is built. It was developed by Gauss in several documents in Latin in the 1820s. They are collected and translated into English as Gauss (1902). The theory considered a two-dimensional surface of varying curvature with coordinates p and q, embedded in a three-dimensional Euclidean space, with coordinates $x$, $y$, and $z$. The distance between neighboring points in the Euclidean space was given by the linear element $\sqrt{dx^2 + dy^2 + dz^2}$ and the corresponding linear element in the curved surface was given by (e.g., 1902, 20, 47):

$$\sqrt{Edp^2 + 2Fdpdq + Gdq^2} \qquad (18)$$

for $E$, $F$ and $G$ suitable functions of $p$ and $q$. Gauss then computed the curvature at each point of the surface from the radii of curvature of sections by normal planes (1902, 15, 97) and also the curves of shortest distance (e.g., 1902, 97). The embedding in the Euclidean space was only an artifice used by Gauss to derive the geometrical self-contained properties of the two-dimensional curved surface, which could be understood without reference to the embedding space.

The extension of Gauss' theory to higher dimensional spaces was undertaken at Gauss' instigation in Riemann's (1854) celebrated *Habilitationsvortrag*, "On the Hypotheses which lie at the Foundations of Geometry." Gauss had embedded his two-dimensional curved surfaces in a higher dimensional Euclidean space. Riemann sought to extend Gauss' analysis without using the artifice of an embedding space. Lacking a serviceable notion in then existing mathematics, Riemann struggled awkwardly to introduce the key concept: an $n$ dimensional manifold of points (*n fach ausgedehnten Mannigfaltigkeit* = $n$-fold extended manifold). He would then turn to the metrical relations defined on these manifolds.

When Einstein was developing general relativity, Gauss' theory had developed into what had then come to be called "differential geometry." The authoritative work was "Bianchi-Lukat" (Lukat 1910). It is Max Lukat's authorized German translation of the second expanded edition of Luigi Bianchi's *Lezioni di geometria differenziale*. While it introduces now familiar analytic mathematics associated with quadratic differential forms, it is a work of synthetic geometry. The analytic methods are secondary to its primary focus, the geometry of spaces and surfaces. What are their metrical properties? What are their curvature and geodetic properties? Which coordinate systems are usefully adapted to them? Which are the conformal transformations of the coordinates? It is an expansive work that continues in this vein for 24 chapters and over 700 pages.

## 6.2 THEORY OF QUADRATIC DIFFERENTIAL FORMS

Through Gauss' introduction of the line element (18), quadratic differential forms became the foundational analytic structure of the geometry of spaces of variable curvature. The mathematics of these differential forms could, however, be investigated independently of their application in geometry. Such investigation provided the analytic component that figured prominently in the mathematics used by Einstein in his general theory of relativity.

---

30     See Reich (1994, 184) for more discussion of how the two programs of research in geometry and analysis remained distinct at the time of Einstein's discovery of general relativity, even though connections between the two programs were then recognized.

A foundational work historically in these investigations is Christoffel's (1869) "On the transformations of homogeneous differential expressions of the second order." The project of the paper is laid out in its first paragraphs.

> In the differential expression
>
> $$F = \sum \omega_{i,k} \partial x_i \partial x_k; \qquad i, k = 1, 2, \dots n,$$
>
> the coefficients $\omega$ are arbitrary functions of the variables $x_1, x_2, \dots x_n$ that are independent of one another. If, instead of the latter, one introduces a system of functions of the new variables $x_1', x_2', \dots x_n'$ that are [also] independent of one another, then $F$ goes over into a new differential expression
>
> $$F' = \sum \omega_{i,k}' \partial x_i' \partial x_k'$$
>
> that is equal to the original by virtue of the substitution carried out.
>
> Conversely if the differential expressions $F$ and $F'$ are given, then one can pose the question of which conditions are necessary, such that the one could be transformed into the other, and in case this is possible, which substitutions bring about the required transformation.

What is notable is what is present and what is not present in this opening statement.

The project *is* formulated fully within analysis in terms of formulae for $F$ and $F'$, expressed as functions of the variables $x_1, \dots x_n$, It is to find substitutions of the variables under which the quantities $F$ and $F'$ are equal, so that $F$ is invariant; and to determine conditions necessary for the existence of the substitutions.

The project is *not* formulated as one in geometry. The variables are not coordinates of a geometric surface; the invariant $F$ is not the metrical line element of such a surface; and the associate coefficients $\omega_{i,k}$ are not coefficients of a metric tensor. They are simply functions of the variables indicated and the project is to ascertain their behavior under substitutions of the variables.

The analysis continues without any explicit connection to geometry. What we now know as the "Christoffel symbols" are introduced. What we now call the "Riemann-Christoffel curvature tensor" is introduced without any special fanfare or name. It is simply a quantity, denoted by four indices "($gkhi$)" (1869, 54), that has a role in determining which transformations leave F unchanged. It is only in the last paragraph of the last page that a connection to geometry is made. Christoffel mentions that his quantity F is also known as the "square of the line element belonging to the space of three dimensions" and he directs his readers for details to Riemann's celebrated, posthumously published *Habilitationsvortrag* of 1854.

This attribution marked the beginning of the now long-standing tradition of associating the origin of this four-index symbol "($gkhi$)" with Riemann's efforts to extend Gauss' theory of curved surfaces in his address of 1854. As readers of Riemann's *Habilitationsvortrag* know, this poses a historical puzzle. The quantity does not appear in the *Habilitationsvortrag*. For a work of such importance in geometry, it is striking for having few analytic expressions. Readers who persist in the quest for Riemann's formulation of this quantity are directed to a later work in thermal physics, Riemann's (1861) *Commentatio*. The quantity that comes to be known as the "Riemann-Christoffel curvature tensor" appears on page 95, in an equation that is labeled "(I)." The vanishing of the quantity in (I) is the condition under which a general quadratic differential form, written as $\sum b_{\iota\iota'} ds_\iota ds_{\iota'}$, can be transformed into a simplified form, written as $\sum (ds_{\iota'})^2$.

While later literature commonly presents this equation (I) as a direct development of the geometric ideas of Riemann's *Habilitationsvortrag*, a study of this literature by Farwell and Knee (1990) finds no sound basis for this interpretation. Rather, they find that the equation is developed independently of the geometry. They locate it in the development of what they call "tensor analysis" and conclude:

> Riemann did not consider the equivalence of forms in relation to geometry, but rather in the context of heat conduction. Riemann's mathematical derivations in the second part of the "Commentatio" contain no reference to heat conduction, but equally they contain no reference to geometry. The one allusion to geometry is an illustration, which is not linked to heat conduction and does not obviously therefore serve as a 'useful addition' (237).

No doubt Riemann recognized the importance of his equation (I) in differential geometry. However, his development of it proceeded independently of the geometry and belongs within the tradition of the analysis of quadratic differential forms.

Ricci and Levi-Civita's (1900) "Methods of the Absolute Differential Calculus and their Applications" continues in the analytic tradition of Christoffel. Its purpose is to present what it calls the "absolute differential calculus" as a method with many applications. Einstein's use of the calculus in general relativity has led to it coming to be known as "tensor calculus" and for it to be connected almost inseparably with Einstein's theory. Hermann's (1975, ix) English translation, for example, renders Ricci and Levi-Civita's "*systèmes covariants et contrevariants*" as "Covariant and Contravariant Tensors" and their "*Quadrique fondamentale*" [fundamental quadratic differential form] as "Riemannian-Metrics." (See *Appendix: Einstein and Grossmann Generalize the Term "Tensor."*)

Ricci and Levi-Civita did not conceive their calculus as a theory of differential geometry. It was conceived as a general, analytic tool, one of whose applications lay in geometry. Modern readers, expecting a work in differential geometry, should pause at the opening sentence:

> Let us denote by T a completely general transformation of variables
>
> (1)                         $x_i = x_i(y_1, y_2, ..., y_n)$
>
> that is bijective and regular in the domain we will consider; ... (1900, 128)

The analysis concerns variables. These $x$'s and $y$'s are not, in general, coordinates. They are only identified as such when later sections apply the methods to geometry. Much of the article reviews applications outside geometry. Chapter III considers applications in analysis. Chapter V treats applications in mechanics, which we would now identify as Lagrangian and Hamiltonian mechanics. Chapter VI is labeled as applications in physics. It investigates various field equations, such as Laplace's equation for a potential, vector formulations involving the divergence and curl, and the differential equations of electrodynamics and heat conduction.

The article does make many connections to geometry. Chapter II "Intrinsic geometry as an instrument of the calculus," seeks to re-express the concepts of the calculus in geometric terms. It starts by identifying the quadratic differential form, expressed in terms of variables, with the line element of a geometry, expressed in terms of coordinates. A connection to Gauss' theory of surfaces, however, is not made until Chapter IV "Geometric Applications." Gauss' theory is treated as something independent of the calculus. Ricci and Levi-Civita lament that the expositions of Gauss' theory lack a unified method. Their goal is to apply their calculus to provide one:

> The absolute differential calculus, on the contrary, leads to [a unified method] without any effort, by giving the theory as simple a form as possible (1900, 165).

We find a similar separation of the mathematics of quadratic differential forms in Wright's treatise on quadratic differential forms. Its first "historical" chapter introduced the notion of a quadratic differential form through the line element of Gauss' theory of surfaces. The chapter concludes by separating the general theory from this one application:

> Thus far it is suggested that the invariants [of quadratic differential forms] are essentially connected with differential geometry. This is by no means the case. They are connected with a certain form, and any interpretation of this form leads to a corresponding interpretation for the invariants (1908, 4).

To illustrate other interpretations, Wright notes that kinetic energy in Lagrangian mechanics is expressed by a quadradic differential form in the mechanics generalized coordinates.

Decades later, after Einstein's general theory of relativity had become the most prominent application of Ricci and Levi-Civita's calculus, Levi-Civita still presented the absolute differential calculus as a method, distinct from differential geometry, which was one of its principal applications. Levi-Civita's (1926) Absolute Different Calculus was given the parenthetic subtitle "(Calculus of Tensors)." This separation is marked by its care in distinguishing the variables of the calculus from the coordinates of a space. Chapter IV is devoted to problems in analysis. It opens with:

> This chapter is devoted to the study of the effect on some analytical entities of a change of variables (61).

It continues with the specification of the notion of a transformation of *variables*:

> Consider $n$ independent variables $x_1$, $x_2$, ... $x_n$, which we shall as usual denote collectively by $x$, and suppose a transformation applied to them which leads to another set of $n$ independent variables $\overline{x}$; ... (61).

That this notion is, in principle, more general than the purely geometric notion, Levi-Civita then explained:

> The geometrical name for this operation is of course *change of co-ordinates*; ..." (1926, Levi-Civita's emphasis).

## 7. EINSTEIN AND GROSSMANN ADOPT THE ABSOLUTE DIFFERENTIAL CALCULUS

We learn from later autobiographical recollections how Einstein came to the mathematical methods used in his general theory of relativity. He had by 1912 an already well-developed program of research that sought a generalization of the principle of relativity and a theory of gravitation. In it, extending coordinate transformations beyond the Lorentz transformation would figure centrally.[31] After Einstein's return to Zurich in August 1912, he approached a friend from his university days, the mathematician Marcel Grossmann, for assistance. Einstein's description (1956, 16) of the problem posed to Grossmann is likely somewhat anachronistic: how could he find generally covariant equations that would govern the metric tensor $g_{ik}$ of spacetime. Grossmann searched the literature and found the work of Riemann, Ricci and Levi-Civita, that, Einstein recalled, extended Gauss' theory of curved surfaces. Most important was what Grossmann soon called the "Riemann differential tensor" (Einstein and Grossmann 1913, 35) and, in slightly variant form, the "Christoffel four-index-symbol of the first kind" (36).

---

31  See Norton (2020) for details.

Einstein and Grossmann's collaboration resulted in the first sketch of the general theory of relativity, the "*Entwurf* …" or "Sketch …" paper, Einstein and Grossmann (1913). It contained the first published exposition of the mathematical methods to be used in the developing theory. Einstein delegated the exposition of these methods to Grossmann's "Mathematical Part." Grossmann made quite clear at the outset that his exposition would be analytic in its orientation, reflecting Ricci and Levi-Civita's approach. He began his exposition by acknowledging his primary sources, Christoffel's (1869) and Ricci and Levi-Civita's paper (1900, 23):

> The mathematical aids for the development of the vector analysis of a gravitational field, which is characterized by the invariance of the line element
>
> $$ds^2 = \sum_{\mu\nu} dx_\mu dx_\nu,$$
>
> goes back to the fundamental paper by Christoffel [(1869)] on the transformation of quadratic differential forms. Proceeding from Christoffel's results, Ricci and Levi-Civita [(1900)] have developed their methods of the absolute differential calculus, i.e. [absolute in the sense] of independent of the coordinate system. It enables the differential equations of mathematical physics to be given an invariant form.

The exposition that follows focused on analytic expressions and how they transform under coordinate transformations. The topics covered are standard and familiar to modern readers: the transformation rules for covariant and contravariant tensors, the conversions among covariant and contravariant tensors, the formation of new covariant quantities by covariant differentiation, and eventually the Riemann tensor.

Once one looks for it, it is striking that there is essentially no explicit mention of any geometrical ideas. The focus is narrowly on the transformation properties of various analytic expressions. The omission was no oversight, as Grossmann made clear, in an admission that modern readers might find startling:

> … I have deliberately set aside geometrical aids, since, in my opinion, they contribute little to the intuitive understanding [*Veranschaulichung*] of the formation of concepts of the vector analysis (1913, 24).

This treatment by Grossmann became the template used by Einstein in his subsequent developments of the mathematical methods needed by his new general theory of relativity. It is used in Einstein's (1914) review of the still incomplete general theory of relativity; his triumphant (1916) review of the completed theory; and his text-book like exposition (1923).

These expositions followed Grossmann's approach of focusing on the transformation properties of analytic expressions and shared its aversion to geometrical aids. Presumably, this reflected Einstein's comfort with the analytic approach. It conforms with the question within the analytic approach that Einstein put to Grossmann at the outset in 1912. He asked after an equation and its transformation behavior:

> Is there a differential equation for the $g_{ik}$ that is invariant under non-linear coordinate transformations? (Einstein 1956, 15).

He did not ask the synthetic question of what spacetime geometry might generalize the principle of relativity and accommodate gravity.

There was an eclectic character to Grossmann's exposition. He sought to use the methods of Ricci and Levi-Civita to subsume an existing literature of the vector analysis of four dimensional spacetimes. This vector analysis originated with Minkowski and was developed by Sommerfeld and Laue. One trace of this subsumption is the idiosyncratic use of the term "tensor" from vector analysis in the "*Entwurf …*" paper. Its use subsequently became standard in both physics and mathematics. (For details, see *Appendix: Einstein and Grossmann Generalize the Term "Tensor."*) In his introductory remarks, Grossmann sought to reassure his readers of the success of the subsumption:

> … the greater generality of the concepts formed in [the new general vector
> analysis] gives it a clarity that the special cases often lack (1913, 23).

That the subsumption was successful may well be correct from Grossmann and Einstein's perspectives. However, it had all but eradicated the synthetic-geometric perspective that Minkowski himself had advanced. In his celebrated popular lecture, "Space and Time," Minkowski sought to suppress the analytic, that is, the emphasis on the transformation properties of the equations. He favored an elevation of the synthetic geometry in the concept of his four-dimensional world of spacetime. He wrote:

> … the word relativity-postulate for the requirement of an invariance with the
> [Lorentz] group $G_c$ seems to me very feeble. Since the postulate comes to mean
> that only the four-dimensional world in space and time is given by phenomena,
> but that the projection in space and in time may still be undertaken with a
> certain degree of freedom, I prefer to call it the *postulate of the absolute world*
> (or briefly, the world-postulate) (1908, 83, his emphasis).

Einstein made clear that he eschewed Minkowski's geometric conception of special relativity in the introduction to his 1916 review article on his general theory. He characterized the advance made by Minkowski as:[32]

> The generalization of the theory of relativity was facilitated greatly through
> the conception the special theory of relativity was given by Minkowski. This
> mathematician first recognized clearly the formal equivalence of the spatial
> coordinates and the time coordinate and made it usable for the construction of
> the theory (Einstein 1916, 769).

What is notable in Einstein's characterization of the advance is Einstein's omissions of the idea of spacetime and its geometry. He has reduced Minkowski's contribution to one that is expressible in the analytic language of "*formal* equivalence" of coordinates.

## 8. EINSTEIN PRIVILEGES THE ANALYTIC

The methods Einstein and Grossmann introduced were an eclectic mix of ideas: geometrical ideas from Gauss' theory of surfaces; analytic ideas from Christoffel's treatment of the invariants of quadratic differential forms; and vectorial ideas from the mathematics used in electrodynamics and special relativity. All are present in one form or another. However, when Einstein had to choose among the ideas, he favored the analytic. In addition to the cases noted above, here are just a few more examples of this favoring that permeates Einstein's treatments of general relativity.

### 8.1 VARIABLES OR COORDINATES?

We noted above that Ricci and Levi-Civita called their $x$'s and $y$'s "variables" when they were part of a general analysis of expressions. It was almost only when they applied their methods to geometry that these variables were relabeled as "coordinates" (*coordonnées*). Einstein's *x*'s and *y*'s were almost always called "coordinates" (*Koordinaten*). On one occasion of critical importance in the physical interpretation of the general relativity, he slipped back into the language of variables. In his review article, Einstein (1916, 776–777) offered his "point-coincidence" argument for the physical basis of a requirement of general covariance. We need not rehearse the content of the argument,[33] but only note how he represented his *x*'s:

> … We allot to the universe four space-time *variables* $x_1, x_2, x_3, x_4$, in such a way that for every point-event there is a corresponding system of values of the *variables* $x_1 … x_4$. To two coincident point-events there corresponds one system of values of the *variables* $x_1 … x_4$, i.e. coincidence is characterized by the identity of the co-ordinates. If, in place of the *variables* $x_1 … x_4$, we introduce *functions* of them, $x_1'$, $x_2', x_3', x_4'$, as a new system of co-ordinates, … (1916a, 117–118, my emphasis).

---

32   This passage is from the first page of Einstein's review article. This page was omitted from Einstein (1916a) in the later, widely read collection, *The Principle of Relativity*.

Correspondingly, the variables are not "transformed" as coordinates usually are said to be. The new variables are introduced as *functions* of the old variables. We find a similar reversion to variable language when Einstein formulates his requirement of general covariance. He wrote in emphasized text:

> *The general laws of nature are to be expressed by equations which hold good for all systems of co-ordinates, this is, are co-variant with respect to any substitutions whatever (generally covariant)* (1916a, 117).

The term "substitutions" connotes the analytic operation of replacing one variable in an equation by another functionally dependent on it.

## 8.2 THE PRINCIPLE OF EQUIVALENCE

Einstein's principle of equivalence played a central role both in Einstein's discovery of his general theory of relativity and in his interpretation of the physical content of the theory. The principle, however, has proven quite recalcitrant to univocal interpretation. There is an expansive literature on the principle both struggling to understand its content and offering alternative versions of it.[34] Here is the principle in Einstein's review article:

> For, if we now assume the special theory of relativity to apply to a certain four-dimensional region with the co-ordinates properly chosen, then the $g_{\sigma\tau}$ have the values given in (4) [diagonal form]. A free material point then moves, relatively to this system, with uniform motion in a straight line. Then if we introduce new space-time co-ordinates $x_1, x_2, x_3, x_4$, by means of any substitution we choose, the $g^{\sigma\tau}$ in this new system will no longer be constants, but functions of space and time. At the same time the motion of the free material point will present itself in the new co-ordinates as a curvilinear non-uniform motion, and the law of this motion will be independent of the nature of the moving particle. We shall therefore interpret this motion as a motion under the influence of a gravitational field. We thus find the occurrence of a gravitational field connected with a space-time variability of the $g_{\sigma\tau}$ (1916a, 120).

This formulation asserts that a Minkowski spacetime in one set of coordinates has no gravitational field, but in another set of coordinates it does. This understanding has proven unintelligible to many later commentators. Synge's lament is now classic. He described his role:

---

33  For details, see Giovanelli (2013).

34  See Norton (1985) and Lehmkuhl (2022) for a variety of attempts to provide a cogent reading of Einstein's formulations.

It is to support Minkowski's way of looking at relativity that I find myself
pursuing the hard path of the missionary (1960, ix).

Einstein's principle of equivalence was singled out for exasperated disdain:

> … I have never been able to understand this Principle. … In Einstein's theory,
> either there is a gravitational field or there is none, according as the Riemann
> tensor does not or does vanish. This is an absolute property; it has nothing to do
> with any observer's world-line. Space-time is either flat or curved, …

Synge's lament gives pithy expression to the understanding of a synthetic geometer who
takes the geometry of the spacetime to be primary: there is just one spacetime. Changing
the coordinate system does not alter it physically.

Einstein's account becomes intelligible if we allow that his was not the synthetic approach.
What mattered to him were the analytic expressions. In one coordinate system, the analytic
expressions represent the motion as a linear relation among the coordinates. In the other
coordinate system, the analytic expressions represent the same motion as a non-linear
relation among the coordinates. Since he was not proceeding in the synthetic approach,
the spacetime geometry has merely a supplementary role and does not control the physical
interpretation of the analytic expressions. Thus, Einstein can assign a different physical
interpretation to each of the these differing analytic expressions. And he does. One
corresponds to a gravitation-free case, the other to the presence of a gravitational field.

That analytic expressions related by a simple coordinate transformation can represent different
physical situations will surely seem untenable if one takes the geometry of the spacetime
as primary. A manuscript on special and general relativity by Einstein (1920), written but
not published in 1920, shows how Einstein conceived the separation physically of the two
expressions. After a lengthy discussion of special relativity, he turned to general relativity.
His introduction of "The Basic Idea of the Theory of General Relativity in Its Original Form"
in Part II, Section 15, took an unexpected turn. He recalled the thought experiment of the
magnet and conductor from his discovery of special relativity. In an inertial frame in which a
magnet is at rest, the magnet is surrounded by a pure magnetic field. If one transforms to an
inertial frame moving with respect to the magnet, a new entity, an electric field, is produced
by the time varying magnetic field in the new inertial frame.[35] He concluded (his emphasis):

> … the existence of the electric field was therefore a relative one, depending
> on the state of motion of the coordinate system used; and only the electric
> and magnetic fields combined could be granted a kind of objective reality,
> aside from the state of motion of the observer or, corr[espondingly], the
> coordinate system.

Einstein then recalled the "happiest thought of [his] life" that, in 1907, inspired the
principle of equivalence and set him on the pathway to general relativity:

> The gravitational field only has a relative existence, in a way similar to the
> electric field produced by magneto-electric induction (1920).

That is, the gravitational field of the accelerating coordinate system should be treated
in the same way as the induced electric field. It has a relative existence depending on
the coordinate system chosen, even if only the perspectives of both coordinate systems
combined should be accorded objective reality.

---

35   For discussion of the importance of this thought experiment in Einstein's discovery of
special relativity, see Norton (2004).

We need not labor too hard to convince ourselves of the cogency of these attributions of reality. All that matters here is that Einstein was convinced of their cogency. He was quite comfortable relating attributions of reality to observers' states of motion. In his defense of the reality of the Lorentz length contraction, he summarized its status as:

> … it [Lorentz contraction] is not "real" in so far as it does not exist for a co-moving observer; however, it is "real" for a non-co-moving observer, i.e. [it is real] in that it can be demonstrated in such a way by physical means (Einstein 1911).

## 8.3 THE RIEMANN-CHRISTOFFEL TENSOR

Einstein's treatment of the Riemann-Christoffel tensor again illustrates his privileging of the analytic expressions and their transformations over the geometry of the spaces. He wrote in his 1916 review article:[36]

> The mathematical importance of this tensor is as follows: If the continuum is of such a nature that there is a co-ordinate system with reference to which the $g_{\mu\nu}$ are constants, then all the [components of the Riemann-Christoffel tensor] $B^{\rho}_{\mu\sigma\tau}$ vanish. If we choose any new system of co-ordinates in place of the original ones, the $g_{\mu\nu}$ referred thereto will not be constants, but in consequence of its tensor nature, the transformed components of $B^{\rho}_{\mu\sigma\tau}$ will still vanish in the new system. Thus the vanishing of the Riemann tensor is a necessary condition that, by an appropriate choice of the system of reference, the $g_{\mu\nu}$ may be constants. In our problem this corresponds to the case in which,*[37] with a suitable choice of the system of reference, the special theory of relativity holds good for a *finite* region of the continuum (1916a, 141, his emphasis).

Here, Einstein's interpretation follows closely the analytic understanding of Christoffel and Riemann in his analytic mode. The quantity provided necessary and sufficient conditions for one analytic expression—a quadratic differential form—to be transformed into another.

It also reaffirms Einstein's interpretation of the principle of equivalence. Special relativity is recovered only after a coordinate system is adopted in which the metric is diagonalized. In other coordinate systems, such as those that are adapted to acceleration, a real gravitational field is present and we have left special relativity.

What matters for our purposes is this privileging of the analytic expressions. What matters also is what is missing: There is no mention of the geometry of spaces. Gauss' treatment of the same result, in the case of a two-dimensional, positive definite geometry, directly expresses it in terms of the geometry of surfaces. It is his "remarkable theorem" (*theorem egregium*) Gauss wrote:[38]

> THEOREM. *If a curved surface can be developed onto any other surface, the measure of curvature at every point remains unchanged.*
> It is also evident *that any finite part of a curved surface, after developing on another surface, will retain the same curvature.*

---

36   Reich (1994) has also emphasized how Einstein's article avoided the geometrical connotations of the Riemann-Christoffel tensor. She remarked, for example, on Einstein's discussion of the tensor (210) "The word curvature does not even appear."

37   Einstein's footnote, his emphasis: "The mathematicians have proved that this is also a *sufficient* condition."

38   My translation corrects the English translation of (Gauss 1902, 47) which combines sentences and drops some text from the original Latin.

A special case, to which geometers have hitherto restricted their investigations, is that of surfaces developable onto a plane. Our theory automatically teaches that the measure of curvature of such surfaces at any point is = 0 (1828, 24–25, his emphasis).

Here, the older expression to be "developed" means to be mapped in a way that preserves metrical properties. A cylinder, for example, can be mapped onto a flat plane by unrolling the cylinder.

Gauss' *theorem egregium* can be promoted to the spacetimes of special and general relativity. A version of it expresses the significance of the vanishing of the Riemann-Christoffel tensor for those who privilege the geometry. An example arises in Eddington's (1923) relativity text in Chapter 3, Section 36 (Eddington's emphasis):

> *Hence the vanishing of the Riemann-Christoffel tensor is a necessary condition for flat space-time.* This condition is also *sufficient*—if the Riemann-Christoffel tensor vanishes space-time must be flat.

Others gave essentially similar formulations at this time, such as in Tolman (1934, 186).

## 8.4 EINSTEIN ON MINKOWSKI

That Einstein's understanding of the Riemann-Christoffel tensor so avoids geometrical conceptions is no aberration of the one example. We saw it already in Einstein's assessment of the advance brought by Minkowski's reformulation of Einstein's 1905 special theory of relativity as a theory of spacetime (1916, 769). This striking assessment was repeated elsewhere.

We would now say that Minkowski's contribution fundamentally reconfigured what we take the theory to be. In Einstein's original formulation, it was a theory of the Lorentz transformation of physical quantities. Minkowski's formulation introduced a new concept, spacetime, and special relativity became the theory of its geometry. That theory introduced geometric structures that were not present in Einstein's formulation. They include the light cones that divide spacetime displacements into timelike, spacelike, and lightlike; a notion of four-dimensional distance in the spacetime, the interval; that moving bodies trace out four-dimensional, timelike worldlines; that hyperbolas are the geometric figures analogous to circles in Euclidean geometry; and so on.

In his popular *Relativity*, Einstein reported these geometric conceptions in a brief 17th chapter, after 16 chapters that recount the theory without them. When that already brief chapter came to report what was important in Minkowski's reformulation of his theory, it identified nothing geometrical in conception. Rather, it identified how Minkowski's work enabled a clarification of the formal expressions used in the theory:

> But the discovery of Minkowski, which was of importance for the formal development of the theory of relativity, does not lie here. It is to be found rather in the fact of his recognition that the four-dimensional space-time continuum of the theory of relativity, in its most essential formal properties, shows a pronounced relationship to the three-dimensional continuum of Euclidean geometrical space. … In order to give due prominence to this relationship, however, we must replace the usual time co-ordinate t by an imaginary magnitude $\sqrt{-1}ct$ proportional to it. Under these conditions, the natural laws satisfying the demands of the (special) theory of relativity assume mathematical forms, in which the time co-ordinate plays exactly the same ro1e as the three space coordinates. Formally, these four co-ordinates correspond exactly to the three space co-ordinates in Euclidean geometry. It must be clear even to the non-mathematician that, as a consequence of this purely formal addition to our knowledge, the theory perforce gained clearness in no mean measure (Einstein 1921, 38–39).

## 8.5 EINSTEIN AGAINST GEOMETRIZATION

One needs only a cursory review of Einstein's writings on general relativity to see that he privileged analytic expressions over geometric conceptions. Geometric curvature is rarely mentioned. The transformations of analytic expression get thorough attention. Thus, it is the "Riemann-Christoffel tensor" whose significance lies in its control over which analytic expressions can be transformed into which others. It is not, as it soon came to be called,[39] the "Riemann-Christoffel *curvature* tensor" that measures the geometric curvature of space or spacetime.

It was plausible that this emphasis in Einstein's writing merely reflected a disinterest in geometric conceptions. Lehmkuhl (2014) has shown that this is not so. In several writings, Einstein took issue with the idea that general relativity had in some sense "geometrized" gravity. For example, Einstein wrote a review of Émile Meyerson's *La deduction relativiste*. In it, Einstein rebuked Meyerson for seeing in relativity a program of "the reduction [*Zurückführung*] of all concepts of the theory to spatial, or rather geometrical, concepts" (Lehmkuhl 2014, 318).[40] On the contrary, Einstein found that the idea of physics tracing back ["*führe…zurück*"] to geometry has no clear meaning. He concluded:

> The fact that the metric tensor is denoted as "geometrical" is simply connected to the fact that this formal structure first appeared in the area of study denoted as "geometry". However, this is by no means a justification for denoting as "geometry" every area of study in which this formal structure plays a role, not even if for the sake of illustration, one makes use of notions which one knows from geometry. Using a similar reasoning Maxwell and Hertz could have denoted the electromagnetic equations of the vacuum as "geometrical" because the geometrical concept of a vector occurs in these equations.

Einstein offered an alternative conception. The goal of his relativistic work is unification, such as the unification of gravity and electricity then attempted by Weyl and Eddington (and of course Einstein's own efforts at a unified field theory). Lehmkuhl illustrated how general relativity itself was regarded by Einstein as unifying earlier ideas of gravity and inertia.

Einstein's treatment of the same issue a year earlier shows his impatience with the idea of geometrization. Hans Reichenbach had been working on Hermann Weyl's efforts to extend general relativity to electromagnetism.[41] In a letter to Einstein of April 24 1926, Reichenbach had argued that the "geometrical representation of electricity" is "nothing more than a graphical representation and amounts, therefore, to nothing physical at all" (Kormos et al. 1918, 412). Einstein agreed and reinforced Reichenbach's deflation in a letter to him of April 8, 1926:

> You are completely right. It is preposterous to believe that "geometrization" means something essential. It is for me a kind of novice aid [*Eselsbrücke*] for the discovery of numerical laws. Whether one attaches a "geometrical" conception to a theory is an inessential private matter. The essential thing with Weyl is that he subjects the formulas to a new condition ("gauge invariance") as well as invariance with respect to [spacetime] transformations (Kormos et al. 2018, 424).

---

39   For example in Bergmann (1942. Part II, Ch. XI).

40   The French text for these passages is in Einstein and Metz (1928, 164–165). Lehmkuhl (2014, 318) has located Einstein's original German text that was translated into French by Metz. My extracts are from Lehmkuhl's translation of this German text.

41   For further details of Reichenbach's negative views about geometrization, see Giovanelli (2022; 2023, §2) and Lehmkuhl (2014).

Einstein's reaction, once again, suppresses the geometry in favor of privileging the transformational behavior of formulae.

Einstein's description of geometrization as an "*Eselsbrücke*" requires some reflection. It is literally "donkey bridge" and its meaning in Einstein's time and now is an artificial aid, such as (presently) a mnemonic. It has a geometric connection that may be important. It is the German translation of the Latin *pons asinorum*. The term designates Theorem 5 in Book 1 of Euclid's *Elements*. It owes its name to the bridge-like form of the figure and its reputation as the theorem whose understanding would first defeat someone without good geometrical abilities. A geometrical donkey could not pass over the bridge to the rest of Book 1. To reflect all these considerations, the term is translated here as "novice aid."

The sense in which Einstein took geometrical considerations to be such a novice aid might be well captured by his use in his 1917 cosmology paper, described above, of a fictional four-dimensional space to introduce the line element of the spherically curved three-dimensional space of his new cosmology. It is an artifice that speeds us to the result, but is no basis for concluding that there is factually a higher dimensional space into which the three-dimensional space curves. He wrote of it:

> The four-dimensional Euclidean space from which we started serves only to define conveniently [*zur bequemen Definition*] our hypersurface (1917, 149).

The same point is made again of this construction in Einstein's Princeton lectures:

> The aid of a fourth dimension has naturally no significance except that of a mathematical artifice (1923, 115, fn).

The notion of geometrization assailed in these exchanges with Meyerson and Reichenbach by Einstein may well be stronger than that of the synthetic conception outlined in this paper. If Einstein's "reduction" [*Zurückführung*] has the same meaning as "reduction" in the more recent literature in philosophy of physics, then the reduction of gravitation to geometry would be akin to Maxwell's reduction of light to electromagnetic radiation. If that notion is what Einstein intended, then he was well within his rights to reject it.

The modern, synthetic conception of geometrization is weaker. It is just that the interval s and the metric tensor $g_{ik}$ are properties of an independently existing structure. It is called "geometrical" since it generalizes the structure that is the subject of Euclidean geometry. Einstein may well be right that a careful examination would find this terminology to be meaningless. But that does not dismiss the idea of the structure itself and that it is the primary object of investigation in general relativity. His exchanges with Meyerson and Reichenbach gave Einstein the opportunity to recognize the structure and endorse the corresponding synthetic viewpoint. He did not. Instead, he recalled analytic expressions and their transformations.

# PART III. EINSTEIN, SINGULARITIES, AND ARBITRARINESS

## 9. SINGULARITIES IN MATHEMATICS AND PHYSICS

The term "singularity" has powerful connotations today. We believe that the entirety of our universe sprang explosively from a singular big bang. The greatest menace offered by our physics is the voracious singularity at the heart of a black hole. At both, we expect physical conditions so extreme as to outstrip all imagination. The term has proliferated in the popular conscience. "The singularity" is foretold as the moment when self-reinforcing technological change transforms human civilization in unforeseeable ways. The idea is yet another touched by the great John von Neumann. According to Stanislaw Ulam (1958, 7), he contemplated "…the ever accelerating progress of technology and changes in the mode of human life, which gives the appearance of approaching some essential singularity in the history of the race beyond which human affairs, as we know them, could not continue."

The apocalyptic or perhaps just radical aura surrounding the notion of a singularity was not present in the mathematics and physics of Einstein's time. In technical contexts,[42] the notion figured primarily in mathematical work. There was no sense of a catastrophic threat or ominous breakdown. Rather a division into singular and non-singular behavior was part of the routine taxonomy of mathematical structures. To be non-singular was merely a frequent antecedent condition in useful theorems.

A result now familiar was also familiar then. As reported in Bocher (1907, 66),[43] in linear algebra, a linear transformation of variables is many-to-one and has no inverse if the determinant of the matrix of transformation coefficients is zero. Then, the transformation and the matrix are both said to be "singular." Of relevance to metrical geometry is a result shown later in Bocher (1907, Ch. X). A quadratic form of rank $r$, $\Sigma_1^n a_{ij} x_i x_j$, can always be reduced to a sum of squares by a non-singular transformation of the variables.

In this benign vein, talk of singularities was pervasive in the literature in differential geometry in the first decades of the twentieth century. Fowler gave a general characterization of them in curves in which the usually applicable, general treatments break down (his emphasis):

> These cases of exception correspond to exceptional points on the curve, usually such that there are only a finite number in any finite region, at which the curve has some peculiar property such as a stationary tangent, a point of inflexion, exceptionally high order contact with its circle of curvature or its envelope, etc. All such points at which the curve possesses peculiar properties may be considered to be *singular points* on the curve… (1920, 80).

He continued to note that it is usual to restrict the term to cases in which certain first derivatives vanish.

One index of the ubiquity of the notion in geometry is provided by Volume 3, Part 3, of the authoritative Teubner Encyclopedia, Meyer and Mohrmann, (1902–1927) that is devoted to differential geometry. The terms "singular" and "singularity" (*singulär, singularität*) appear 162 times in a volume of over 800 pages.[44] They are used to describe singular points and their higher dimensional analogs, singular lines and surfaces.

Discussion of singularities was almost entirely restricted to the mathematical literature. While they arose also in physical applications in field theories, their presence commonly produced little if any comment. When they were recognized, they were not found to be so troublesome as to require any remedial action. Chapter IX on Spherical Harmonics of Maxwell's *Treatise* begins with a lengthy section "On Singular Points at which the Potential becomes Infinite." In the simplest case, Maxwell (1873, 157) considered the potential $V$ at a place $r$ distant from electricity $e$ condensed to a point with spatial coordinates $(a, b, c)$. The potential $V$ is given by his equation "(1)" and is $V = e/r$. Maxwell reported the singular behavior:

> At the point $(a, b, c)$ the potential and all its derivatives become infinite, but at every other point they are finite and continuous, and the second derivatives of $V$ satisfy Laplace's equation (1873, 157).

---

42   The most common use of the term, however, was in non-technical discourse where it just designated a form of uniqueness. For example, Taylor's (1813, 261) inventory of synonyms records it in the context of human personality types as: "The word singularity is applied by a natural hyperbole to any rare form of behaviour; to any unusual degree of peculiarity."

43   The condition that matrices be non-singular is so important that the term "non-singular" appears 472 times in the text; and the term "singular" appears 142 times.

44   The volume consists of article spanning two decades. They are dated 1902, 1902, 1903, 1903, 1903, 1914, 1920, 1921 and 1923.

He then reflected on the troublesome character physically of this result:

> Hence, the value of $V$, as given by equation (1), may be the actual value of the potential in the space outside a closed surface surrounding the point $(a, b, c)$, but we cannot, except for purely mathematical purposes, suppose this form of the function to hold up to and at the point $(a, b, c)$ itself. For the resultant force close to the point would be infinite, a condition which would necessitate a discharge through the dielectric surrounding the point, and besides this it would require an infinite expenditure of work to charge a point with a finite quantity of electricity (1873, 157).

Maxwell's solution was, in effect, to ignore any oddity at the infinite point itself and confine consideration to the region surrounding it and approaching arbitrarily closely to it (Maxwell's emphasis):

> We shall call a point of this kind an infinite point of degree zero. The potential and all its derivatives at such a point are infinite, but the product of the potential and the distance from the point is ultimately a finite quantity e when the distance is diminished without limit. This quantity $e$ is called the *charge* of the infinite point.

Maxwell's tame treatment of singularities seems, however, to be the exception. They were otherwise largely ignored in the German electrodynamics literature. "Abraham-Foeppl" (1918) was the standard German language textbook on Maxwell's electrodynamics in Einstein's time. It made no mention of singular points in its fields. In its treatment of point sources (*Quellpunkte*) in Chapter 14, it noted that the radially-directed, vector force due to a point source of charge $e$ diminishes with distance $r$ from the source as $e/r^2$. The text remarks that this quantity "becomes infinite, if one enters the source point," but says nothing more. There is no suggestion that this infinity might be troublesome. Sommerfeld (1904–1922) is the volume of the Teubner Encyclopedia that treats electrodynamics. Articles from 1902 to 1906 occupy the first five hundred pages.[45] The term "singular" appears just twice in incidental roles.

I have found one exception from Einstein's time. Mie (1912/1913) developed a theory of matter based on a modification of Maxwell's electrodynamics. A major goal was to find a solution of the theory's equations for the static, spherically symmetric field for an electron. The bulk of the second part of Mie's paper is devoted to examining the singularities that appear in the static spherically symmetric solutions, especially at radial coordinate $r = 0$ and $r = \infty$. His goal, according to Pauli (1958, 192) was an everywhere regular solution, which he failed to recover.

---

45   The next article is Pauli's review of relativity theory of 1920 and then, finally, a 1921 article on the electron theory of metals.

Einstein surely knew of this work of Mie, a prominent physicist whose work was taken up by David Hilbert. Einstein entered into a correspondence with him in 1917 and 1918.[46] Mie's work is discussed briefly in Einstein's (1919, 349) early attempt to use gravitation theory in a theory of elementary particles and again in Einstein and Grommer (1927, 3). Einstein judged Mie's theory as unsatisfactory. I have found no evidence that Mie's treatment of singularities played any role in Einstein's thought. In any case, Einstein had no need of Mie's theory to suggest to him the idea of representing particles as singularities. Einstein (1909, 824–825) had already speculated that light quanta could be incorporated into electromagnetic theory if the quanta were themselves to singularities in the field.

## 10. EINSTEIN AGAINST AND FOR SPACETIME SINGULARITIES

Einstein's treatments of singularities poses two puzzles. The first—"against"—arises from the discussion of the last section. Until Einstein's work on general relativity, singularities had little presence in physical theorizing. They were certainly not the theoretical anathema portrayed by Einstein. Why did Einstein choose to vilify them?

The second puzzle—"for"—is why, in spite of his vilification of singularities, they made several appearances in his theorizing. This second puzzle attracted comment from Earman and Eisenstaedt:

> … it should be emphasised how flexible and adaptable, how much of an opportunist Einstein was with respect to singularities. At first singularities were to be regarded with such horror that the presence of a singularity in the De Sitter solution was sufficient to disqualify it from serious consideration. But shortly thereafter singularities that correspond to mass concentrations are to be welcomed (at least on a provisional basis), for then De Sitter's solution can be accommodated without compromising the precious Mach principle (1999, 194).

Elsewhere (1999, 186), they describe Einstein's use of singularities in his collaborative treatment of the problem of motion in Einstein, Infeld and Hoffmann (1938). Einstein is, they report, "always an opportunist" and the result is an unsuccessful "pact with the devil."

This section will offer solutions to both puzzles. To the first, Einstein's antipathy to singularities derived from an enduring and deep-seated presumption, fundamental to all his work in relativity. It is the requirement that arbitrariness must be eliminated from physical theory. To the second, we recall that general relativity was, for Einstein, only an intermediate theory on the way to his unified field theory. General relativity, by itself, did not represent matter adequately. Its singularities can mark where general relativity fails and the unified field would be needed to represent matter. Until that unified field theory was secured, the singularities could serve as a surrogate for that matter. Even then, Einstein's decision to admit singularities provisionally was made within the confines of the ultimate goal of eliminating arbitrariness. We shall see that the limited use of singularities was, in Einstein's estimation, the least arbitrary of the options available.

These issues are developed mostly within Einstein's search for his unified field theory. For helpful surveys of Einstein work on his unified field theory, see Sauer (2014) and van Dongen (2010).

---

46    Schulmann et al. 1998, Docs. 346, 348, 407, 410, 416, 421, 456, 460, 465, 470, 488, 493, 532.

# 10.1 AGAINST SINGULARITIES

Einstein did not initially harbor an aversion to singularities. Rather, his attitude agreed with the generally untroubled acceptance noted above in his contemporary literature that radially symmetric fields have a singularity at the source point. Such acceptance was implicit in Einstein's (1909, 824–825) speculation that light quanta may themselves be singularities of the field. Einstein himself made the connection to this earlier literature in a letter to Henrik Lorentz of May 23, 1909 (Klein et al. 1993, Doc 163):

> Rather I believe that light is grouped around a singular point in a way similar to how we are used to assuming for electrostatic fields.

He glossed the supposition to Jakob Laub in a letter of May 19, 1909 (Klein et al. 1993, Doc 161) as "Linear differential equations with singularities." The supposition of linearity is important. Einstein's light quantum hypothesis was that, thermodynamically, in the high frequency Wien regime, light quanta are mutually independent. This independence is directly recoverable if the field equations are linear, since multiple solutions can be superposed by simple addition.[47]

A change in Einstein's attitude coincided with the completion of his general theory of relativity and the move towards a unified field theory. It changed in time to ground Einstein's criticism of de Sitter's solution to his $\lambda$-augmented gravitational field equations. Prior to the singularity's recharacterization as a "mass-horizon," its presence was sufficient grounds for Einstein to dismiss the solution. His letter to de Sitter of July 31, 1917 (Schulmann et al. 1998, Doc. 366) began:

> However I may conceive it, I cannot ascribe any physical possibility to your solution. The difficulty has to do with the fact that in the (naturally measured) finite the $g_{\mu\nu}$ assume singular values.

We saw above that, in 1922, in response to a query from Hadamard, Einstein characterized the realization of the singular Schwarzschild radius as "an unimaginable misfortune [*malheur*] for theory…" In their joint paper, Einstein and Rosen (1935, 73) deplored singularities as nullifying the laws of a theory. (Their formulation is quoted in Subsection 10.3 below.) The search for singularity-free solutions was a prominent, stable feature of Einstein's attempts at a unified field theory. He wrote,

> The most important question related to the (strict) field equations is that of the existence of singularity-free solutions, which could represent electrons and protons (1930, 23).

Similarly, Einstein and Mayer wrote:

> According to our conviction, a satisfactory field theory must manage with a singularity-free description of the total field, therefore also of the field inside corpuscles (1932, 130).

In his popular, "Physics and Reality," Einstein expressed it this way:

> What appears certain to me, however, is that, in the foundations of any consistent field theory, the particle concept must not appear in addition to the field concept. The whole theory must be based solely on partial differential equations and their singularity-free solutions (1936, 306–307).

---

47    As Einstein and Infeld (1949, 210) note, the failure of linearity becomes important in Einstein's later attempts to represent particles by singularities, for otherwise interactions between particles would be impossible.

48    The series of dashes is an older German convention to mark the end of a section. I include the German here since the standard translation in the Schilpp volume is weak. It is: "The question is: What are the everywhere regular solutions of these equations?- - -"

We can gauge the importance to Einstein of these singularity-free solutions by his choice of the question that closes the final page of his "Autobiographical Notes" (Einstein 1949, 94).[48]

> Die Frage ist: Was für im ganzen Raume singularitatsfreie Losungen dieser Gleichungen gibt es?- - -
>
> The question is: what sort of singularity-free solutions are there of these equations?- - -

It is a poignant question, since it is written with the understanding that his decades of work towards his unified field theory were incomplete.

## 10.2 EINSTEIN AGAINST ARBITRARINESS

*That* Einstein opposed singularities in field theories is clear. What remains to be answered is the question of *why* he opposed them. The answer comes from an enduring supposition in Einstein's thought. We should be striving for physical theories free of arbitrariness, he held. Einstein's aversion to singularities is simply part of that project.

This supposition may seem less important compared to the more visible demands Einstein made on physical theories: the principles of relativity, Mach's principle and determinism in quantum systems. However, a closer look at all of these more visible demands shows that they share a common aversion to arbitrariness. This commonality is not emphasized in Einstein's earlier writings and it is quite plausible that he did not recognize its importance or even its presence. It emerges explicitly in Einstein's later writings. Presumably it did so since it served his needs in addressing the physical issues of his later theorizing. The abandoning of Mach's principle in his later years also made it possible for the aversion to arbitrariness that lay behind it to become a more explicit concern.

This aspect of Einstein's thought has attracted less attention from historians. An exception is Don Howard's (1992) study of the *Eindeutigkeit* [uniqueness] principle in philosophical work contemporary with Einstein. This demand for uniqueness, Howard argued, underlies Einstein's insistence that we not admit a violation of determinism in general relativity in the context of the "hole argument." A failure of uniqueness, we might note, amounts to the introduction of arbitrariness into our physical theorizing.

Perhaps the most explicit formulations came in his "Autobiographical Notes." Einstein listed two conditions for the evaluation of physical theories. The first was compatibility with empirical facts. The second he found much harder to articulate. We favor "naturalness" and "logical simplicity," where he used quotation marks for both terms in deference to the difficulty of precisely defining them. The related "inner perfection" of a theory, Einstein noted, is enhanced if the theory is not chosen arbitrarily:

> The following I reckon as also belonging to the "inner perfection" of a theory: We prize a theory more highly if, from the logical standpoint, it is not the result of an arbitrary choice among theories which, among themselves, are of equal value and analogously constructed (1949, 23).

A later version revealed that Einstein's discomfort with arbitrariness was so deeply rooted that he could provide no deeper foundation for it. We have come to the deepest of his convictions:

> … I would like to state a theorem which at present can not be based upon anything more than upon a faith in the simplicity, i.e., intelligibility, of nature: there are no *arbitrary* constants of this kind; that is to say, nature is so constituted that it is possible logically to lay down such strongly determined laws that within these laws only rationally completely determined constants occur (not constants, therefore, whose numerical value could be changed without destroying the theory)- - - (63, Einstein's emphasis).

Once we know to look for it, we find the elimination of arbitrariness figuring in many places in Einstein's thought. Here are some examples that, retrospectively, can be seen to share the commonality, although we cannot assume that this was explicitly recognized by Einstein at the time.

Einstein used the elimination of arbitrariness in characterizing the motivation for the 1905 special theory of relativity. Its key innovation was the recognition of the relativity of simultaneity; and that was possible, Einstein recalled,[49] only after an arbitrariness in earlier theories were identified:

> … all attempts to clarify this paradox [of the chasing of light] satisfactorily were condemned to failure as long as the axiom of the absolute character of time, viz., of simultaneity, unrecognizedly was anchored in the unconscious. Clearly to recognize this axiom and its arbitrary character really implies already the solution of the problem (1949, 53).

What became Mach's principle was, for Einstein, a way of eliminating the idea of inertial systems of reference or inertial spaces as primitive posits in physical theories prior to general relativity. They were, as Einstein put it, "*bloss* fingierte Ursache"—"*merely* fictitious causes" (1916, 771, his emphasis). Earlier, Einstein explained his concern as,

> What is unsatisfactory is that it remains unexplained how inertial systems could be distinguished from all other systems (1913, 1260).

That is, their designation is an arbitrary element in these prior theories. In a letter to Lorentz of January 15, 1915, he was forceful in his insistence on the elimination of this arbitrariness:

> Of two things $K_1$ and $K_2$ that are equally justified in their definitions, one is distinguished without a physical basis (that is in principle accessible to observation)—My confidence in the logical consistency of natural events bristles against this most strongly [*sträubt sich … aufs kräftig[st]e*]. In my view, a world picture that does without this sort of arbitrariness is preferable (Schulmann et al. 1998, Doc. 47).

This form of unacceptable arbitrariness arose in Einstein's cosmological reflections through the possibility of stipulating boundary conditions for the metric field at spatial infinity. It is worth examining his concerns here in more detail since he will elsewhere treat the arbitrariness of singularities as being of the same type.

In his cosmology paper (Einstein 1917), he considered the case in which the metric field might be stipulated to adopt special relativistic limiting values as spatial infinite is approached. This might be the case, for example, if the cosmology consisted of a central collection of masses in an otherwise empty space. Such boundary conditions, Einstein continued, would violate the principle of relativity, in so far as they introduce a preferred reference system; and it would be unstable according to statistical physics (1917, 147). The objection that attracted most discussion, however, was that it would violate what he would soon call Mach's principle:

> … we fail to comply with the requirement of the relativity of inertia. For the inertia of a material point of mass *m* (in natural measure) depends upon the $g_{\mu\nu}$; but these differ but little from their postulated values, as given above, for spatial infinity. Thus inertia would indeed be *influenced*, but would not be *conditioned* by matter (present in finite space). If only one single point of mass were present, according to this view, it would possess inertia, and in fact an inertia almost as great as when it is surrounded by the other masses of the actual universe (1917a, 183, Einstein's emphasis).

---

49   Complaints about the arbitrariness of theories appear elsewhere in Einstein's (1949) recollections, such as 31, 37 (arbitrariness of older field theories) and 55 (arbitrariness in assumption of existence of rigid bodies).

In his Princeton lectures, Einstein had put a quantitative measure on how much arbitrariness is introduced by stipulating these boundary conditions:

> The hypothesis that the universe is infinite and Euclidean at infinity, is, from the relativistic point of view, a complicated hypothesis. In the language of the general theory of relativity it demands that the Riemann tensor of the fourth rank $R_{iklm}$ shall vanish at infinity, which furnishes twenty independent conditions, while only ten curvature components $R_{\mu\nu}$, enter into the laws of the gravitational field. It is certainly unsatisfactory to postulate such a far-reaching limitation without any physical basis for it (1923, 110).

Einstein then gave a terser formulation:

> If we think these ideas consistently through to the end we must expect the *whole* inertia, that is, the *whole* $g_{\mu\nu}$ -field, to be determined by the matter of the universe, and not mainly by the boundary conditions at infinity (1923, 114, his emphasis).

## 10.3 AGAINST THE ARBITRARINESS OF SINGULARITIES

Why did Einstein so oppose singularities in his theories? Einstein and Rosen gave the answer. After recalling the presence of singularities in present theories, they wrote in a revealing passage:[50]

> … writers have occasionally noted the possibility that material particles might be considered as singularities of the field. This point of view, however, we cannot accept at all. For a singularity brings so much arbitrariness into the theory that it actually nullifies its laws. … Every field theory, in our opinion, must therefore adhere to the fundamental principle that singularities of the field are to be excluded (1936, 73).

Thus, elsewhere in his popular "Physics and Reality," after discussion the prospects of matter-free theories, Einstein continued:

> How are we to proceed from this point in order to obtain a complete theory of atomically constituted matter? In such a theory, singularities must certainly be excluded, since without such exclusion the differential equations do not completely determine the total field (1936, 312).

---

50    The ellipses contain an embarrassing oversight. Einstein and Rosen justify their conclusion by recalling a solution of the gravitational field equations with two singularities at relative rest and thus not able to represent particles that attract. It was an embarrassing oversight since Einstein had earlier determined that the solution's singularity filled an axis connecting the points. For details see Lehmkuhl (2019 , §5, fn. 61). This oversight may have strengthened Einstein and Rosen's resolve in choosing the strong verbiage of "nullify." However, the aversion to singularities did not weaken in later work, after Einstein presumably had the occasion to recognize the oversight.

Several pages later, he asserted the positive result:

> As a matter of fact, up to now we have never succeeded in a field-theoretical
> description of corpuscles free of singularities, and we can, *a priori*, say nothing
> about the behavior of such entities. *One thing*, however, is certain: if a field
> theory results in a representation of corpuscles free of singularities, then the
> behavior of these corpuscles in time is determined solely by the differential
> equations of the field (320, his emphasis).

If we reflect on familiar examples of singularities in spacetime theories, Einstein's point is
straightforward. Most simply, singular solutions are associated with parameters that must
be set externally. In modern cosmology, for example, to recover a specific matter content
of the universe, we need to augment a big bang cosmology by fixing the early density of
matter emerging from the initial singularity.[51] That fixing is done empirically by projecting
backwards from present conditions.

While singularities may introduce some arbitrariness,[52] uncontrolled by the theory's
differential equations, how did Einstein and Rosen come to conclude the far stronger claim
that they *nullify* the laws of a theory? The answer is that Einstein equated the introduction
of singularities into a theory with the arbitrary stipulation of boundary conditions. The
connection was made explicitly by Einstein in a later appendix on the theory of the non-
symmetric field, added to his Princeton lectures, *Meaning of Relativity*. The appendix is
poignant, since it concludes with a recognition by Einstein in his final years, that then
present mathematical methods were simply not adequate to developing his long-sought
goal: a theory of particles whose exact solutions are free from singularities. In the midst
of this concession, Einstein equated singularities with boundary conditions, which he
regarded as then ineliminable:

> A field theory is not yet completely determined by the system of field equations.
> Should one admit the appearance of singularities? Should one postulate
> boundary conditions? As to the first question, it is my opinion that singularities
> must be excluded. It does not seem reasonable to me to introduce into a
> continuum theory points (or lines, etc.) for which the field equations do not hold.
> Moreover, the introduction of singularities is equivalent to postulating boundary
> conditions (which are arbitrary from the point of view of the field equations) on
> 'surfaces' which closely surround the singularities. Without such a postulate the
> theory is much too vague. In my opinion the answer to the second question is
> that the postulation of boundary conditions is indispensable (1956a, 164).

First, we can see how this equation can be sustained. Consider the Schwarzschild line
element above

$$ds^2 = -\frac{1}{1 - \frac{2m}{r}} dr^2 - r^2(d\theta^2 + \sin^2\theta d\phi^2) + \left(1 - \frac{2m}{r}\right) dt^2. \tag{5}$$

---

51    Einstein, as we might expect, took a dim view of the singularity (1956a, 124): "The
introduction of a such a new singularity seems problematical in itself."

52    Perhaps Einstein also drew on the result that, in general, including singularities in a
field greatly increases the size of the solution space of its governing differential equation and
thus the extent of stipulations needed to identify the solution sought. The simplest example is
that Laplace's equation $\nabla^2\varphi = 0$, for a potential $\varphi$, has a unique solution $\varphi = 0$ if we assume $\varphi$
vanishes at spatial infinity. Non-trivial solutions are only possible if we introduce singularities
that play the role of charges. Einstein certainly knew this elementary result and used it in his
solving of his weak field gravitational field equations in Einstein (1913, 1259).

It represents a one-parameter family of line elements satisfying the usual set of symmetry conditions and boundary conditions. If the parameter $m$ is set to zero, we recover the line element of special relativity. Some other stipulation is needed to designate a different solution. We would now just specify a non-zero value for $m$. We could alternatively proceed as Einstein indicated. We could stipulate as another boundary condition that the sphere at radius $r + \varepsilon$ for all $t$, $\varepsilon > 0$, has the metrical component $g_{tt} = (1 - 2m/(r + \varepsilon))$.

Another way to see the similarity that Einstein does not give is merely to replace the radial coordinate $r$ by an impact coordinate $u = 1/r$. Then the Schwarzschild line element becomes[53]

$$ds^2 = -\frac{1}{u^4(1 - 2mu)}du^2 - \frac{1}{u^2}(d\theta^2 + \sin^2\theta d\phi^2) + (1 - 2mu)dt^2,$$

The usual boundary condition is that as $r \to \infty$, $g_{tt} \to 1$ and $g_{rr} \to -1$. This same boundary condition is now relocated to the origin at $u = 0$, where it reappears as a singularity in which $g_{uu} \to -\infty$.

If Einstein treated the introduction of a singularity in the same way as the boundary conditions at infinity in cosmology, we can see why the introduction of such singularities might nullify the laws, for that nullification occurred in the cosmology. If we posit special relativistic boundary conditions at infinity, in Einstein's example, that arbitrary boundary condition all but completely determines the inertial properties of bodies in the spacetime.

## 10.4 FOR SINGULARITIES: SINGULARITIES AS SURROGATES FOR MATTER

Given Einstein's view of the corrupting role of singularities, we must ask why, in some contexts, he seemed quite comfortable with singularities. Einstein's tolerance for these singularities is explained by two factors. First—to be discussed in this subsection—they were not to be taken seriously, physically, but merely marked where the source-free theory of spacetime failed and an unarticulated matter theory would take over. Second—to be discussed in the next subsection—Einstein tolerated singularities temporarily as the less arbitrary choice on the way to his long-term goal, a singularity free, unified field theory.

The use of singularities as surrogates for matter is a repeated device in Einstein's work. Einstein and Rosen (1936) identified the singularity in a uniformly accelerated coordinate system as representing matter. Einstein reconciled himself to the singularity in the de Sitter solution by identifying it as a matter concentration. Elsewhere, Einstein sought to recover the equations of motion of free, massive particles without positing the geodesic equation by investigating the motions associated with singularities according to his source-free gravitational field equations. An earlier attempt was in his collaboration with Jakob Grommer (Einstein 1927; Einstein and Grommer 1927). A later attempt came in his collaboration with Leopold Infeld and Banesh Hoffmann (Einstein, Infeld and Hoffmann 1938).

The justification for treating singularities as a surrogate for matter was simple: As long as the matter was spatially concentrated in a small volume, then the space outside this matter filled volume would be governed by Einstein's source-free gravitational field equations. The plausible supposition was that collapsing an already concentrated matter distribution into a singular point would not greatly alter the surrounding spacetime. Thus, the spacetime surrounding a singularity could be used as a serviceable approximation for the spacetime surrounding a concentrated matter distribution.

---

53    Using $dr = -(1/u^2)du$.

This justification was given by Hilbert in his lectures from the summer semester of 1916. He wrote:

> These mathematical difficulties [of solving the gravitational field equations] impede us, already f[or] e[xample] in the construction of a single, neutral point mass. If we could construct such a neutral mass and if we could know the course of the $<g_{\mu\nu}>$ in the neighborhood of this position, and if we let the neutral mass degenerate ever more towards a point mass, then the $g$ would turn out to be a *singularity* at this point. We would have to regard such [a thing] as *allowed* in the sense that the $g_{\mu\nu}$ outside the immediate vicinity of the singularity correctly returns the process truly realized in nature. Now we must have before us such a singularity in [line element (5)]. Furthermore, we can say that the construction of a neutral point mass, even if it becomes possible later, will turn out to be so complicated that, for the applications in which one does not consider the immediate vicinity of the point mass suffering from a singularity, it has become possible already now to calculate with sufficient accuracy the approximately correct gravitational potentials (1916, 253, his emphasis).

Hilbert included a similar, briefer assessment in his influential 1916 "Foundations of Physics":

> Although, according to my understanding, only regular solutions represent immediately the reality of the physical, fundamental equations, solutions with non-regular points are still directly an important mathematical means for approximating characteristically regular solutions. In this sense, according to the processes of Einstein and Schwarzschild, the non-regular measure [line element (5)] at $r = 0$ and $r = [2m]$ is to be seen as the expression for the gravitation of a centrally symmetrically distributed mass in the vicinity of the null point. … In the same sense, a point mass is also to be conceived as the boundary case of a certain distribution of electricity around a point. Indeed, I foresee in this place deriving the equations of motion of them from my physical fundamental equations (1916a, 70–71).

In keeping with Hilbert's view, it became standard in early expositions of general relativity to treat the singularity at the Schwarzschild radius as sufficiently problematic physically as to be physically impossible. The routine response was that real masses must always have a size larger than this radius. For example, after announcing that the singularity at the Schwarzschild radius cannot be eliminated by a coordinate transformation, Laue continued:

> One must indeed conclude from this that each mass $m$, if of spherical form, necessarily has a radius greater than that corresponding to the value $r = 2Cmc^2$ [Schwarzschild radius]. And in fact we so far know of no contradictory case, also not with atomic nuclei (1921, 215).

Eddington's text drew a similar conclusion:

> There is a singularity at $r = 2m$, so that the particle must have a finite perimeter not less than $4\pi m$ (1923, 186).

Einstein's attempts, in his collaboration with Grommer and with Infeld and Hoffmann, clearly depend on this orientation.[54] That is, they suppose that a singularity will behave sufficiently like a free mass to enable the recovery of the equations of motion of the free mass. Lehmkuhl (2017; 2019) has given an extensive analysis of Einstein and Grommer's collaborative work and of further issues surrounding it. He concluded that:

> … the singularity should be interpreted to signify a placeholder or a blind spot of the theoretical treatment, rather than something that should be interpreted literally, as referring and approximately true (2017, 212).

---

54    The cogency of Einstein's approach has been debated. Tamir (2012) has elaborated the critique that it is nonsensical to talk of the motion of a singularity, since a singularity is not a locus of events in the spacetime. Lehmkuhl (2017) has sought to vindicate Einstein's approach by arguing that the motions traced in Einstein's constructions do not need to be that of a true singularity but only a highly concentrated matter distribution.

Einstein, Infeld, and Hoffmann proceeded explicitly with the same understanding. They recognized that the surrogacy of singularities for masses is an empirical assumption that may fail:

> It is of significance that our equations of motion do not restrict the motion of the singularities more strongly than the Newtonian equations, but this may be due to our simplifying assumption that matter is represented by singularities, and it is possible that it would not be the case if we could represent matter in terms of a field theory from which singularities were excluded (1938, 66).

They identified the security of their results in the security of the theory in the surrounding matter- and singularity-free regions:

> For, although the equations of the field are undefined at the singularities, their validity in the regular region is sufficient to determine the motion of these singularities (80).

We see the same attitude elsewhere in Einstein's reaction to what became known as the "big bang" singularity in expanding universe cosmologies. Einstein and de Sitter (1932) had presented a simple expanding universe cosmology. In it, the scale factor increases with cosmic time $t$ as $t^{2/3}$, so that there is a curvature and matter density singularity at $t = 0$. In their paper, Einstein and de Sitter made no mention of this time dependence or of this initial singularity. Presumably, it did not attract their serious attention. The time dependency is only given later by Einstein in a little-known 1933 article on cosmology, reported by O'Raifeartaigh et al. (2015). There, Einstein presumed that a breakdown in the assumption of the uniform distribution of matter would mean that their equations no longer apply close to the initial time $t = 0$.

In an appendix to the 1937 second edition of his Princeton lectures, Einstein now included a more reflective assessment: The present general theory of relativity would likely break down in the early universe so that a singularity is not assured:[55]

> The theoretical doubts are based on the fact that for the time of the beginning of the expansion the metric becomes singular and the density, $\rho$, becomes infinite. In this connexion the following should be noted: The present theory of relativity is based on a division of physical reality into a metric field (gravitation) on the one hand, and into an electromagnetic field and matter on the other hand. In reality space will probably be of a uniform character and the present theory be valid only as a limiting case. For large densities of field and of matter, the field equations and even the field variables which enter into them will have no real significance. One may not therefore assume the validity of the equations for very high density of field and of matter, and one may not conclude that the 'beginning of the expansion' must mean a singularity in the mathematical sense. All we have to realize is that the equations may not be continued over such regions (Einstein 1956a, 129).

---

55    A footnote a few pages earlier gave a terser assessment (Einstein 1956a, 124): "It may be plausible that the theory is for this reason inadequate for very high density of matter. It may well be the case that for a unified theory there would arise no singularity."

## 10.5 SINGULARITIES AS THE LESSER ARBITRARINESS

In using singularities as surrogates for matter, Einstein had not abandoned his enduring goal of eliminating arbitrariness from his physical theorizing. His goal remained a unified field theory of matter in which particles were represented within regular solutions, free from the arbitrariness of singularities. We saw, above, through the 1936 proposal of Einstein-Rosen bridges, that Einstein was even willing, at least temporarily, to alter his source-free gravitational field equations if it would lead to a singularity-free account of particles.

The enduring difficulty for Einstein was that this prized unified field remained tantalizingly beyond his reach. If he could not realize a theory free of arbitrariness, then, as a practical matter, he could seek to reduce the arbitrariness in his theories. His treatment of singularities conformed with this practice.

Einstein's assimilation of the singularity in de Sitter's cosmological solution as a mass horizon lay loosely within this conception. It was not a problem he had sought. De Sitter had pressed it on him. The singularity arose, we saw, through Einstein seeking a static gravitational field within de Sitter's solution. Then, through the concept of the mass horizon, Einstein could conform the singularity with Mach's principle and thereby to his goal of limiting arbitrariness, even if the de Sitter solution itself was not Einstein's choice for a cosmology.

The project of a more sustained investigation of singularities that was chosen by Einstein concerned the use of singularities to recover the motion of free bodies. There is some evidence that Einstein came to this use of singularities as surrogates for matter after an early reluctance. In the text of his Princeton lectures of 1921, he reflected on the failure of then present theories to produce a serviceable theory of the inner structure of charged particles. A footnote abjured the use of singularities to represent the charged particles:

> It has been attempted to remedy this lack of knowledge by considering the charged particles as proper singularities. But in my opinion this means giving up a real understanding of the structure of matter. It seems to me much better to give in to our present inability rather than to be satisfied by a solution that is only apparent (Einstein 1923, 55).

These hesitations were gone by the time of his collaboration with Grommer.[56] Their joint paper, Einstein and Grommer (1927, 3) described the state of electrodynamics and other field theories as disturbing ("*störend*") and blemished ("*Schönheitsfehler*") in having two foundational laws. A partial differential equation governs the fields and a total differential equation governs the motions of particles. While arbitrariness is not mentioned explicitly, the goal is to reduce the arbitrariness by eliminating one of the foundational laws.

Einstein and Grommer then consider whether the same "duality," as they call it, is to be found in general relativity. They described three approaches. The first was just to duplicate the duality by positing, as an independent law, that material points move on geodesics. The second was to introduce an energy tensor for matter. It follows from the gravitational field equations that such a tensor must obey a conservation law in the form of its vanishing divergence. They abandoned this approach, they reported, since it had been unable to account for elementary particles using continuous fields; and for other unspecified reasons. The third approach was to conceive of elementary particles as singularities. In his review of them, Lehmkuhl called the second and third the "*T* approach" and the "vacuum approach" (2017, 1204).

---

56    Lehmkuhl (2019, §4) attributes a leading role in the transition to Einstein's correspondence with the physicist, Yuri Rainich.

This third approach is the one Einstein and Grommer pursued. They summarized their project in emphasized text:

> *However, it has turned out that the law of motion of the singularities is completely determined by the field equations and by the character of the singularities,* without needing supplementary assumptions. To show this is the goal of the present investigation (Einstein and Grommer 1927, 4).

If Einstein and Grommer succeed in this goal, they have reduced the arbitrariness in the duality of fundamental laws.

A little over a decade later, Einstein's collaborative efforts with Infeld and Hoffmann proceeded on the same basis. This time, they rejected Lehmkuhl's "*T* approach" because of the uncontrolled arbitrariness it introduces. The paper states this in its first sentences:

> In this paper we investigate the fundamentally simple question of the extent to which the relativistic equations of gravitation determine the motion of ponderable bodies.
>
> Previous attacks on this problem … have been based upon gravitational equations in which some specific energy-momentum tensor for matter has been assumed. Such energy-momentum tensors, however, must be regarded as purely temporary and more or less phenomenological devices for representing the structure of matter, and their entry into the equations makes it impossible to determine how far the results obtained are independent of the particular assumption made concerning the constitution of matter (Einstein, Infeld, Hoffmann 1938, 65).

That is, choosing a specific stress-energy tensor means that the final results are not general, but depend on the particular form of matter represented; and these particulars may in turn not be fundamental, but merely reflect the superficial appearance of the matter at hand ("phenomenological"). This hesitation over the *T* approach had already been reported by Einstein in his 1936 "Physics and Reality." There he expressed it in the form of a celebrated metaphor:

> It [general relativity with a stress energy tensor *T*] is sufficient-as far as we know-for the representation of the observed facts of celestial mechanics. But it is similar to a building, one wing of which is made of fine marble (left [geometric] part of the equation), but the other wing of which is built of low-grade wood (right [matter] side of equation). The phenomenological representation of matter is, in fact, only a crude substitute for a representation which would do justice to all known properties of matter (1936, 311).

In his "Autobiographical Notes," Einstein later summarized how the project had eliminated an arbitrariness from general relativity that had to be removed:

> In the relativistic theory of gravitation, it is true, the law of motion (geodetic line) was originally postulated independently in addition to the field-law equations. Afterwards, however, it became apparent that the law of motion need not (and must not) be assumed independently, but that it is already implicitly contained within the law of the gravitational field (Einstein 1949, 79).

Einstein both sought to eliminate singularities from his theorizing and, at the same time, found them useful in theorizing. We may well wonder if Einstein recognized that this could give his theorizing a capricious appearance. He did recognize the problem and we have his answer, again in his "Autobiographical Notes":

> Now it would of course be possible to object: If singularities are permitted at the positions of the material points, what justification is there for forbidding the occurrence of singularities in the rest of space? This objection would be justified if the equations of gravitation were to be considered as equations of the total field. [Since this is not the case], however, one will have to say that the field of a material particle may the less be viewed as a *pure gravitational field* the closer one comes to the position of the particle. If one had the field-equation of the total field, one would be compelled to demand that the particles themselves would *everywhere* be describable as singularity-free solutions of the completed field-equations. Only then would the general theory of relativity be a *complete* theory (Einstein 1949, 81, Einstein's emphasis).

Lehmkuhl's concluding remarks give a quite serviceable recounting of Einstein's answer, drawing on Einstein's earlier work:

> Finally, we now understand how Einstein could have allowed for singularities to account for matter in GR, yet be adamant that no singularities were allowed to occur in the sought-after unified field theory. It was the hybrid character of GR that allowed for this double standard: it was exactly because it was not supposed to be an adequate theory of matter that it was acceptable to allow for singularities as place-holders of matter. But it was not acceptable to allow for singularities in the domain about which GR was supposed to be fundamental, correct: regions of spacetime with only gravitational fields, free of matter. Accordingly, a unified field theory of gravity and matter would have to live up to these latter, stricter standards: no singularities anywhere (2019, 189).

## 11. CONCLUSION

In their synoptic analysis of Einstein's treatments of spacetime singularities, Earman and Eisenstaedt, in a moment of exasperation, prescribe the challenge taken up in this paper: "But as we will see, much of the early work on spacetime singularities seemingly defies explanation" (1999, 193). What I hope this analysis has demonstrated is that Einstein's treatment of singularities was not capricious and ill-informed. Rather, it was part of a well thought out research program that did not turn out to be productive.

Einstein followed an established tradition in mathematics that took analytic expressions to be the primary subjects of analysis; geometric notions play only an heuristic role. A geometric structure could suggest corrections to the analytic expressions, but those corrections need only to be accepted if they conformed with other physical facts. In both cases of the singularities Einstein identified at the Schwarzschild radius and in the de Sitter solution, the geometry suggested a correction in which events distinct in the analytic expressions were to be identified. From the geometric perspective, the coordinate systems used had "gone bad." However, both corrections required Einstein to give up something he believed was required by the physical facts of the two cases, the static character of his line elements. The analytic expressions prevailed and the singularities remained.

In retrospect, the more productive approach would have been to accept the geometric correction and discard the static requirement as Einstein had formulated it. That this approach would prove to be more productive could not be known a priori. It is only with the later advances in physics and astrophysics that its superiority becomes apparent. Contrary to Einstein's supposition, there is no pathology in a Schwarzschild spacetime at the Schwarzschild radius. Rather, that radius merely marks an event horizon. The physics within it has attracted considerable attention. Similarly, de Sitter's solution turns out not to represent a static spacetime. The full de Sitter hyperboloid recovered from the geometry is the simplest case of an expanding universe, whose expansion is driven by the cosmological constant $\lambda$. In present cosmologies, it may prove to be the fate of our universe in the distant future.

Puzzlement over how Einstein could neglect these geometrical directives has obscured what may well be the more interesting question historically: How did singularities play a role in Einstein's research? There was little concern prior to Einstein about singularities in physical theories. It was Einstein who identified them as troublesome and developed a program of research that demanded their elimination. His goal was distinctively and even idiosyncratically Einstein: an elimination of arbitrariness from physical theory. His fiercely defended and then abandoned Mach's principle was just one instrument serving this larger goal. Singularities, Einstein argued, introduced unacceptable arbitrariness into physical theories, comparable to that introduced by arbitrary boundary conditions. They had to go.

While never abandoning the hope for a unified field theory, free of singularities, Einstein sought to minimize the arbitrariness in his physical theories as an intermediate towards that final theory. This minimization underwrote Einstein's use of singularities to represent point masses in his attempts to identify the equations of motion that governed them. While these singularities would be unacceptable in his final, unified field theory, Einstein judged them to introduce less arbitrariness into the analysis than the available alternatives: simply positing the geodesic law as an independent law; or seeking an energy tensor to represent matter separately from the metric field.

## APPENDIX A: EINSTEIN AND GROSSMANN GENERALIZE THE TERM "TENSOR."

Where Ricci and Levi-Civita (1900) had used the terms "covariant and contravariant systems" (*systèmes covariants et contrevariants*), Einstein and Grossmann used the term "tensor" as a generic replacement for "system" and applied it in the now familiar way to all systems transforming linearly under first derivatives of the coordinates (1913, 25).[57] That this was a novel use of the term "tensor" was noted in contemporary literature, such as Budde (1914, 246) and later by Veblen (1927, 28). How Einstein and Grossmann chose this idiosyncratic substitution remains an open question. The term tensor was used by Hamilton in his work on quaternions as a measure of the unsigned magnitude of quantities. It was chosen by Hamilton to reflect the idea that a length in space is extended if a tension is applied (1853, 56–58). This seems to have been the common use of the term in mathematics in the nineteenth century. Föppl's introduction to Maxwell's electrodynamics followed Hamilton's usage (Föppl 1894, 6). "Tensor" designated the scalar magnitude of vectors. Since this scalar magnitude is of great importance in Maxwell's theory, it is not surprising that the term appears 72 times in Föppl's volume. Given the youthful

---

57    Reich (1994) is an expansive and detailed study of the development of what became known as the "tensor calculus" and also of the introduction and use of the term "tensor."

Einstein's early fascination with Maxwell's theory, it is quite likely that he had contacted the volume. It was the premier German language exposition of Maxwell's theory during Einstein's youth.[58] This prolific use of the term "tensor" did not survive long in treatises on electrodynamics. In Abraham's (1904) "completely reworked" (*vollständig umgearbeitet*) second edition of Föppl's 1894 volume, this use of tensor is replaced by the more familiar "scalar" (*Skalar*) (4–6). The term "tensor" played only a minor role in the new exposition through the notion of a "tensor triple" (*Tensortripel*), which corresponds to the modern notion of a 3x3 Cartesian tensor. It figured in rotational mechanics (§14) and fluid flow (§17). This novel use of the term "tensor" had been introduced explicitly by Voigt (1898) in his work on the mechanics of crystals. He recognized (vi, 20) the prior use of the term in the theory of quaternions but felt that the extension to the multi-component quantity was admissible, since this quantity would be used to represent stresses and strains in materials three-dimensional materials.

## APPENDIX B: A FANCIFUL FIELD THEORY

The line element

$$ds^2 = dr^2 + r^2 d\theta^2, \tag{B1}$$

with $0 \leq r < \infty$ and $0 \leq \theta < 2\pi$, where we assume the expression cyclic in $\theta$, such that, for each $r$, $\theta$ and $\theta + 2\pi$ designate the same point. It is one of the simplest cases in which we routinely judge that a coordinate system has "gone bad." For, when this line element represents the metrical geometry of a Euclidean surface, the one point at the origin at $r = 0$ is assigned all the angular coordinates $0 \leq \theta < 2\pi$. Correspondingly, we have singular behavior in the coefficients of the metric tensor. While $g_{\theta\theta} = 0$, we have the more troublesome $g^{\theta\theta} = \infty$. The singularity is dismissed as a mere coordinate artefact.

Are we compelled to this dismissal? Might $r = 0$ identify an infinity of distinct points, according to the different values of $0 \leq \theta < 2\pi$? The example of the Euclidean surface is so familiar that it is easy to assume that such an identification is simply a novice blunder in mathematics whenever the analytic expression (B1) appears. Familiarity is not the same as mathematical necessity. That the coordinate system in an analytic expression of the form (B1) has "gone bad" at $r = 0$ is not a mathematical necessity. Whether it has gone bad depends on how the variables are interpreted physically and, in this case, the physical assumption that a Euclidean surface is represented.

To see that the physical interpretation of the variables really does control whether $r = 0$ is a singularity in expression (B1), we merely need to concoct a physical scenario in which $r = 0$ does represent a singularity in the physics. The following fanciful physical theory is not intended as a serious physical theory but merely to demonstrate that, under some interpretations, $r = 0$ in (B1) is singular.

The theory is set in a two-dimensional space that is topologically $\mathbb{R}^{\geq 0} \times \mathbb{S}$, that is, a truncated cylinder, shown in Figure 11. No presumption is made about the spatial metric for the space. It will not be needed. What matters is that each coordinate pair $(r, \theta)$ identifies a unique point in the space, including when $r = 0$. There are point charges in the space that interact through retarded potentials. That is, the action of one charge upon another is delayed by a time that is a function of the distance between the charges. In such theories, the retardation time is commonly isotropic in distance and is given by (distance)/$c$.

---

58    Frank's (1979, 38) biography asserts positively that Einstein did read the Föppl volume.

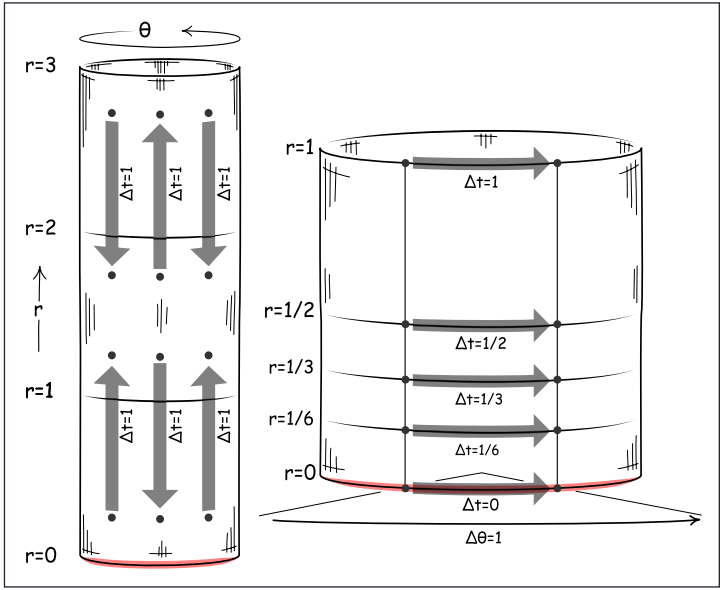

In this fanciful theory, the retardation times are defined in terms of coordinate differences and are anisotropic. In the *r* coordinate direction, the retardation time *dt* for a coordinate differential *dr*, with *dθ* = 0, is just *dt* = *dr*. In the *θ* coordinate direction, the retardation time *dt* for a coordinate differential *dθ*, with *dr* = 0, is just *dt* = *rdθ*. We *stipulate* that the general case is governed by the line element of the same form as (B1)[59]

$$dt^2 = dr^2 + r^2 d\theta^2. \tag{B2}$$

This line element specifies a temporal distance between points in the space that mimics a spatial, metrical distance.

As *r* approaches 0, the retardation times exhibit extreme behavior. That is, if we pick two points separated by $\Delta\theta > 0$, with $\Delta r = 0$, the retardation time decreases with *r* as

$$\Delta t = r\Delta\theta$$

as shown in Figure 11 for $\Delta\theta = 1$.

At *r* = 0, the retardation time is $\Delta t = 0$. The line element (B2) has become singular. As before, the associated metrical coefficients are $g_{\theta\theta} = 0$ and $g^{\theta\theta} = \infty$. However, we cannot escape the singularity by declaring that the coordinates (*r*,*θ*) have "gone bad" at *r* = 0. By stipulation, they have not. Indeed, the singularity is an essential part of the physics. For, while is it mathematically inconvenient, the singularity has a natural physical interpretation: At *r* = 0, the interaction has become an instantaneous action at a distance with zero retardation time.

## ACKNOWLEDGMENTS

My special thanks are owed to Harrison Payne, who read and commented extensively on drafts of this text. I thank Diana Kormos Buchwald, Michel Janssen, and Dennis Lehmkuhl for helpful comments.

---

59    This formula is not the same as assuming that the velocity components of the retarded action sum by the Pythagorean formula.

## COMPETING INTERESTS

The author has no competing interests to declare.

## AUTHOR AFFILIATIONS

**John D. Norton**  orcid.org/0000-0003-0936-5308
Department of History and Philosophy of Science, University of Pittsburgh, US

## REFERENCES

**Abraham, Max**, and **August Föppl**. 1904. *Theorie der Elektrizität*. Vol. 1. *Einführung in die Maxwellsche Theorie des Elektrizität*. 2nd ed. Leipzig: B. G. Teubner.

**Bergmann, Peter G.** 1942. *Introduction to the Theory of Relativity*. Englewood Cliffs, NJ: Prentice-Hall.

**Bocher, Maxime.** 1907. *Introduction to Higher Algebra*. New York: MacMillan.

**Budde, Emil.** 1914. *Tensoren und Dyaden im Dreidimensionalen Raum*. Braunschweig: Friedr. Vieweg & Sohn. DOI: https://doi.org/10.1007/978-3-663-07329-1

**Christoffel, Elwin Bruno.** 1869. "Ueber die Transformation der homogenen Differentialausdrücke zweiten Grades." *Journal für die reine und angewandte Mathematik* 70: 46–70. DOI: https://doi.org/10.1515/crll.1869.70.46

**de Sitter, Willem.** 1917. "On the Relativity of Inertia. Remarks Concerning Einstein's Latest Hypothesis." *Koninklijke Nederlandsche Akademie van Wetenschappen, Proceedings* 19: 1217–1225.

**de Sitter, Willem.** 1917a. "On Einstein's Theory of Gravity, and its Astronomical Consequences. Third Paper." *Monthly Notices of the Royal Astronomical Society* 78: 3–28. DOI: https://doi.org/10.1093/mnras/78.1.3

**Droste, Johannes.** 1916. "The Field of a Single Centre in Einstein's Theory of Gravitation, and the Motion of a Particle in that Field." *Koninklijke Nederlandsche Akademie van Wetenschappen, Proceedings* 19(1917): 197–215 (Communicated May 27, 1916).

**Earman, John**, and **Jean Eisenstaedt.** 1999. "Einstein and Singularities." *Studies in History and Philosophy of Modern Physics* 30: 185–235. DOI: https://doi.org/10.1016/S1355-2198(99)00005-2

**Eddington, Arthur.** 1923. *The Mathematical Theory of Relativity*. Cambridge: Cambridge University Press.

**Einstein, Albert.** 1909. "Uber die Entwickelung unserer Anschauungen über das Wesen und die Konstitution der Strahlung." *Physikalische Zeitschrift* 10: 817–825.

**Einstein, Albert.** 1911. "Zum Ehrenfestschen Paradoxon." *Physikalische Zeitschrift* 12: 509–510.

**Einstein, Albert.** 1913. "Zum gegenwärtigen Stande des Gravitationsproblems." *Physikalische Zeitschrift* 14: 1249–1262.

**Einstein, Albert.** 1914. "Die formale Grundlage der allgemeinen Relativitätstheorie." *Königlich Preussischen Akademie der Wissenschaften, Sitzungsberichte* 1914: 1030–1085.

**Einstein, Albert.** 1915. "Erklärung der Perihelbewegung des Merkur aus der allgemeinen Relativitätstheorie." *Königlich Preussische Akademie des Wissenschaften*. Berlin. 1915: 831–839.

**Einstein, Albert.** 1916. "Die Grundlage der allgemeinen Relativitätstheorie." *Annalen der Physik* 49: 769–822. DOI: https://doi.org/10.1002/andp.19163540702

**Einstein, Albert.** 1916a. "The Foundation of the General Theory of Relativity." In *The Principle of Relativity*, by H. A. Lorentz, A. Einstein, H. Minkowski, and H. Weyl, 111–164. Translated by W. Perrett and G. B. Jeffrey. Methuen, 1923. Dover, n. d.

**Einstein, Albert.** 1917. "Kosmologische Betrachtungen zur allgemeinen Relativitätstheorie." *Sitzungsberichte der Preussischen Akademie der Wissenschaften* 1917: 142–152.

**Einstein, Albert.** 1917a. "Cosmological Considerations on the General Theory of Relativity." In *The Principle of Relativity*, by H. A. Lorentz, A. Einstein, H. Minkowski, and H. Weyl, 177–188. Translated by W. Perrett and G. B. Jeffrey. Methuen, 1923. Dover, n. d.

**Einstein, Albert.** 1918. "Prinzipielles zur allgemeinen Relativitätstheorie." *Annalen der Physik* 55: 241–244. DOI: https://doi.org/10.1002/andp.19183600402

**Einstein, Albert.** 1918a. "Kritisches zu einer von Hrn. De Sitter gegebenen Lösung der Gravitationsgleichungen." *Sitzungsberichte der Preussischen Akademie der Wissenschaften* 1918: 270–272.

**Einstein, Albert.** 1919. "Spielen Gravitationsfelder im Aufbau der materiellen Elementarteilchen eine wesentliche Rolle?" *Sitzungsberichte der Preussische Akademie der Wissenschaften* 1919: 349–356.

**Einstein, Albert.** 1920. "Fundamental Ideas and Methods of the Theory of Relativity, Presented in Its Development." In *The Collected Papers of Albert Einstein. Vol. 7: The Berlin Years: Writings, 1918–1921*, edited by M. Janssen, R. Schulmann, J. Illy, C. Lehner, and D. Kormos Buchwald. Princeton: Princeton University Press, 2002.

**Einstein, Albert.** 1921. *Relativity: The Special and the General Theory*. Translated by R. W. Lawson. New York: Henry Holt & Co.

**Einstein, Albert.** 1923. *The Meaning of Relativity*. Princeton: Princeton University Press.

**Einstein, Albert.** 1927. "Allgemeine Relativitätstheorie und Bewegungsgesetz." *Sitzungsberichte der Preussische Akademie des Wissenschaften*. Phys.-Math. Klasse. 1927: 235–245.

**Einstein, Albert.** 1930. "Die Kombatibilität der Feldgleichungen in der einheitlichen Feldtheorie." *Sitzungsberichte der Preussische Akademie des Wissenschaften*. Phys.-Math. Klasse, 1930: 18–23.

**Einstein, Albert.** 1936. "Physics and Reality." In *Ideas and Opinions*, edited by Carl Seelig, translated and edited by Sonja Bergmann, 290–323. New York: Crown, 1954.

**Einstein, Albert.** 1939. "On a Stationary System With Spherical Symmetry Consisting of Many Gravitating Masses." *The Annals of Mathematics* 40(4): 922–936. DOI: https://doi.org/10.2307/1968902

**Einstein, Albert.** 1949. "Autobiographical Notes," pp. 2–94 and "Remarks Concerning the Essays Brought Together in this Co-operative Volume," pp. 665–88. In *Albert Einstein: Philosopher-Scientist*, edited by P. A. Schilpp, New York: MJF Books.

**Einstein, Albert.** 1956. "Autobiographische Skizze," In *Helle Zeit - Dunkle Zeit: In Memoriam Albert Einstein*, edited by Carl Seelig, 9–17. Braunschweig/Wiesbaden: Friedr. Vieweg & Sohn, reprint 1986. DOI: https://doi.org/10.1007/978-3-322-84225-1_2

**Einstein, Albert.** 1956a. *Appendix II: Relativistic Theory of the Non-Symmetric Field*." In *Meaning of Relativity*, 5th edition, 133–166. Princeton: Princeton University Press.

**Einstein, Albert**, and **Jakob Grommer**. 1927. "Allgemeine Relativitätstheorie und Bewegungsgesetz," *Sitzungsberichte der Preussische Akademie des Wissenschaften*. Phys.-Math. Klasse 1927: 2–13.

**Einstein, Albert**, and **Marcel Grossmann**. 1913. *Entwurf einer Verallgemeinerten Relativitätstheorie und einer Theorie der Gravitation*. Leipzig u. Berlin: B. G. Teubner.

**Einstein, Albert**, and **Leopold Infeld**. 1949. "On the Motion of Particles in General Relativity Theory." *Canadian Journal of Mathematics* 1: 209–241. DOI: https://doi.org/10.4153/CJM-1949-020-8

**Einstein, Albert**, **Leopold Infeld**, and **Banesh Hoffmann**. 1938. "The Gravitational Equations and the Problem of Motion." *Annals of Mathematics* 39: 65–100. DOI: https://doi.org/10.2307/1968714

**Einstein, Albert**, and **Walther Mayer.** 1932. "Einheitliche Theorie von Gravitation und Elektrizität," *Sitzungsberichte der Preussische Akademie des Wissenschaften*. Phys.-Math. Klasse 1932: 130–137.

**Einstein, Albert**, and **André. Metz.** 1928. "A propos de La Déduction Relativiste de M. Émile Meyerson." *Revue Philosophique de la France et de l'Étranger* 105: 161–166.

**Einstein, Albert**, and **Nathan Rosen.** 1935. "The Particle Problem in the General Theory of Relativity." *Physical Review* 48: 73–77. DOI: https://doi.org/10.1103/PhysRev.48.73

**Einstein, Albert**, and **Willem de Sitter.** 1932. "On the Relations Between the Expansion and the Mean Density of the Universe." *Proceedings of the National Academy of Sciences* 18: 213–214. DOI: https://doi.org/10.1073/pnas.18.3.213

**Eisenstaedt, Jean.** 1989. "The Early Interpretation of the Schwarzschild Singularity." In *Einstein and the History of General Relativity. Einstein Studies*, Vol. 1, edited by D. Howard and J. Stachel, 213–233. Boston: Birkhäuser.

**Eisenstaedt, Jean.** 1993. "Lemaître and the Schwarzschild Solution." In *The Attraction of Gravitation: New Studies in the History of General Relativity*, edited by J. Earman, M. Janssen and J. D. Norton, 353–388. Boston: Birkhäuser.

**Farwell, Ruth**, and **Christopher Knee.** 1990. "The Missing Link: Riemann's 'Commentatio,' *Geometry and Tensor Analysis.*" *Historia Mathematica* 17: 223–255. DOI: https://doi.org/10.1016/0315-0860(90)90002-U

**Finkelstein, David.** 1958. "Past-Future Asymmetry of the Gravitational Field of a Point Particle." *Physical Review*, 110: 965–967. DOI: https://doi.org/10.1103/PhysRev.110.965

**Föppl, August.** 1894. *Einführung in die Maxwellsche Theorie des Elektricität*. Leipzig: B. G. Teubner.

**Fowler, Ralph H.** 1920. *The Elementary Differential Geometry of Plane Curves*. Cambridge: Cambridge University Press.

**Frank, Philipp.** 1979. *Einstein: Sein Leben und seine Zeit*. Braunschweig/Wiesbaden: Friedr. Vieweg & Sohn. DOI: https://doi.org/10.1007/978-3-322-99011-2

**Gauss, Karl Friedrich.** 1828. *Disquisitiones Generales circa Superficies Curvas*. Gottingae: Typis Dieterichianis.

**Gauss, Karl Friedrich.** 1902. *General Investigations of Curved Surfaces of 1827 and 1825*. Translated by J. C. Morehead and A. M. Hiltebeitel. Princeton: Princeton University Library.

**Giovanelli, Marco.** 2013. "Erich Kretschmann as a proto-logical-empiricist: Adventures and misadventures of the point-coincidence argument." *Studies in History and Philosophy of Modern Physics*, 44: 115–134. DOI: https://doi.org/10.1016/j.shpsb.2012.11.004

**Giovanelli, Marco.** 2022. "Geometrization vs. Unification: the Reichenbach–Einstein Quarrel About the Fernparallelismus Field Theory." *Synthese* 200(213): 1–44. DOI: https://doi.org/10.1007/s11229-022-03531-2

**Giovanelli, Marco.** 2023. "Coordination, Geometrization, Unification. An Overview of the Reichenbach-Einstein Debate on the Unified Field Theory Program.' In *Philosophers and Einstein Relativity: The Earl Philosophical Reception of the Relativistic Revolution*, edited by Chiara Russo Krauss and Luigi Laino, 147–192. Springer. DOI: https://doi.org/10.1007/978-3-031-36498-3_6

**Hamilton, William.** 1853. *Lectures on Quaternions*. Dublin: Hodges and Smith.

**Hermann, Robert.** 1975. *Ricci and Levi-Civita's Tensor Analysis Paper*. Brookline, MA: Math Sci Press.

**Hilbert, David.** 1916. "Die Grundlagen der Physik II: Vorlesung, Göttingen, Sommersemester 1916." In *David Hilbert's Lectures on the Foundations of Physics 1915–1927*, edited by T. Sauer and U. Majer, Chapter 2. Berlin: Springer, 2009.

**Hilbert, David.** 1916a. "Die Grundlagen der Physik: zweite Mitteilung." *Nachrichten von der Königlichen Gesellschaft der Wissenschaften zu Göttingen, Mathematisch-physikalische Klasse* 1917: 55–76. (Presented 23 December 1916).

**Howard, Don.** 1992. "Einstein and Eindeutigkeit: A Neglected Theme in the Philosophical Background to General Relativity." In *Studies in the History of General Relativity: Einstein Studies*, Volume 3, edited by J. Eisenstaedt and A. J. Kox, 154–243. Boston: Birkhäuser.

**Janssen, Michel.** 2014. "'No Success Like Failure …': Einstein's Quest for General Relativity, 1907–1920." In *The Cambridge Companion to Einstein*, edited by M. Janssen and C. Lehner, Chapter 6. Cambridge: Cambridge University Press. DOI: https://doi.org/10.1017/CCO9781139024525

**Klein, Felix.** 1909. *Elementarmathematik vom höheren Standpuncte aus*. Teil II. Geometrie. Leibzig: B. G. Teubner.

**Klein, Felix.** 1918. "7. Mai. Klein, Über Einsteins kosmologische Ideen von 1917 (Mitteilung an die Berliner Akademie vom 8. Februar)" and "11. Juni. Klein, Fortsetzung des Vortrags vom 7. Mai," *Jahresbericht der deutschen Mathematiker-Vereinigung*. 27: 42–44.

**Klein, Felix.** 1918a. "Über die Integralform der Erhaltungsgesetze und die Theorie der räumlich- geschlossenen Welt." *Nachrichten von der Gesellschaft der Wissenschaften zu Göttingen, Mathematisch-Phylikalische Klasse* 1918: 394–423.

**Klein, Felix.** 1919. "Bemerkungen über die Beziehungen des de Sitter'schen Koordinatensystems B zu der allgemeinen Welt positiver Krümmung." *Koninklijke Akademie van Wetenschappen te Amsterdam, Proceedings* 21: 614–615.

**Klein, Felix.** 1921. *Gesammelte mathematische Abhandlungen*. Volume 1. Berlin: Julius Springer. DOI: https://doi.org/10.1007/978-3-642-51960-4

**Klein, Felix.** 1939. *Elementary Mathematics from an Advanced Standpoint: Geometry*. Translated by E. R. Hendrick and C. A. Noble, 3rd edition. New York: MacMillan; Dover, 2004.

**Klein, Martin J., A. J. Kox,** and **Robert Schulmann.** 1993. *The Collected Papers of Albert Einstein. Volume 5. The Swiss Years: Correspondence, 1902–1914*. Princeton: Princeton University Press.

**Kormos Buchwald, Diana, József Illy, Ze'ev Rosenkranz,** and **Tilman Sauer.** (eds.) 2012. *The Collected Papers of Albert Einstein: The Berlin Years: Writings & Correspondence, January 1922–March 1923*, Volume 13. Princeton: Princeton University Press.

**Kormos Buchwald, Diana, József Illy, A. J. Kox, Dennis Lehmkuhl, Ze'ev Rosenkranz,** and **Jennifer Nollar James.** 2018. *The Collected Papers of Albert Einstein: Volume 15: The Berlin Years: Writings & Correspondence, June 1925–May 1927*. Princeton: Princeton University Press.

**Kruskal, Martin.** 1960. "Maximal Extension of the Schwarzschild Metric." *Physical Review* 119: 1743–1745. DOI: https://doi.org/10.1103/PhysRev.119.1743

**Lanczos, Kornel.** 1921. "Bemerkung zu de Sitterschen Welt." *Physikalische Zeitschrift*. 23: 539–543.

**Landsman, Klaas.** 2021. *Foundations of General Relativity: From Einstein to Black Holes*. Nijmegen, Netherlands: Radboud University Press. DOI: https://doi.org/10.54195/EFVF4478

**Laue, Max.** 1921. *Die Relativitätstheorie: Zweiter Band: Die allgemeine Relativitätstheorie und Einsteins Lehre von der Schwerkraft*. Braunschweig: Friedr. Vieweg & Sohn.

**Lehmkuhl, Dennis.** 2014. "Why Einstein Did Not Believe that General Relativity Geometrizes Gravity." *Studies in History and Philosophy of Modern Physics* 46: 316–326. DOI: https://doi.org/10.1016/j.shpsb.2013.08.002

**Lehmkuhl, Dennis.** 2017. "Literal versus Careful Interpretations of Scientific Theories: The Vacuum Approach to the Problem of Motion in General Relativity." *Philosophy of Science* 84: 1202–1214. DOI: https://doi.org/10.1086/694398

**Lehmkuhl, Dennis.** 2019. "General Relativity as a Hybrid Theory: The Genesis of Einstein's Work on the Problem of Motion." *Studies in History and Philosophy of Modern Physics* 67: 176–190. DOI: https://doi.org/10.1016/j.shpsb.2017.09.006

**Lehmkuhl, Dennis.** 2022. "The Equivalence Principle(s)." In *Routledge Companion to the Philosophy of Physics*, edited by E. Knox and A. Wilson, Chapter 9. New York and London: Routledge. DOI: https://doi.org/10.4324/9781315623818-14

**Lemaître, Georges.** 1933. "L'Univers en Expansion." *Annales de la Société Scientifique de Bruxelles*. Series A, Vol. LIII, 51–85.

**Levi-Civita, Tullio.** 1926. *The Absolute Differential Calculus. (Calculus of Tensors).* Translated by M. Long. London and Glasgow: Blackie and Son.

**Lukat, Max.** 1910. *Luigi Bianchi: Vorlesungen über Differentialgeometrie.* 2nd ed. Leipzig and Berlin: B. G. Teubner.

**Maxwell, James Clerk.** 1873. *A Treatise on Electricity and Magnetism*. Volume 1. Oxford: Clarendon.

**Meyer, W. Fr.**, and **H. Mohrmann.** 1902–1927. *Encyklopädie der Mathematischen Wissenschafter mit Einschluss ihrer Anwendungen. Vol. 3. Geometrie. Part 3. Differentialgeometrie.* Leipzig: B. G. Teubner.

**Mie, Gustav.** 1912/1913. "Grundlagen einer Theorie der Materie." *Annalen der Physik* 37 (1912), 511–534; 39 (1912), 1–40; 40 (1913), 1–66. DOI: https://doi.org/10.1002/andp.19133450102

**Minkowski, Hermann.** 1908. "Space and Time," In *The Principle of Relativity*, by H. A. Lorentz, A. Einstein, H. Minkowski, and H. Weyl, 75–91. Translated by W. Perrett and G. B. Jeffrey.Methuen, 1923. Dover, n. d.

**Nordmann, Charles.** 1922. "Einstein: Expose et Discute sa Théorie." *Revue des deux Mondes* 9: 129–166.

**Norton, John D.** 1985. "'What was Einstein's Principle of Equivalence?' Studies in History and Philosophy of Science." (Reprint) In *Einstein and the History of General Relativity: Einstein Studies* Vol. 1, edited by D. Howard and J. Stachel, 5–47. Boston: Birkhauser, 1989.

**Norton, John D.** 1993. "General Covariance and the Foundations of General Relativity: Eight Decades of Dispute." *Reports on Progress in Physics* 56: 791–858. DOI: https://doi.org/10.1088/0034-4885/56/7/001

**Norton, John D.** 1999. "Geometries in Collision: Einstein, Klein and Riemann." In *The Symbolic Universe*, edited by J. Gray, 128–144. Oxford: Oxford University Press. DOI: https://doi.org/10.1093/oso/9780198500889.003.0008

**Norton, John D.** 2004. "Einstein's Investigations of Galilean Covariant Electrodynamics prior to 1905." *Archive for History of Exact Sciences* 59: 45–105. DOI: https://doi.org/10.1007/s00407-004-0085-6

**Norton, John D.** 2020. "Einstein's Conflicting Heuristics: The Discovery of General Relativity." In *Thinking about Space and Time: 100 Years of Applying and Interpreting General Relativity*. Einstein Studies, Volume 15, edited by C. Beisbart, T. Sauer and C. Wüthrich, 17–48. Cham, Switzerland: Birhäuser/Springer Nature. DOI: https://doi.org/10.1007/978-3-030-47782-0_2

**O'Raifeartaigh, Cormac, Michael O'Keeffe, Werner Nahm,** and **Simon Mitton.** 2015. "Einstein's Cosmology Review of 1933: a New Perspective on the Einstein-de Sitter Model of the Cosmos." *European Physical Journal H* 40: 301–335. DOI: https://doi.org/10.1140/epjh/e2015-50061-y

**Pauli, Wolfgang.** 1921. *Relativitätstheorie*. Leipzig, Berlin: B. G. Teubner. DOI: https://doi.org/10.1007/978-3-663-15829-5

**Pauli, Wolfgang.** 1958. *Theory of Relativity*. Translated by G. Field. London: Pergamon.

**Reich, Karin.** 1994. *Die Entwicklung des Tensorkalküls: vom absoluten Differentialkalkül zur Relativitätstheorie*. Basel: Springer. DOI: https://doi.org/10.1007/978-3-0348-8486-0

**Ricci Curbastro, Gregorio**, and **Tullio Levi-Civita.** 1900. "Méthodes de calcul différentiel absolu et leurs applications." *Mathematische Annalen* 54: 125–201. DOI: https://doi.org/10.1007/BF01454201

**Riemann, Bernhard.** 1854. "Ueber die Hypothesen, welche der Geometrie zu Grunde liegen," *Abhandlungen der Mathematischen Classe der Königlichen Gesellschaft der Wissenschaften zu Göttingen* 13(1868): 133–150.

**Riemann, Bernhard.** 1861. *Commentatio mathematica, qua respondere tentatur quaestioni ab Illma Academia Parisiensi propositae*. Ch. XXII, pp. 370–399, in *Bernhard Riemann's Gesammelte Mathematische Werke und Wissenschaftlicher Nachlass*. Leipzig: B. G. Teubner, 1876.

**Rindler, Wolfgang.** 1977. *Essential Relativity: Special, General and Cosmological*. 2nd edition. New York: Springer Verlag.

**Sauer, Tilman.** 2014. "Einstein's Unified Field theory Program." In *The Cambridge Companion to Einstein*, edited by M. Janssen and C. Lehner, Chapter 9. Cambridge: Cambridge University Press. DOI: https://doi.org/10.1017/CCO9781139024525.011

**Schwarzschild, Karl.** 1916. "Über das Gravitationsfeld eines Massenpunktes nach der Einsteinschen Theorie" *Preussische Akademie des Wissenschaften* 1916: 189–196.

**Schulmann, Robert, A. J. Kox, Michel Janssen**, and **József Illy.** 1998. *The Collected Papers of Albert Einstein*. Volume 8. *The Berlin Years: Correspondence, 1914–1918 Part A: 1914–1917. Part B: 1918*. Princeton: Princeton University Press.

**Smeenk, Christopher.** 2014. "Einstein's Role in the Creation of Relativistic Cosmology." In *The Cambridge Companion to Einstein*, edited by M. Janssen and C. Lehner, Chapter 7. Cambridge: Cambridge University Press. DOI: https://doi.org/10.1017/CCO9781139024525.009

**Sommerfeld, Arnold.** 1904–1922. *Encyklopädie der Mathematischen Wissenschafter mit Einschluss ihrer Anwendungen*. Vol. 5. *Physik*. Part 2. *Elektrizität und Optik*. Leipzig: B. G. Teubner. DOI: https://doi.org/10.1007/978-3-663-16016-8

**Synge, John L.** 1960. *Relativity: The General Theory*. Amsterdam: North-Holland.

**Szekeres, George.** 1959. "On the Singularities of a Riemannian Manifold." *Publicationes Mathematicae Debrecen* 7: 285–301. DOI: https://doi.org/10.5486/PMD.1960.7.1-4.26

**Tamir, Michael.** 2012. "Proving the Principle: Taking Geodesic Dynamics too Seriously in Einstein's Theory." *Studies in History and Philosophy of Modern Physics* 43: 137–154. DOI: https://doi.org/10.1016/j.shpsb.2011.12.002

**Taylor, William.** 1813. *English Synonyms Discriminated*. London: W. Pople.

**Tolman, Richard C.** 1934. *Relativity, Thermodynamics and Cosmology*. Oxford: Clarendon.

**Ulam, Stanislaw.** 1958. "John von Neumann." *Bulletin of the American Mathematical Society* 64: 1–49. DOI: https://doi.org/10.1090/S0002-9904-1958-10189-5

**Van Dongen, Jeroen.** 2010. *Einstein's Unification*. Cambridge: Cambridge University Press.

**Veblen, Oswald.** 1927. *Invariants of Quadratic Differential Forms*. Cambridge: Cambridge University Press.

**Voigt, Woldemar.** 1898. *Die fundamentalen physikalischen Eigenschaften der Krystalle in elementarer Darstellung*. Leipzig: Verlag von Veit & Comp. DOI: https://doi.org/10.1515/9783112439449

**Wald, Robert.** 1984. *General Relativity*. Chicago: University of Chicago Press.

**Weyl, Hermann.** 1918. *Raum-Zeit-Materie*. Berlin: Julius Springer. DOI: https://doi.org/10.1007/978-3-662-43111-5

**Weyl, Hermann.** 1919. "Über die statischen kugelsymmetrischen Lösungen von Einsteins 'kosmologischen' Gravitationsgleichungen." *Physikalische Zeitschrift*, 20: 31–34.

**Weyl, Hermann.** 1921. *Raum-Zeit-Materie.* 4th ed. Berlin: Julius Springer. DOI: https://doi.org/10.1007/978-3-662-02044-9

**Wright, J. Edmund.** 1908. *Invariants of Quadratic Differential Forms.* Cambridge: Cambridge University Press.